\documentclass{article} 

\usepackage[utf8]{inputenc}
\usepackage{longtable} 

\usepackage{eurosym}

\usepackage{url}
\usepackage{amssymb}
\usepackage{pifont}
\usepackage[square,numbers]{natbib}
\usepackage{fontawesome5}
\usepackage{colortbl}
\usepackage{multirow}
\usepackage{lipsum}
\usepackage{amsthm}
\usepackage{mathtools}
\usepackage{xfrac}
\usepackage[export]{adjustbox}
\usepackage{longtable}
\usepackage{makecell}
\usepackage{booktabs}
\usepackage[table,xcdraw]{xcolor}
\usepackage{amssymb}
\usepackage{fancyhdr} 

\usepackage[]{easyReview}

\definecolor{cai_primary}{HTML}{4C9A99}  

\definecolor{cai_secondary}{HTML}{307FE2}  
\definecolor{cai_accent}{HTML}{1D8348}  
\definecolor{cai_dark}{HTML}{3F4444}  
\definecolor{cai_light}{HTML}{F5F5F5}  
\definecolor{cai_purple}{HTML}{8A4FFF}  

\usepackage{graphicx}
\usepackage{subcaption}
\usepackage{verbatim}
\usepackage{placeins}
\usepackage{mdframed}
\usepackage{hyperref}
\hypersetup{
    colorlinks=true,
    urlcolor=cai_secondary,
    linkcolor=cai_secondary,    
    filecolor=cai_accent,      
    citecolor=cai_secondary,
}
\usepackage{fancyvrb}
\usepackage{fixltx2e}
\usepackage{bera}
\usepackage{pdfpages}

\usepackage{pgf-umlsd}
\usepackage{tikz} 
\usetikzlibrary{arrows.meta,positioning,shapes.geometric,calc,patterns} 
\usepackage{pgfplots} 
\pgfplotsset{compat=1.16} 
\usepackage{pgfplotstable} 
\usepackage{pgf-pie} 

\pgfplotsset{
  sharpbar/.style={
    ybar,
    draw=black,
    line width=1.2pt,
    rounded corners=0.5pt,
    preaction={
      transform canvas={xshift=4pt,yshift=-2pt},
      draw=none,
      fill=black!60,
      rounded corners=0.5pt
    }
  },
  sharpbarh/.style={
    xbar,
    draw=black,
    line width=1.2pt,
    rounded corners=0.5pt,
    preaction={
      transform canvas={xshift=4pt,yshift=-2pt},
      draw=none,
      fill=black!60,
      rounded corners=0.5pt
    }
  },
  sharpbarstacked/.style={
    ybar stacked,
    draw=black,
    line width=1.2pt,
    rounded corners=0.5pt,
    preaction={
      transform canvas={xshift=4pt,yshift=-2pt},
      draw=none,
      fill=black!60,
      rounded corners=0.5pt
    }
  }
}
\usepackage[draft=true]{minted}

\usepackage{fancyhdr}
\renewcommand{\headrulewidth}{0.4pt}
\renewcommand{\footrulewidth}{0.4pt}
\renewcommand{\headrule}{\hbox to\headwidth{\color{cai_primary}\leaders\hrule height \headrulewidth\hfill}}
\renewcommand{\footrule}{\hbox to\headwidth{\color{human_color}\leaders\hrule height \footrulewidth\hfill}}

\usepackage{dirtree} 
\usepackage{forest} 
\usepackage{wrapfig} 

\newcommand\timecolor[1]{\textcolor{gray!70}{#1}}

\usepackage{algorithm} 
\usepackage{listings}
\usepackage{geometry}
\usepackage{algpseudocode} 

\makeatletter
\newcommand\fs@nobottomruled{%
  \def\@fs@cfont{\bfseries}\let\@fs@capt\floatc@ruled
  \def\@fs@pre{\hrule height.8pt depth0pt \kern2pt}%
  \def\@fs@post{}
  \def\@fs@mid{\kern2pt\hrule\kern2pt}%
  \let\@fs@iftopcapt\iftrue}
\floatstyle{nobottomruled}
\restylefloat{algorithm}
\makeatother

\usepackage{caption,setspace}
\DeclareCaptionFormat{listing}{\rule{\dimexpr0.9\columnwidth+17pt\relax}{0.4pt}\par\vskip1pt\textcolor{cai_primary}{\faCode\ #1#2#3}}
\usepackage{titlesec}
\titleformat{\section}
  {\normalfont\Large\bfseries\color{cai_primary}}  
  {\thesection}  
  {1em}  
  {}  
  [\titlerule]  

\titleformat{\subsection}
  {\normalfont\large\bfseries\color{human_color}}
  {\thesubsection}
  {1em}
  {}

\titleformat{\subsubsection}
  {\normalfont\normalsize\bfseries\color{cai_dark}}
  {\thesubsubsection}
  {1em}
  {}

\newcounter{code}



\DeclareMathVersion{sans}
\SetSymbolFont{operators}{sans}{OT1}{cmbr}{m}{n}
\SetSymbolFont{letters}  {sans}{OML}{cmbrm}{m}{it}
\SetSymbolFont{symbols}  {sans}{OMS}{cmbrs}{m}{n}

\lstnewenvironment{sflisting}[1][]
  {\lstset{#1}\mathversion{sans}}{}

\usepackage[normalem]{ulem}

\definecolor{grayalias}{HTML}{3F4444}
\definecolor{bluealias}{HTML}{307FE2}
\definecolor{cai_color}{HTML}{4C9A99}  
\definecolor{caicolor}{HTML}{4C9A99}  

\definecolor{agentsred}{HTML}{FF6A4C}
\definecolor{agentsorange}{HTML}{F99244}
\definecolor{agentsblue}{HTML}{2D55CC}

\definecolor{agentsred2}{HTML}{993333}
\definecolor{agentsorange2}{HTML}{E67E22}
\definecolor{agentsblue2}{HTML}{2C3E50}

\definecolor{human_color}{HTML}{173C47}  
\definecolor{speed_color}{HTML}{00BCA2}  

\definecolor{cai_string}{HTML}{2E8B57}    
\definecolor{cai_comment}{HTML}{708090}   
\definecolor{cai_keyword}{HTML}{008080}   
\definecolor{cai_background}{HTML}{F5FFFA} 
\definecolor{cai_identifier}{HTML}{20B2AA} 
\definecolor{cai_number}{HTML}{2F4F4F}     
\definecolor{cai_frame}{HTML}{4C9A99}      

\definecolor{cai_string_muted}{HTML}{3D7A5F}    
\definecolor{cai_comment_muted}{HTML}{7F8C8D}   
\definecolor{cai_keyword_muted}{HTML}{4C9A99}   
\definecolor{cai_background_muted}{HTML}{F8FBFB} 

\definecolor{graph_teal}{HTML}{1ABC9C}      
\definecolor{graph_lightcyan}{HTML}{A8D5D5}  
\definecolor{graph_gray}{HTML}{E8E8E8}      
\definecolor{graph_navy}{HTML}{2C3E50}      
\definecolor{cai_identifier_muted}{HTML}{5F9EA0} 
\definecolor{cai_number_muted}{HTML}{45545E}     
\definecolor{cai_frame_muted}{HTML}{4C9A99}      

\usepackage{authblk}

\renewcommand\Affilfont{\small\normalfont}
\definecolor{cai_affil_color}{HTML}{3F8984} 

\makeatletter
\renewcommand\AB@affilsepx{\\\protect\Affilfont}
\let\orig@maketitle\maketitle
\renewcommand{\maketitle}{%
  \orig@maketitle%
  \vspace{-1.5em}%
  {\color{cai_color!30}\hrule height 0.5pt}%
  \vspace{1em}%
}
\makeatother

\title{\LARGE\textcolor{cai_primary}{\textbf{Cybersecurity AI: A Game-Theoretic AI \\ for \emph{Guiding} Attack and Defense}}}

\author[1]{Víctor Mayoral-Vilches}
\author[1]{María Sanz-Gómez}
\author[1]{Francesco Balassone}
\author[2]{Stefan Rass}
\author[1]{Lidia Salas-Espejo}
\author[2]{Benjamin Jablonski}
\author[1]{Luis Javier Navarrete-Lozano}
\author[1]{Maite del Mundo de Torres}
\author[1]{Cristóbal R. J. Veas Chavez}

\affil[1]{
    {\normalfont\textcolor{cai_color}{\textbf{Alias Robotics}}, Vitoria-Gasteiz, Álava, Spain\\
    {\tt\footnotesize\textcolor{cai_color}{\faEnvelope}~research@aliasrobotics.com}}
}

{
\makeatletter
\renewcommand\AB@affilsepx{ \quad} 
\makeatother

\affil[2]{\normalfont Johannes Kepler University Linz.}
}

\makeatletter
\renewcommand\AB@affilnote[1]{}
\makeatother

\affil[*]{
    {\normalfont{\faGithub}~{\tt\footnotesize \href{https://github.com/aliasrobotics/cai}{https://github.com/aliasrobotics/cai}}} \\
    {\normalfont{\faDiscord}~{\tt\footnotesize \href{https://discord.gg/fnUFcTaQAC}{https://discord.gg/fnUFcTaQAC}}}
}

\begin{document}

\pagestyle{fancy}
\fancyhf{} 
\fancyhead[L]{\textit{\leftmark}} 
\renewcommand{\sectionmark}[1]{\markboth{#1}{}}


\definecolor{RedGCTR}{HTML}{B30000}      
\definecolor{BlueGCTR}{HTML}{3366AA}     
\definecolor{PurplePlain}{HTML}{8B63BE}  
\definecolor{PurpleCoop}{HTML}{663D8C}   
\definecolor{PurpleMerged}{HTML}{4A2C6B} 
\definecolor{NoGCTR}{HTML}{CCCCCC}       

\date{}
\maketitle
\vspace{-1em}

\begin{abstract}

AI-driven penetration testing now executes thousands of actions per hour but still lacks the strategic intuition humans apply in competitive security. To build cybersecurity superintelligence--Cybersecurity AI exceeding best human capability—such strategic intuition must be embedded into agentic reasoning processes. We present Generative Cut-the-Rope (G-CTR), a game-theoretic guidance layer that extracts attack graphs from agent's context, computes Nash equilibria with effort-aware scoring, and feeds a concise digest back into the LLM loop \emph{guiding} the agent's actions. Across five real-world exercises, G-CTR matches 70--90\% of expert graph structure while running 60--245$\times$ faster and over 140$\times$ cheaper than manual analysis. In a 44-run cyber-range, adding the digest lifts success from 20.0\% to 42.9\%, cuts cost-per-success by 2.7$\times$, and reduces behavioral variance by 5.2$\times$. In Attack-and-Defense exercises, a shared digest produces the \textcolor{PurpleMerged}{\textbf{Purple}} agent, winning roughly 2:1 over the LLM-only baseline and 3.7:1 over independently guided teams. This closed-loop guidance is what produces the breakthrough: it reduces ambiguity, collapses the LLM's search space, suppresses hallucinations, and keeps the model anchored to the most relevant parts of the problem, yielding large gains in success rate, consistency, and reliability.
\end{abstract}


\section{Introduction}


Cybersecurity is being revolutionized by the use of AI and particularly Large Language Models (LLMs). In 2023, PentestGPT \cite{deng2024pentestgptllmempoweredautomaticpenetration,mayoral2025offensive} pioneered results showing that LLMs could help guide humans in penetration testing. In early 2025, Cybersecurity AI (CAI) \cite{mayoralvilches2025caiopenbugbountyready} demonstrated that LLMs could yield penetration testing results at expert human levels, while faster and cheaper. The next frontier is “cybersecurity superintelligence”: cybersecurity AI exceeding best human capability. Agents that not only act faster than humans but also reason strategically the way humans do when they mentally “\emph{play the game}”. Just as a chess grandmaster evaluates the board and imagines possible attacker/defender lines before committing to a move, AI agents must internalize that intuition to choose optimal cyber actions under pressure.

\begin{figure}[h!]
  \centering
  \begin{tikzpicture}[scale=1.0, every node/.style={scale=0.85, font=\sffamily},
      datanode/.style={circle, draw=graph_lightcyan!90!black, fill=graph_lightcyan, minimum size=1.3cm, align=center, font=\scriptsize, line width=0.8pt, drop shadow={opacity=0.15, shadow xshift=0.08cm, shadow yshift=-0.08cm}},
      keynode/.style={circle, draw=graph_teal!90!black, fill=graph_teal, text=white, minimum size=1.7cm, align=center, font=\small\bfseries, line width=1pt, drop shadow={opacity=0.25, shadow xshift=0.12cm, shadow yshift=-0.12cm}},
      supportnode/.style={rectangle, draw=graph_lightcyan!80!black, fill=graph_lightcyan, minimum width=1.6cm, minimum height=0.45cm, align=center, font=\tiny, rounded corners=2pt, line width=0.6pt},
      altnode/.style={rectangle, draw=graph_gray!80!black, fill=graph_gray, minimum width=1.5cm, minimum height=0.45cm, align=center, font=\tiny, rounded corners=2pt, line width=0.6pt},
  ]

  \fill[graph_gray!15, rounded corners=6pt] (-7.8, -2.2) rectangle (-3.2, 2.2);
  \node[font=\scriptsize\itshape, text=graph_navy!60] at (-5.5, 2.5) {\textsc{Phase 1: Game-Theoretic AI Analysis (G-CTR)}};

  \fill[graph_teal!8, rounded corners=6pt] (-2.8, -2.2) rectangle (3.2, 2.2);
  \node[font=\scriptsize\itshape, text=graph_navy!60] at (0.2, 2.5) {\textsc{Phase 2: Game-Theoretic AI Guidance (Digest Generation)}};

  \fill[graph_lightcyan!15, rounded corners=6pt] (3.6, -2.2) rectangle (7.8, 2.2);
  \node[font=\scriptsize\itshape, text=graph_navy!60] at (5.7, 2.5) {\textsc{Phase 3: Agent Execution (ReAct)}};
  \node[keynode] (graph) at (-6.6, -1.3) {\small Attack\\Graph\\Generation};
  \node[datanode] (nash) at (-5.2, 0) {\scriptsize Nash\\Equilibrium};
  \node[altnode] (data) at (-4.0, 1.3) {\scriptsize G-CTR\\Results};
  
  \node[datanode] (algo) at (-1.2, 1.1) {\scriptsize Algorithmic\\ digest};
  \node[keynode] (llm) at (-1.2, -1.1) {\small LLM \\digest};
  \node[altnode, minimum size=1.0cm] (digest) at (1.5, 0) {\scriptsize Strategic\\\scriptsize Interpretation};
  
  \node[keynode] (agent) at (4.5, 0) {\small Plan\\ \scriptsize(LLM)};
  \node[datanode, minimum width=2.0cm] (actions) at (6.6, 1.0) {\small Act\\ \scriptsize(Tools)};
  \node[datanode] (observe) at (6.6, -1.0) {\scriptsize Scan \&\\Update\\ (Context)};
  
  \draw[-{Stealth[scale=1.2]}, line width=1.1pt, graph_navy]
      (graph.south east) to[bend left=-100] (nash.east);
  \draw[-{Stealth[scale=1.2]}, line width=1.1pt, graph_navy, rounded corners=4pt]
      (nash.west) to[bend left=100] (data.north west);
  
  \draw[-{Stealth[scale=1.2]}, line width=1.1pt, graph_navy, rounded corners=4pt]
      (data.east) .. controls +(1.5,0) and +(-1.2,0.6) .. (algo.west);
  \draw[-{Stealth[scale=1.2]}, line width=1.1pt, graph_navy, rounded corners=4pt]
      (data.east) .. controls +(1.5,0) and +(-1.2,-0.6) .. (llm.west);
  
  \draw[-{Stealth[scale=1.2]}, line width=1.1pt, graph_navy, rounded corners=4pt]
      (algo.east) .. controls +(0.8,0) and +(-0.8,0.5) .. (digest.north west);
  \draw[-{Stealth[scale=1.2]}, line width=1.1pt, graph_navy, rounded corners=4pt]
      (llm.east) .. controls +(0.8,0) and +(-0.8,-0.5) .. (digest.south west);
  
  \draw[-{Stealth[scale=1.2]}, line width=1.1pt, graph_navy]
      (digest.east) -- (agent.west);
  \draw[-{Stealth[scale=1.2]}, line width=1.1pt, graph_navy, rounded corners=4pt]
      (agent.north east) .. controls +(0.6,0.4) and +(-0.6,0) .. (actions.west);
  \draw[-{Stealth[scale=1.2]}, line width=1.1pt, graph_navy, rounded corners=4pt]
      (observe.west) .. controls +(-0.6,0) and +(0.6,-0.4) .. (agent.south east);
  
  \draw[-{Stealth[scale=1.2]}, line width=1.1pt, graph_navy, rounded corners=6pt]
      (actions.south) .. controls +(0,-0.4) and +(0,0.4) .. (observe.north);
  
  \draw[-{Stealth[scale=1.2]}, line width=1.1pt, dashed, graph_navy, rounded corners=8pt]
      (observe.south) .. controls +(0,-1.2) and +(0,-1.2) .. (graph.south)
      node[pos=0.5, below, font=\scriptsize, inner sep=10pt, rounded corners=1pt] {every 5 interactions ($\sim$80 tools)};
  
  \node[font=\tiny, text=black!80!black, fill=white, inner sep=1.5pt, rounded corners=1pt, draw=graph_teal!30, line width=0.3pt]
      at (-5.2, 0.5) {$<$5ms};
  \node[font=\tiny, text=graph_teal!80!black, fill=white, inner sep=1.5pt, rounded corners=1pt, draw=graph_teal!30, line width=0.3pt]
      at (-6.6, -0.5) {$\approx$20s};

  \node[font=\tiny, text=black!80!black, fill=white, inner sep=1.5pt, rounded corners=1pt, draw=graph_navy!20, line width=0.3pt]
      at (-1.2, 1.6) {$<$10ms};
  \node[font=\tiny, text=graph_teal!80!black, fill=white, inner sep=1.5pt, rounded corners=1pt, draw=graph_navy!20, line width=0.3pt]
      at (-1.2, -0.6) {$\approx$28.3s};
  \node[font=\tiny, text=graph_teal!80!black, fill=white, inner sep=1.5pt, rounded corners=1pt, draw=graph_navy!20, line width=0.3pt]
      at (4.5, 0.5) {$\approx$10s};

  \node[font=\tiny, text=black!80!black, fill=white, inner sep=1.5pt, rounded corners=1pt, draw=graph_navy!20, line width=0.3pt]
      at (6.6, 1.5) {$\approx$60s};

  \draw[line width=0.8pt, graph_navy!40] (3.3, -1.8) -- (3.3, 1.8);
  \draw[line width=0.8pt, graph_navy!40] (3.4, -1.8) -- (3.4, 1.8);



  \draw[line width=1.2pt, graph_teal!70, rounded corners=2pt] (-7.8, -3.6) -- (-7.8, -3.9) -- (3.3, -3.9) -- (3.3, -3.6);
  \node[font=\scriptsize\bfseries, text=white, fill=graph_teal, inner sep=3pt, rounded corners=2pt, draw=graph_teal!80!black, line width=0.6pt]
      at (-2.25, -3.9) {Time Budget: $\approx$50s};

  \draw[line width=1.2pt, graph_lightcyan!90!black, rounded corners=2pt] (3.4, -3.6) -- (3.4, -3.9) -- (7.8, -3.9) -- (7.8, -3.6);
  \node[font=\scriptsize\bfseries, text=white, fill=graph_lightcyan!90!black, inner sep=3pt, rounded corners=2pt, draw=graph_lightcyan!80!black, line width=0.6pt]
      at (5.6, -3.9) {Time Budget: $\approx$70s};

  \end{tikzpicture}
  \caption{Game-Theoretic architecture for guiding attack and defense actions in Cybersecurity AI through closed-loop strategic feedback obtained by applying the G-CTR  method. The system operates in three phases: (1) Game-Theoretic AI Analysis generates attack graphs and computes Nash equilibria via G-CTR to identify optimal attack/defense strategies, (2) Strategic interpretation transforms equilibrium data into actionable guidance for both attackers and defenders, (3) AI agent execution performs security testing with continuous graph refinement every $n$ interactions ($\sim$80 tool calls). This architecture enables real-time strategic adaptation for AI security operations, with Phases 1-2 operating within a $\approx$50s time budget parallel to Phase 3's $\approx$70s execution cycles, providing minimal computational overhead while maximizing strategic impact.}
  \label{fig:system_architecture_init}
\end{figure}
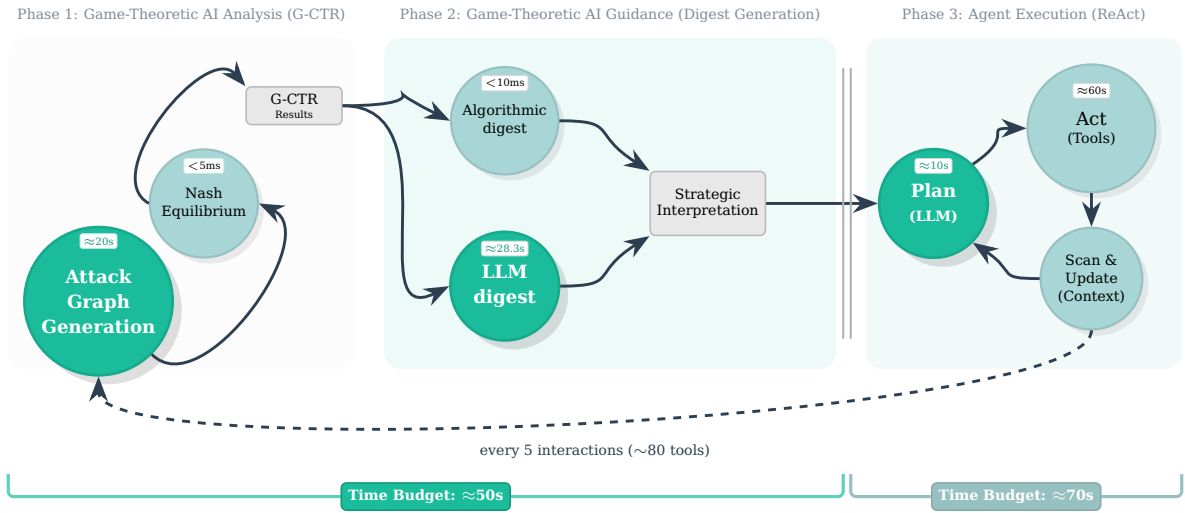

Achieving that intuition requires more than faster execution; it requires embedding a reasoning substrate that weighs attacker and defender payoffs at every step. Humans implicitly apply game theory when playing chess, Go, or when solving competitive cybersecurity challenges such as Attack and Defense CTFs (A\&D): we evaluate the current “board,” imagine opponent responses, and choose the move that maximizes long-term advantage. Our game-theoretic layer does not invent new attacks; it restructures the AI's own actions into an attack graph, computes equilibria, and returns a digest—akin to a chess engine highlighting the strongest lines—that steers the LLM toward statistically advantageous continuations. This closed-loop guidance reduces ambiguity, collapses the search space, and suppresses hallucinations by anchoring the model to what is actually unfolding in the environment, which explains the 5.2$\times$ variance reduction and 2$\times$ success gains reported later. We argue that to surpass human capability, we need AI agents that perform the same mental chess, but at machine scale and speed.

AI-driven penetration testing tools like the open source \emph{Cybersecurity AI} (CAI) \cite{cai2025github} have demonstrated they can discover vulnerabilities 3,600$\times$ faster than humans, yet their outputs (dense logs of automated findings) overwhelm security teams. Meanwhile, game-theoretic frameworks like Cut-the-Rope (CTR) \cite{ctr_paper1,ctr_paper2,rass2025poster} compute optimal defense strategies but require manually-constructed attack graphs as input. This disconnect between rapid AI discovery and strategic defense planning represents a critical gap in modern cybersecurity, raising the fundamental question: can we bridge this gap by automatically transforming AI security outputs into strategic game-theoretic models that then guide both attack and defense operations? In other words, our research here attempts answering the following: \emph{RQ1: can we make Cybersecurity AI agents more effective by guiding them strategically with game theory?}




This paper addresses this question by presenting a game-theoretic AI guidance architecture that bridges the gap between AI-driven security testing and strategic reasoning, providing actionable guidance for both attack and defense operations. Through closed-loop feedback, the system continuously refines its understanding of the threat landscape and adapts its strategic recommendations. The architecture integrates three phases as depicted in Figure \ref{fig:system_architecture_init}: (1) a game-theoretic AI analysis powered by the novel Generative-CTR (G-CTR) which automatically performs attack graph extraction from AI security logs using LLMs and performs Nash equilibrium computation to identify optimal attack/defense strategies, (2) a strategic interpretation that transforms game-theoretic insights into actionable guidance for AI agents, and (3) an open-source AI security framework implementing ReAct agents where both aforementioned methods are integrated within the agent's \emph{Planning} phase. Phases 1-2 operate in parallel with Phase 3, providing minimal computational overhead while maximizing strategic impact.

Empirically, the architecture delivers three key results. First, across five real-world exercises, G-CTR produces 6--15 node graphs with 70--90\% node correspondence to expert annotations (provided by two externally hired domain professionals; see Appendix \ref{appendix_annotation}) while running 60--245$\times$ faster and over 140$\times$ cheaper than manual workflows. Second, in a 44-run cyber-range benchmark targeting the Shellshock CVE-2014-6271 vulnerability~\cite{cve2014-6271,weimer2014cve2014-6271}, 
the LLM digest doubles success probability (20.0\% $\rightarrow$ 42.9\%), boosts cost-per-success by 2.7$\times$, and reduces tool-use variance by 5.2$\times$. Third, in realistic Attack and Defense cybersecurity exercises, sharing a single graph between red and blue agents yields the \textcolor{PurpleMerged}{\textbf{Purple G-CTR$_{\mathbf{merged}}$}} configuration, which defeats the LLM-only baseline roughly 2:1 and outperforms independently guided dual teams 3.7:1. These results foreshadow the broader promise of cybersecurity superintelligence: autonomous agents that blend machine-scale speed with game-theoretic foresight.

Our empirical evaluation on five real-world exercises shows that G-CTR generates graphs 60-245 times faster than manual analysis while achieving 70-90\% node correspondence with expert annotations. More importantly, we demonstrate how this game-theoretic guidance enhances pentesting effectiveness---agents guided by strategic feedback achieve 3.21$\times$ higher success rates while reducing cost-per-finding by 96\%.

The key insight is that LLMs can transform unstructured security narratives into structured models suitable for game-theoretic analysis, enabling a closed-loop system where AI-driven testing and strategic reasoning mutually reinforce each other. This represents a step toward autonomous cybersecurity systems that not only discover vulnerabilities but also reason strategically about optimal exploitation paths and critical defensive positions, guiding both red team operations and blue team prioritization.

\subsection{State of the Art}

AI-powered penetration testing has advanced rapidly, research lines like Cybersecurity AI (CAI) \cite{mayoralvilches2025caiopenbugbountyready, mayoralvilches2025caiopenbugbountyready,mayoral2025cybersecurity, mayoral2025cai, mayoral2025cai_hacking_ai_hackers,mayoral2025cybersecurity_humanoid_robots,mayoral2025the_cybersecurity_of_a_humanoid_robot,balassone2025cybersecurity} report achieving 3,600× speedup over humans and frameworks like PentestGPT \cite{deng2024pentestgptllmempoweredautomaticpenetration}, AutoPT \cite{wu2024autopt}, and VulnBot \cite{kong2025vulnbot} automating complex assessments. However, these tools produce overwhelming amounts of unstructured data that security teams struggle to interpret and act upon.

Attack graphs provide structured representations of attack paths \cite{ammann2002automated}, with traditional tools like MulVAL \cite{ou2005mulval} using logic-based reasoning over vulnerability databases. While recent work explores machine learning enhancements \cite{kaynar2016taxonomy,lallie2020review}, current approaches remain disconnected from live security operations and require extensive manual effort.

Game theory offers strategic analysis of attacker-defender interactions, with CTR \cite{ctr_paper1,ctr_paper2} computing optimal defense strategies on attack graphs. Despite proven effectiveness against APTs \cite{liu2023recent}, game-theoretic approaches assume pre-existing graphs rather than generating them from operational data.


The critical gap lies in the integration: AI security tools generate logs without prioritization, humans process logs and manually create models (attack graphs), game-theoretic frameworks analyze pre-built models without operational integration, and neither provides actionable guidance for both attackers and defenders. No existing system closes the loop by automatically extracting attack graphs from AI operations, computing Nash equilibria to identify optimal strategies, and transforming these game-theoretic insights into real-time guidance that directs subsequent security testing. This disconnect prevents game-theoretic reasoning from enhancing AI-driven red team operations and blue team defensive prioritization, limiting the practical impact of both technologies in modern cybersecurity workflows.

While AI and LLMs have seen growing adoption in cybersecurity, especially in automating penetration testing, several critical challenges remain unresolved:

\begin{itemize}
  \item \emph{Limited Scalability of Attack Graphs.} Existing attack graph methodologies rely heavily on manual curation or static generation approaches, which struggle to scale with the complexity and dynamism of modern network environments. This limits their practical use in continuous, large-scale cybersecurity operations.

  \item \emph{Lack of Comprehensive Evaluation of LLMs in Cybersecurity.} Despite the rapid development of LLMs, their capabilities for understanding and modeling cybersecurity exercises remain poorly characterized. There is an absence of standardized, gold-standard datasets or benchmarks to rigorously assess LLM-driven attack graph generation and reasoning.

  \item \emph{Insufficient Integration of Game-Theoretic Models with AI Automation.} Game theory offers powerful frameworks for risk assessment \cite{ctr_paper1,ctr_paper2, rass2025poster, ou2005mulval} and strategic defense in cybersecurity, yet its fusion with LLM-based automation has not been thoroughly explored or validated in practical tooling.

  \item \emph{Gap Between Fast-Evolving AI Capabilities and Human Annotation Workflows.} The accelerating pace of AI-driven cybersecurity tasks challenges traditional human annotation and analysis methods, creating a need for automated systems that can keep up without sacrificing accuracy or interpretability.
\end{itemize}

Addressing these gaps is key to advancing scalable and autonomous cybersecurity operations. To address these limitations, our work introduces a game-theoretic AI architecture that provides strategic guidance for both attack and defense through closed-loop feedback. Built on top of the CAI framework and inspired by the Cut-The-Rope (CTR) algorithm \cite{ctr_paper1, ctr_paper2}, our architecture  transforms game-theoretic insights into actionable guidance. This enables dynamic attack graph generation coupled with effort-based vulnerability scoring, allowing both defenders to prioritize responses based on cost and path complexity, and attackers to focus on high-probability exploitation paths. By integrating an agentic LLM-driven security automation with game-theoretic reasoning within a unified closed-loop framework, this approach bridges the gap between reactive automation and strategic foresight, providing real-time guidance for red team operations and blue team defensive prioritization.

\subsection{Research Contributions}
This work presents a game-theoretic AI architecture for guiding both attack and defense in cybersecurity through closed-loop feedback. Our contributions include:

\begin{enumerate}
  \item \textbf{Generative Cut-The-Rope (G-CTR)}: We introduce G-CTR (Section \ref{sub:game_theoretic_analysis}), an extension of the original CTR framework that leverages LLMs to automatically generate attack graphs and Nash equilibrium computations from unstructured cybersecurity logs and exercise narratives. G-CTR introduces two key innovations that enable automated game-theoretic analysis in cybersecurity:

  \begin{itemize}
    \item[\textcolor{cai_primary!50}{\textbf{(a)}}] \textcolor{cai_primary!50}{\textbf{Automated Graph Extraction from AI Security Logs.}} We demonstrate that LLMs can automatically extract structured attack graphs from CAI penetration testing logs (Section \ref{subsub:automated_graph_extraction}) with 70-90\% node correspondence to expert-generated graphs, achieving 60-245$\times$ time speedups and 225-450$\times$ cost improvements over manual analysis.

    \item[\textcolor{cai_primary!50}{\textbf{(b)}}] \textcolor{cai_primary!50}{\textbf{Effort-Based Scoring for LLM-Generated Graphs.}} We introduce a practical adaptation of CTR's probabilistic model to handle the unique characteristics of LLM-generated attack graphs (Section~\ref{sub:score_computation}). Our effort score combines message distance, token count, and cost metrics to quantify attack difficulty in the absence of traditional probability estimates, enabling game-theoretic analysis on automatically generated graphs.
  \end{itemize}

  \item \textbf{Strategic Digest Generation for Guiding Attack and Defense:} We introduce a digest generation pipeline that transforms Nash equilibrium computations into actionable strategic guidance (Section \ref{section:strategic_feedback_loop}), closing the loop between G-CTR analysis and CAI execution. In 44 cyber-range penetration tests, LLM-based digests improved success rates from 20.0\% to 42.9\% (2.15$\times$), cut cost-per-success from \$0.32 to \$0.12 (2.7$\times$), and reduced tool-usage variance by 5.2$\times$, demonstrating that LLM interpretation of game-theoretic outputs materially enhances autonomous pentesting.

  \item \textbf{Practical Validation Framework:} We evaluate our approach on five real-world security exercises, demonstrating that LLM-generated graphs (6--15 nodes, up to four vulnerable stages) maintain 70--90\% node correspondence with expert baselines while delivering 60--245$\times$ time savings and $>$140$\times$ cost reduction. These studies establish that automated graph generation plus effort-aware scoring can prioritize pentesting effort and defender investments without human curation.

  \item \textbf{Breakthrough Multi-Agent Guidance:} We extend the architecture to Attack and Defense CTFs and introduce the \textcolor{PurpleMerged}{\textbf{Purple G-CTR$_{\mathbf{merged}}$}} team configuration, where red and blue agents share a single G-CTR graph and situational context. This GenAI+game-theory agent wins \texttt{pingpong} matches 52.4\% vs 28.6\% (\textasciitilde1.8:1) against the LLM-only baseline and defeats independent dual guidance 55\% vs 15\% (\textasciitilde3.7:1) in \texttt{cowsay}, establishing a new state of the art for AI-driven Attack and Defense play.

\end{enumerate}

The remainder of this paper is organized as follows: Section \ref{section:Background} provides background on CAI and game--theoretic approach relevant to our work. Section \ref{section:Methodology} presents our proposed game-theoretic AI architecture. Section \ref{section:Results} presents our experimental results. Section \ref{section:Discussion} discusses the implications of our findings. Finally, we conclude the paper with a summary of contributions and key insights.

\section{Background}\label{section:Background}

\subsection{Cybersecurity AI}
Artificial Intelligence (AI) is transforming cybersecurity, yet many AI solutions remain proprietary and inaccessible to small and medium-sized enterprises. To address this gap, open and transparent frameworks have been developed in which CAI \cite{mayoralvilches2025caiopenbugbountyready, aliasrobotics2025cai} stands at the fore front.
CAI is a modular, autonomous system with Human-In-The-Loop oversight that achieves up to 3,600$\times$ faster performance than humans and excelling in global CTF competitions \cite{palisaderesearch_ai_vs_human}. AI’s speed and complexity create significant interpretability challenges, highlighting the need for tools that can effectively interpret and analyze its behaviour. Attack Graphs provide a human--interpretable framework for this purpose. 
There is a critical need not only to observe AI actions but to systematically interpret, visualize, and strategically analyze them for actionable insights. Attack graphs address this by modeling vulnerabilities, exploits, and defenses in a human--interpretable framework. To fully leverage their strategic potential, we leverage the Cut--the--Rope (CTR) model and expand it to build a game-theoretic AI architecture atop CAI.

\subsection{Game Theory in Cybersecurity: Cut-the-Rope} \label{sub:Cut_the_Rope}
Cybersecurity inherently involves adversarial interactions where perfect protection is unattainable due to unknown vulnerabilities and system complexity. The Cut-the-Rope (CTR) model \cite{ctr_paper1} provides a rigorous game-theoretic framework for this domain by modeling systems as directed acyclic graphs $G = (V, E)$, where nodes $V$ represent system components and edges $E$ denote exploitable transitions. This model provides security games that have has been used in various subsequent studies including the assessment of effective defenses in robots \cite{ctr_paper2}, and to estimate the zero-day risks \cite{rass2025poster}.

CTR formalizes a zero-sum asynchronous game where an attacker, already present at unknown node $\theta \in V \setminus \{v_0\}$, advances stochastically toward target $v_0$ following a Poisson process with rate $\lambda$. The defender inspects nodes from admissible set $AS_1 \subseteq V \setminus \{v_0\}$ according to mixed strategy $\sigma_d \in \Delta(AS_1)$. The probability of catching the attacker is:

\begin{equation}
U_D(\sigma_d, \pi, \theta) = \sum_{c \in AS_1} \sigma_d(c) \cdot P_\pi(c \mid \theta, \lambda)
\end{equation}

where $P_\pi(c \mid \theta, \lambda) = \frac{f_{\text{Pois}}(d_\pi(\theta, c); \lambda)}{\sum_{x \in V(\pi)} f_{\text{Pois}}(d_\pi(\theta, x); \lambda)}$ represents the likelihood of the attacker being at node $c$ given path $\pi$ and distance $d_\pi(\theta, c)$. The Nash equilibrium $(\sigma_d^\star, \pi^\star, \theta^\star)$ solves the minimax problem:

\begin{equation}
\sigma_d^\star = \arg \min_{\sigma_d} \max_{\pi \in AS_2, \theta \in V(\pi)} \left( 1 - U_D(\sigma_d, \pi, \theta) \right)
\end{equation}

This yields the optimal randomized defense policy that minimizes the worst-case attack success probability, representing a stable state where neither player benefits from unilateral deviation. For complete mathematical derivation, see Appendix~\ref{appendix_game_theory_ctr}.

In all experiments we instantiate the Poisson attacker process with a rate of $\lambda = 2$ transitions per defender inspection window ($\lambda_a = 2$, $\lambda_d = 1$), i.e., the attacker advances an expected two edges along its chosen path between successive digest-triggered observations.


\section{Game-Theoretic AI for Guiding Attack and Defense}\label{section:Methodology}


\subsection{Game-Theoretic AI Analysis via Generative Cut-The-Rope (G-CTR)} \label{sub:game_theoretic_analysis}
This section details how the CTR model, originally for static attack graphs, is extended for its integration atop CAI. We introduce \textbf{Generative Cut-The-Rope (G-CTR)}, an extension of the original CTR framework that leverages LLMs to automatically generate attack graphs from unstructured cybersecurity logs and exercise narratives. Modifications enable CTR to process LLM--inferred data from cybersecurity logs, supporting dynamic reasoning without losing its theoretical foundation.

\subsubsection{Automated Graph Extraction from AI Security Logs}\label{subsub:automated_graph_extraction}

AI-driven penetration testing generates vast quantities of unstructured logs documenting tool executions, vulnerability discoveries, and exploitation attempts. Traditional attack graph construction requires security experts to manually analyze these logs, identify attack paths, and encode relationships---a process that can take hours or days per exercise \cite{ammann2002automated}. LLMs offer a transformative alternative: by leveraging their understanding of cybersecurity concepts and contextual reasoning capabilities, LLMs can automatically parse security logs, extract relevant entities (vulnerabilities, exploits, system components), and construct structured attack graphs that capture the progression of penetration testing activities. This automation enables real-time graph generation that keeps pace with AI-driven security testing, addressing the bottleneck between rapid vulnerability discovery and strategic analysis.


The main difference between the original CTR library and G-CTR lies in how the Attack Graph is built. The original CTR uses a graph manually created by human experts, based on known threats and fixed probabilities. In contrast, G-CTR uses LLM-generated graphs that emerge from automated analysis of security logs. Detailed comparison of these two approaches is presented in Table~\ref{tab:ctr_vs_ctrincai}.

\begin{longtable}{>{\raggedright\arraybackslash}p{3.2cm}>{\raggedright\arraybackslash}p{5.5cm}>{\raggedright\arraybackslash}p{5.8cm}}
\caption{Key differences between the original CTR implementation and G-CTR.}
\label{tab:ctr_vs_ctrincai} \\
\toprule
\textcolor{caicolor}{\textbf{Feature}} & \textcolor{caicolor}{\textbf{CTR (Original)}} & \textcolor{caicolor}{\textbf{G-CTR (LLM-Driven)}} \\
\midrule
\endfirsthead

\multicolumn{3}{l}{\textbf{Table \thetable\ (continued)}} \\
\toprule
\textcolor{caicolor}{\textbf{Feature}} & \textcolor{caicolor}{\textbf{CTR (Original)}} & \textcolor{caicolor}{\textbf{G-CTR (LLM-Driven)}} \\
\midrule
\endhead

\bottomrule
\endfoot

\rowcolor{gray!5}
\textbf{\textcolor{caicolor}{Graph Semantics}} &
Success rate of exploiting a given node along a specific attack path. &
Effort required to reach the first vulnerable node along a path (Section~\ref{sub:score_computation}).\\
\addlinespace[0.2cm]

\textbf{\textcolor{caicolor}{Graph Generation}} &
Manually constructed by cybersecurity experts. All nodes, vulnerabilities, and paths are hand-annotated. &
Automatically inferred by LLMs generating structured \texttt{JSONL} output from security logs (Appendix~\ref{appendix_annotation}). Two-step adaptation: (1) preprocessing---merge entry points to node 1, add artificial edges (Figure~\ref{fig:fake_graph}); (2) postprocessing---prune non-vulnerable leaves, add artificial leaf nodes to remark vulnerable ones (with probability: 100\%).\\
\addlinespace[0.2cm]

\rowcolor{gray!5}
\textbf{\textcolor{caicolor}{Node Semantics}} &
System components or entities identified by human experts. &
Elements identified by LLM as relevant. Single \texttt{message\_{}id} may yield multiple nodes if LLM extracts distinct semantic information.\\
\addlinespace[0.2cm]

\textbf{\textcolor{caicolor}{Node Count}} &
Determined by human expert analysis. &
Capped at percentage of total log messages (Section \ref{subsub:understanding_gctr}). Limit prevents hallucinations while maintaining coverage.\\
\addlinespace[0.2cm]

\rowcolor{gray!5}
\textbf{\textcolor{caicolor}{Vulnerable Nodes}} &
Only leaf nodes (end states of successful exploitation in closed environment). &
Any node (intermediate or leaf) can be vulnerable. Not all graphs contain vulnerable nodes---depends on LLM identification.\\
\addlinespace[0.2cm]

\textbf{\textcolor{caicolor}{Edge Values}} &
\textit{Probability} of reaching target node \(v_0\) along path. Derived from probabilistic models (Poisson/geometric distributions). &
\textit{Effort score} (0–1) to reach vulnerable node. 0 = unreachable, 1 = same \texttt{message\_{}id} as vulnerability. Lower scores = higher effort (Section~\ref{sub:score_computation}).\\
\addlinespace[0.2cm]

\rowcolor{gray!5}
\textbf{\textcolor{caicolor}{Score/Probability Computation}} &
Probabilistic models capturing attacker behavior and system structure (Poisson/geometric distributions). &
Quantitative formulae combining token count, message distance to a vulnerability, and cost to next vulnerable node (Section~\ref{sub:score_computation}).\\
\addlinespace[0.2cm]

\textbf{\textcolor{caicolor}{Entry Points}} &
Multiple entry points merged into synthetic root node (node 0) for unified structure. &
Single entry point: first node (node 1) corresponding to first \texttt{message\_id}. Disconnected components reconnected to root.\\
\addlinespace[0.2cm]

\rowcolor{gray!5}
\textbf{\textcolor{caicolor}{Modeling Scope}} &
One cybersecurity exercise or system at a time. &
Can model and merge multiple exercises across different systems and targets.\\

\end{longtable}

\begin{figure}[h]
    \includegraphics[width=0.8\textwidth]{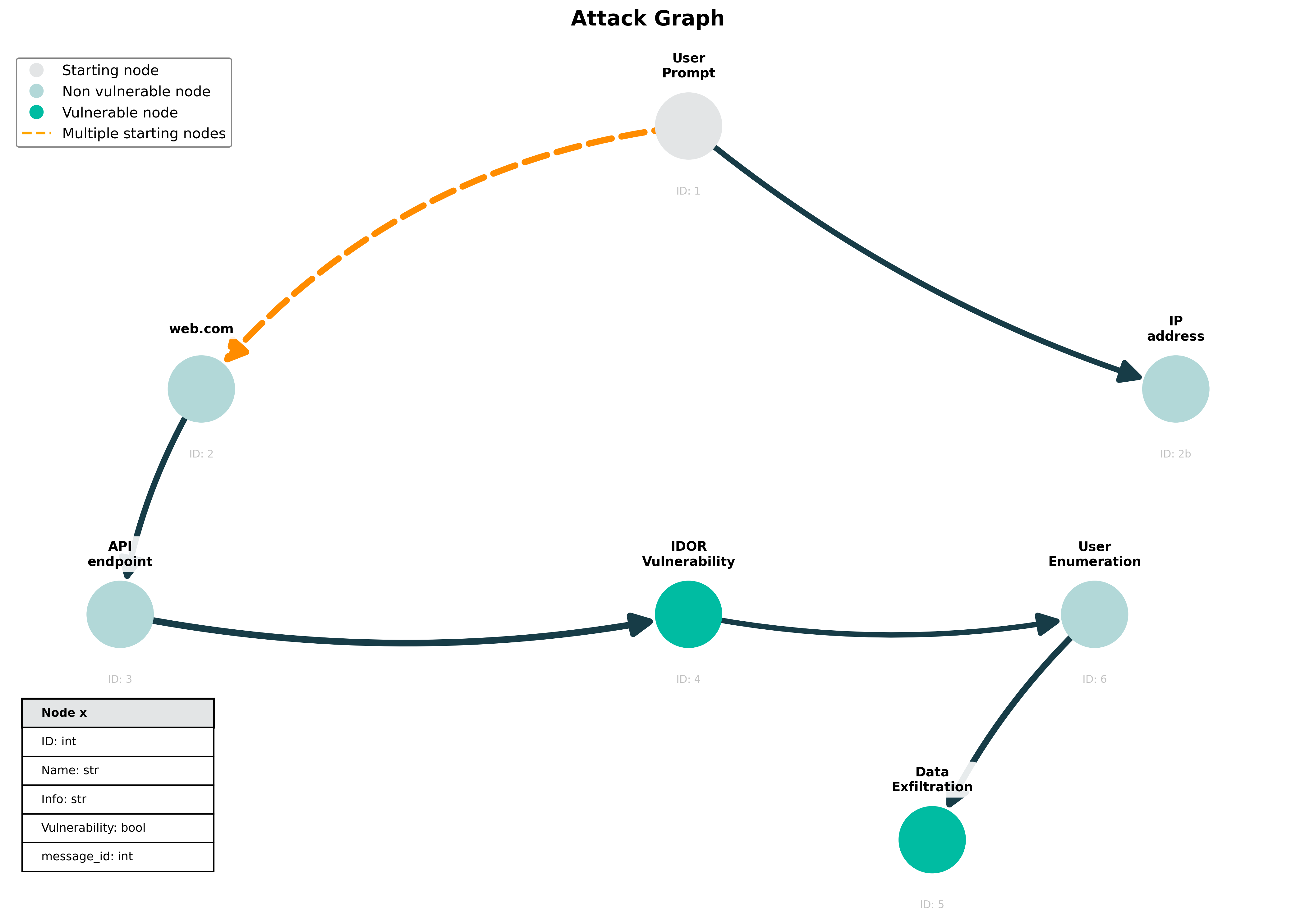}
    \centering
    \caption{\textbf{Attack Graph example}: nodes represent structured information extracted from logs, each containing attributes such as name, message ID, vulnerability status, and additional context. They are categorized into three types: starting nodes (gray), non-vulnerable nodes (light teal), and vulnerable nodes (dark teal), based on their risk level. The orange dashed arrow shows an alternative entry point inferred by the LLM, which is merged into the graph for G-CTR computation because we suppose that everything begins at the first user prompt recorded (\textit{message\_id = 1}). In this figure, the attack starts at the ``User Prompt'' (ID: 1) and progresses through intermediate nodes like ``web.com'' to reach a key vulnerability-``IDOR Vulnerability'' (ID: 4)-which lead to ``Data Exfiltration'' (ID: 5).}
    \label{fig:fake_graph}
\end{figure}

Figure \ref{fig:fake_graph} shows an example of an attack graph constructed within the G-CTR framework. Each node represents a semantically meaningful element from a cybersecurity exercise log, such as domains, IPs, ports, APIs, databases, issues, misconfigurations, and more. Directed edges indicate possible attacker progressions through the scenario. Unlike traditional CTR graphs, this graph reflects an LLM’s interpretation of context and threat progression.

Using LLMs to infer attack graphs, such as the one shown in Figure~\ref{fig:fake_graph}, provides automation flexibility, but also poses challenges. In bounded attack scenarios, cycles or repeated node visits are not allowed, whereas LLM-generated graphs may unintentionally include cycles despite careful prompting. A theoretical solution is to run the LLM multiple times and apply a geometric distribution to assign decreasing probabilities to attack paths with loops. However, this approach is limited in practice because LLMs produce different graphs on each execution due to their inherent probabilistic behavior.

Our G-CTR implementation handles LLM-generated graphs with potential cycles and structural issues through multiple post-processing steps: (1) uses NetworkX's \cite{hagberg2008exploring} \texttt{all\_simple\_paths} function to extract only acyclic paths, automatically excluding any cycles; (2) recursively removes non-vulnerable leaf nodes to ensure only vulnerable nodes appear as terminal states; (3) adds artificial "leaf\_X" nodes with 100\% transition probability to each vulnerable node to satisfy game-theoretic requirements that all targets must be leaf nodes; (4) connects disconnected starting nodes to a single entry point (minimum node ID) via artificial edges; and (5) removes any incoming edges to the designated starting node to ensure a single attack origin---all while prompting the LLM to avoid cycles in the first place (via the system prompt).

\subsubsection{Attack Graphs and Threat Modeling} 

While attack graphs are sometimes informally equated with threat models, they are more accurately understood as \textit{structured artifacts within the threat modeling process}~\cite{haque2017evolutionary}. Threat modeling is the comprehensive activity of identifying, characterizing, and prioritizing potential threats to a system, typically answering questions about what can go wrong, who might attack, what assets are at risk, and what mitigations exist. Attack graphs and attack trees serve as formalization and visualization techniques within this broader process---they help enumerate attack paths, model adversarial capabilities, and communicate security risks to stakeholders~\cite{ammann2002automated}. However, a complete threat model encompasses additional dimensions: asset identification, trust boundaries, attacker profiles, business impact analysis, and mitigation strategies that extend beyond graph structure alone.

In this context, G-CTR's automated extraction of attack graphs from live penetration testing logs represents a novel contribution to threat modeling practice. Rather than relying on manually constructed, static threat scenarios developed during design phases, G-CTR generates \textit{evidence-based, dynamic threat representations} grounded in actual security testing activities. This approach bridges offensive and defensive security: the attack graphs capture real exploitation attempts, tool chains, and vulnerability dependencies observed during AI-driven pentesting, transforming operational data into strategic threat intelligence. While these graphs do not constitute complete threat models on their own, they provide empirically grounded attack path analysis that can inform threat prioritization, defense allocation, and risk assessment---key outputs of comprehensive threat modeling frameworks. Furthermore, by automating graph extraction at the pace of AI-driven security testing, G-CTR enables continuous threat model refinement, addressing a long-standing challenge in traditional threat modeling: keeping threat representations synchronized with evolving system configurations and emerging attack techniques.

\subsubsection{Effort-Based Scoring for LLM-Generated Graphs} \label{sub:score_computation}


Unlike CTR's probabilistic distributions over static graphs, G-CTR quantifies attacker effort through a composite score reflecting the computational resources required to reach vulnerable nodes in dynamically constructed attack graphs.

Let $\textcolor{cai_purple}{\mathcal{E}_i}$ denote the effort to transition from node $i$ to the next vulnerable node. We define the normalized effort score as a convex combination of three metrics:

\begin{equation}
\label{eq:heuristic_gctr}
\textcolor{cai_purple}{\mathcal{E}_i} = \sum_{k \in \{\text{msg}, \text{tok}, \text{cost}\}} \textcolor{cai_accent}{w_k} \cdot \textcolor{cai_primary}{\phi_k}(i)
\end{equation}

where $\textcolor{cai_accent}{\mathbf{w}} = (\textcolor{cai_accent}{w_\text{msg}}, \textcolor{cai_accent}{w_\text{tok}}, \textcolor{cai_accent}{w_\text{cost}}) \in \Delta^3$ forms a probability simplex, and $\textcolor{cai_primary}{\phi_k}: \mathbb{N} \rightarrow [0,1]$ are normalized effort functions.

\vspace{0.3em}
\noindent\textbf{Message Distance ($\textcolor{cai_primary}{\phi_\text{msg}}$).} For the log file $\ell$ 
with $\textcolor{cai_dark}{J_\ell}$ total messages, let $\textcolor{cai_dark}{m_i}$ denote messages between node $i$ and its successor vulnerable node. The normalized message effort is:

\begin{equation}
\label{eq:heuristic_gctr_p_message}
\textcolor{cai_primary}{\phi_\text{msg}}(i) = 1 - \frac{\textcolor{cai_dark}{m_i} - 1}{\textcolor{cai_dark}{J_\ell} - 1}
\end{equation}

\noindent\textbf{Token Complexity ($\textcolor{cai_primary}{\phi_\text{tok}}$).} Let $\textcolor{cai_dark}{t_i}$ represent tokens consumed along edge $i$, approximated via a standardized tokenizer (\texttt{Qwen/Qwen1.5-0.5B-Chat}). For total tokens $\textcolor{cai_dark}{T_\ell} = \sum_{j=1}^{J_\ell} t_j$ in the log file $\ell$:

\begin{equation}
\label{eq:heuristic_gctr_p_tokens}
\textcolor{cai_primary}{\phi_\text{tok}}(i) = 1 - \frac{\textcolor{cai_dark}{t_i}}{\textcolor{cai_dark}{T_\ell}}
\end{equation}

\noindent\textbf{Computational Cost ($\textcolor{cai_primary}{\phi_\text{cost}}$).} Given actual token count $x$ with cost $C_x$, we estimate cost $\hat{C}_y$ for heuristic count $y$ as $\hat{C}_y = (y/x) \cdot C_x$. Let $\textcolor{cai_dark}{c_i}$ denote the estimated cost for transition $i$. With total cost $\textcolor{cai_dark}{C_\ell} = \sum_{j=1}^{J_\ell} c_j$:

\begin{equation}
\label{eq:heuristic_gctr_p_cost}
\textcolor{cai_primary}{\phi_\text{cost}}(i) = 1 - \frac{\textcolor{cai_dark}{c_i}}{\textcolor{cai_dark}{C_\ell}}
\end{equation}

The formulation ensures $\textcolor{cai_purple}{\mathcal{E}_i} \in [0,1]$, where higher values indicate greater effort required for exploitation. This metric replaces CTR's Poisson-based transition probabilities with empirically grounded computational complexity measures suitable for LLM-generated attack graphs.\footnote{Cost-token correlation is addressed via weight adjustment, e.g., $(\textcolor{cai_accent}{w_\text{cost}}, \textcolor{cai_accent}{w_\text{tok}}) = (0.3, 0.4)$ when using cloud APIs.}

\subsubsection{Understanding G-CTR Outputs} \label{subsub:understanding_gctr}
The output of G-CTR represents the solution to a security game between an attacker and a defender. The defender protects selected nodes with limited resources, while the attacker aims to reach a target undetected. G-CTR computes optimal mixed strategies for both, solving the game-theoretic equilibrium described in Section~\ref{sub:Cut_the_Rope}.

The output included three main components:
\begin{itemize}
    \item \textbf{Defender Strategy Table:} A probability distribution over intermediate defendable 
    nodes, indicating where to allocate defenses to minimize attack success.
    \item \textbf{Attacker Strategy Table:} A list of possible attack paths, each with a probability reflecting how likely a rational attacker is to choose it.
    \item \textbf{Game Equilibrium:} The success probability both players can guarantee defenders aim to keep it below this threshold; attackers aim to reach or exceed it.
\end{itemize}

\begin{figure}[H]
    \centering
    \includegraphics[width=\textwidth]{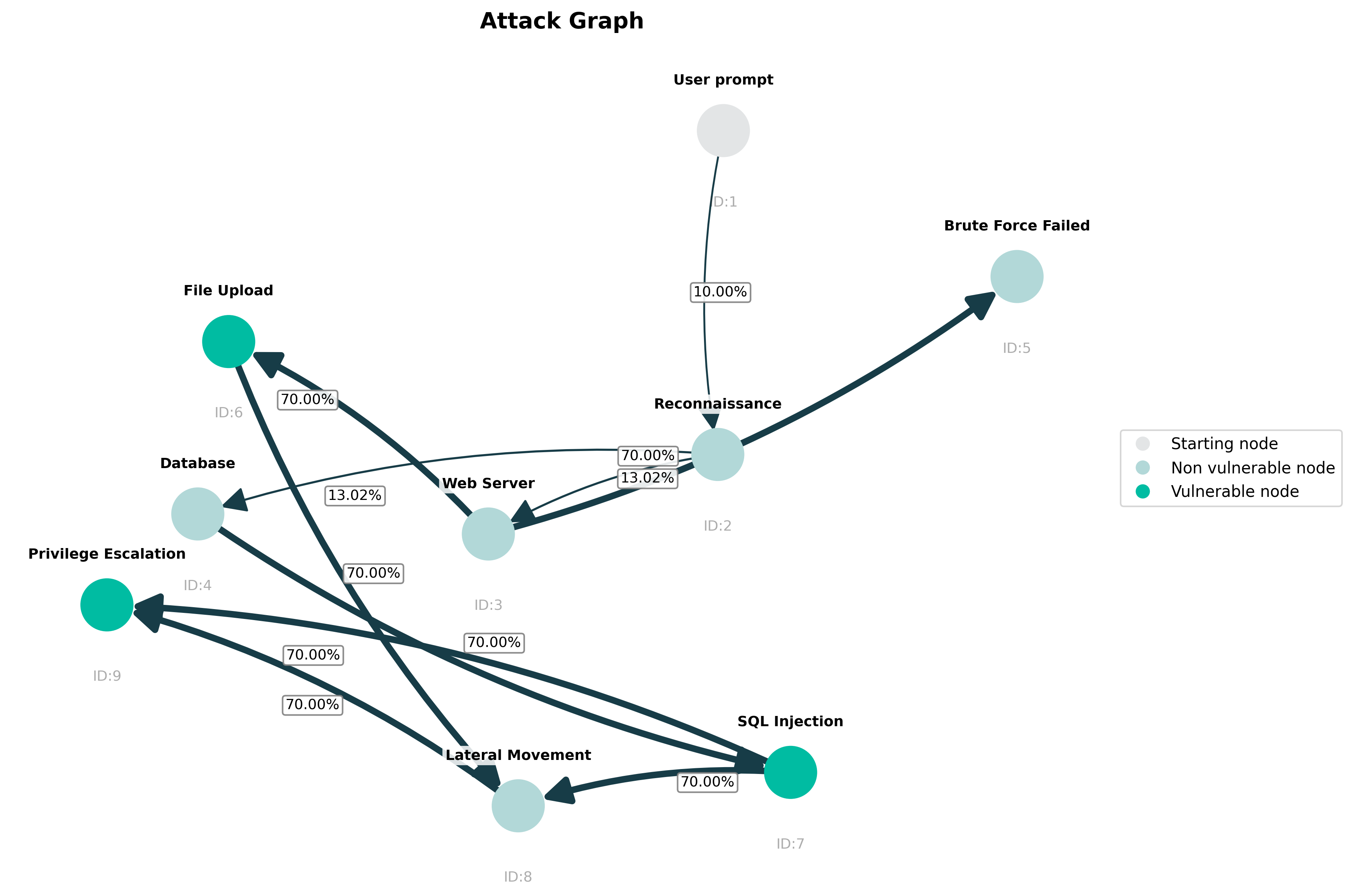}
    \caption*{(a) Attack Graph Example}

    \vspace{0.5cm}

    \begin{minipage}{0.45\textwidth}
        \centering
        \renewcommand{\arraystretch}{1.4}
        \begin{tabular}{cc}
        \hline
        \textcolor{cai_primary}{\textbf{Node ID}} & \textcolor{cai_primary}{\textbf{Probability}} \\
        \hline
        8 & 0.673528 \\
        4 & 0.326472 \\
        2 & 0.000000 \\
        3 & 0.000000 \\
        \hline
        \end{tabular}
        \caption*{(b) Optimal Defense Strategy}
    \end{minipage}
    \hfill
    \begin{minipage}{0.50\textwidth}
        \centering
        \renewcommand{\arraystretch}{1.4}
        \begin{tabular}{ccc}
        \hline
        \textcolor{cai_primary}{\textbf{Path ID}} & \textcolor{cai_primary}{\textbf{Path Sequence}} & \textcolor{cai_primary}{\textbf{Probability}} \\
        \hline
        5 & 1 $\textcolor{cai_primary}{\rightarrow}$ 2 $\textcolor{cai_primary}{\rightarrow}$ 4 $\textcolor{cai_primary}{\rightarrow}$ 7 $\textcolor{cai_primary}{\rightarrow}$ 9 & 0.673528 \\
        1 & 1 $\textcolor{cai_primary}{\rightarrow}$ 2 $\textcolor{cai_primary}{\rightarrow}$ 3 $\textcolor{cai_primary}{\rightarrow}$ 6 $\textcolor{cai_primary}{\rightarrow}$ 8 $\textcolor{cai_primary}{\rightarrow}$ 9 & 0.326472 \\
        3 & 1 $\textcolor{cai_primary}{\rightarrow}$ 2 $\textcolor{cai_primary}{\rightarrow}$ 4 $\textcolor{cai_primary}{\rightarrow}$ 7 & 0.000000 \\
        2 & 1 $\textcolor{cai_primary}{\rightarrow}$ 2 $\textcolor{cai_primary}{\rightarrow}$ 4 & 0.000000 \\
        4 & 1 $\textcolor{cai_primary}{\rightarrow}$ 2 $\textcolor{cai_primary}{\rightarrow}$ 4 $\textcolor{cai_primary}{\rightarrow}$ 7 $\textcolor{cai_primary}{\rightarrow}$ 8 $\textcolor{cai_primary}{\rightarrow}$ 9 & 0.000000 \\
        \hline
        \end{tabular}
        \caption*{(c) Attacker Strategy}
    \end{minipage}

    \vspace{0.5cm}

    \begin{minipage}{\textwidth}
        \centering
        \colorbox{cai_light}{%
            \parbox{0.85\textwidth}{%
                \centering
                \vspace{1.2ex}
                \textcolor{cai_dark}{\textbf{Game Equilibrium}}\\[0.5ex]
                \textcolor{cai_primary}{Defender can keep attacker success below:} \textbf{0.035280}\\[0.3ex]
                \textcolor{cai_primary}{Attacker can guarantee success probability of:} \textbf{0.035280}
                \vspace{1.2ex}
            }%
        }
        \caption*{(d) Game Equilibrium}
    \end{minipage}

    \caption{Attack Graph and Equilibrium analysis }
    \label{fig:Graph_attack_example}
\end{figure}

Figure \ref{fig:Graph_attack_example} illustrates a graph modeling potential attack steps from an initial entry point (ID: 1) to key targets: File Upload (ID: 6), Privilege Escalation (ID: 9), and SQL Injection (ID: 7). Each node represents an attack stage, with edges indicating the score or cost of transitioning between them.

Using this graph, G-CTR computes the optimal mixed strategies for both the defender and the attacker. This results are shown in Figure~\ref{fig:Graph_attack_example}. In this case, the defender's strategy is expressed as a probability distribution over defendable nodes where defensive resources can be allocated. The results indicate that the defender should allocate approximately 67.35\% of their defensive effort to the Lateral Movement node (ID: 8) and 32.65\% to the Database node (ID: 4), leaving nodes such as Reconnaissance (ID: 2) and Web Server (ID: 3) undefended due to their lower strategic value.
The G-CTR identifies the most likely attack paths based on graph structure and vulnerabilities. The dominant path (67.35\%) goes through Reconnaissance \textcolor{cai_primary}{$\to$} Database \textcolor{cai_primary}{$\to$} SQL Injection \textcolor{cai_primary}{$\to$} Privilege Escalation (1 \textcolor{cai_primary}{$\to$} 2 \textcolor{cai_primary}{$\to$} 4 \textcolor{cai_primary}{$\to$} 7 \textcolor{cai_primary}{$\to$} 9), while a secondary path (32.65\%) follows File Upload and Lateral Movement (1 \textcolor{cai_primary}{$\to$} 2 \textcolor{cai_primary}{$\to$} 3 \textcolor{cai_primary}{$\to$} 6 \textcolor{cai_primary}{$\to$} 8 \textcolor{cai_primary}{$\to$} 9). All other paths are unlikely due to low success or high detection risk.

The final equilibrium outcome of this game-theoretic interaction reports a mutual success probability of 3.53\%. This value represents the optimal balance point: the attacker cannot guarantee more than a 3.53\% chance of reaching the target regardless of their path choice, and the defender cannot reduce this probability further without additional resources. 

By analyzing this example, the Cut-the-Rope framework demonstrates its ability to derive strategic insights that go beyond simple vulnerability analysis, enabling informed, resource-efficient cybersecurity decision-making based on adversarial reasoning.

There are also some special cases that affect the outcome of the defender and attacker probabilities. The following consider our implementation of the optimization problem:
\begin{itemize}
    \item Starting nodes cannot be defended. If the starting node has direct access to a vulnerable node (i.e., there are no adversary intermediate locations to defend) the optimization will return 0. 
    \item A vulnerable node represents a compromised system state or exploited condition that has already been achieved by the attacker. These nodes model the successful exploitation of a vulnerability rather than the vulnerability itself. In the CTR game-theoretic model, defense actions are modeled as inspections along attack paths rather than patching of individual vulnerabilities. While real-world vulnerabilities may indeed be patchable, the model focuses on path-based defense strategies: defenders allocate inspection resources to intermediate nodes to detect and intercept attackers before they reach critical assets. Non-vulnerable nodes represent system components or network positions where such defensive inspections can be deployed to prevent attackers from progressing toward vulnerable states.
    
    
    \item If there is only one possible exploitation path that the attacker can take, the probability assigned to that path will be 100\% and the game equilibrium will be 0.

    \item The G-CTR implementation enforces adaptive graph complexity bounds through a piecewise-linear scaling heuristic. This heuristic limits the maximum number of nodes in the generated attack graph based on the total number of messages in the conversation log. Specifically, for shorter conversations (<70 messages), the node limit is set to 12--16\% of the message count; for medium-length conversations (70--199 messages), the limit decreases to 6--12\%; and for longer conversations ($\geq$200 messages), it is further reduced to 3.5--5\%. These percentages were empirically tuned to balance graph expressiveness against computational cost: shorter logs benefit from higher node-to-message ratios to capture sufficient detail, while longer logs require stricter limits to prevent excessive graph complexity. Additionally, the resulting node count is clamped to an absolute range of [4, 25] nodes (not percentages) to ensure that very short conversations still produce meaningful graphs and that very long conversations remain computationally tractable.
\end{itemize}


\subsection{Game-Theoretic AI Digest for \emph{Guiding} Attack and Defense} \label{section:strategic_feedback_loop}

\algrenewcommand{\alglinenumber}[1]{\tiny \textcolor{cai_primary}{#1:}}  

\begin{figure}[h!]
  \centering
  \begin{tikzpicture}[scale=1.0, every node/.style={scale=0.85, font=\sffamily},
      datanode/.style={circle, draw=graph_lightcyan!90!black, fill=graph_lightcyan, minimum size=1.3cm, align=center, font=\scriptsize, line width=0.8pt, drop shadow={opacity=0.15, shadow xshift=0.08cm, shadow yshift=-0.08cm}},
      keynode/.style={circle, draw=graph_teal!90!black, fill=graph_teal, text=white, minimum size=1.7cm, align=center, font=\small\bfseries, line width=1pt, drop shadow={opacity=0.25, shadow xshift=0.12cm, shadow yshift=-0.12cm}},
      supportnode/.style={rectangle, draw=graph_lightcyan!80!black, fill=graph_lightcyan, minimum width=1.6cm, minimum height=0.45cm, align=center, font=\tiny, rounded corners=2pt, line width=0.6pt},
      altnode/.style={rectangle, draw=graph_gray!80!black, fill=graph_gray, minimum width=1.5cm, minimum height=0.45cm, align=center, font=\tiny, rounded corners=2pt, line width=0.6pt},
  ]


  \node[font=\tiny\itshape, text=cai_primary!70] (attack_graph_label) at (-5.5, 6.2)
    {Example Generated Attack Graph with Probabilities};
  \node[anchor=north, below=0.02cm of attack_graph_label] (attack_graph) {
    \includegraphics[width=4.2cm]{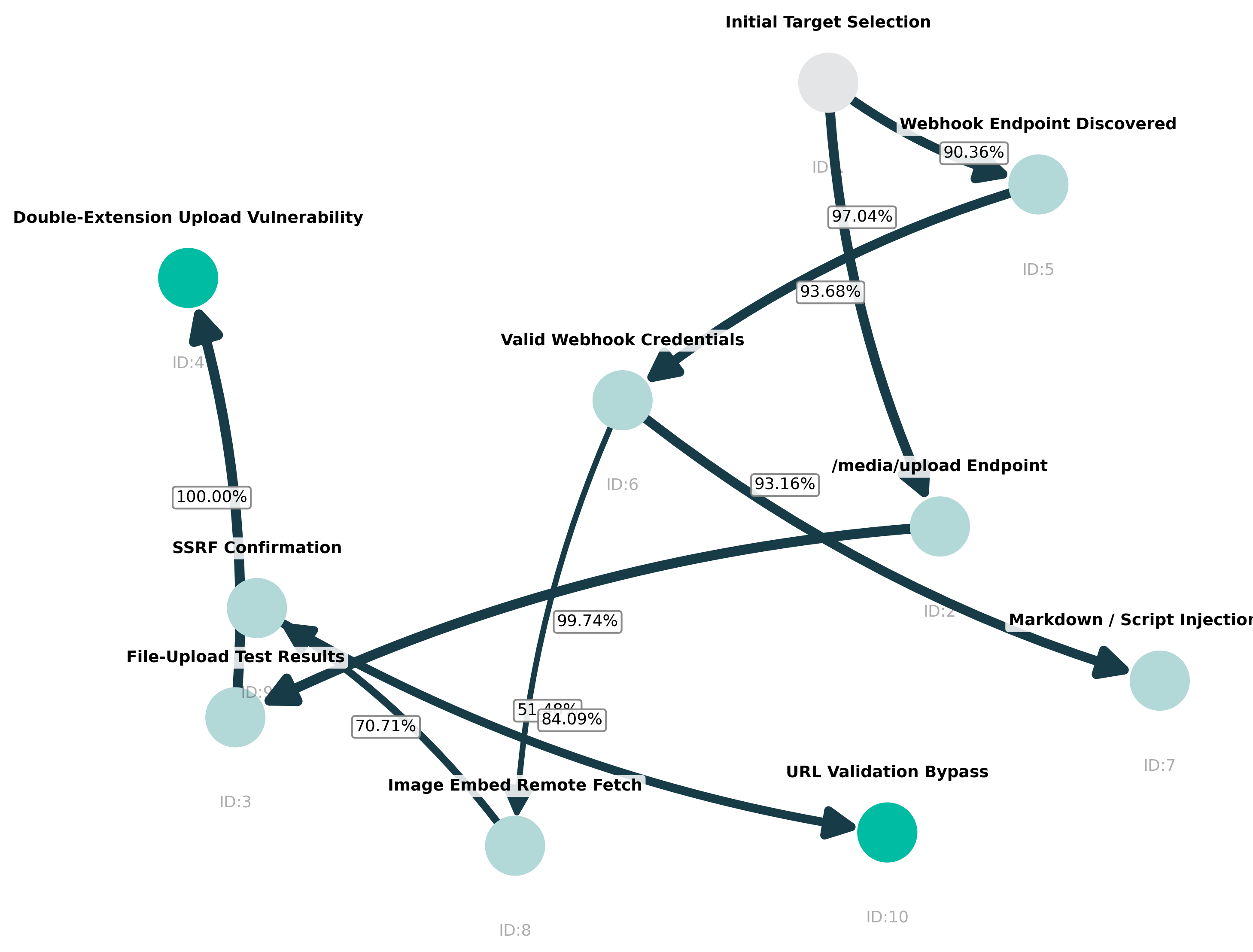}
  };

  \node[font=\tiny\itshape, text=cai_primary!70] (digest_label) at (0.2, 6.2)
    {Example Guidance Digest added to the Agent's System Prompt};
  \node[anchor=north, text width=5.5cm, font=\tiny\ttfamily, align=left,
        draw=graph_teal!40, fill=graph_teal!5, inner sep=4pt, rounded corners=3pt,
        line width=0.4pt, below=0.02cm of digest_label] (digest_example) {
      \textcolor{cai_primary}{\textbf{G-CTR Security Analysis (added to the System Prompt)}}\\[2pt]
      \textbf{Identified Attack Paths:}\\[1pt]
      Path 1: CTF Challenge $\rightarrow$[66\%] Nmap Scan $\rightarrow$[80\%] Open Ports Discovery $\rightarrow$[85\%] FTP Service (Port 21) $\cdots\rightarrow$[0\%] Target (FTP Service (Port 21))\\[1pt]
      Path 2: CTF Challenge $\cdots\rightarrow$[30\%] FTP Download Attempt 1 $\rightarrow$[57\%] FTP Service (Port 21) $\cdots\rightarrow$[0\%] Target (FTP Service (Port 21))\\[1pt]
      Path 3: CTF Challenge $\cdots\rightarrow$[17\%] FTP Download Attempt 2 $\cdots\rightarrow$[3\%] FTP Service (Port 21) $\cdots\rightarrow$[0\%] Target (FTP Service (Port 21))\\[2pt]
      
      \textbf{Critical Bottlenecks} (Attack Weaknesses):\\[1pt]
      - \texttt{FTP Download Attempt 2 -> FTP Service (Port 21)}: 3.1\% success rate\\[1pt]
      - \texttt{...}: ...\\[1pt]
      
  };

  \node[font=\tiny\itshape, text=cai_primary!70] (cai_label) at (5.7, 6.2)
    {Cybersecurity AI (CAI) Agent Prompt};
  \node[anchor=north, below=0.02cm of cai_label] (cai_logo) {
    \includegraphics[width=5.0cm]{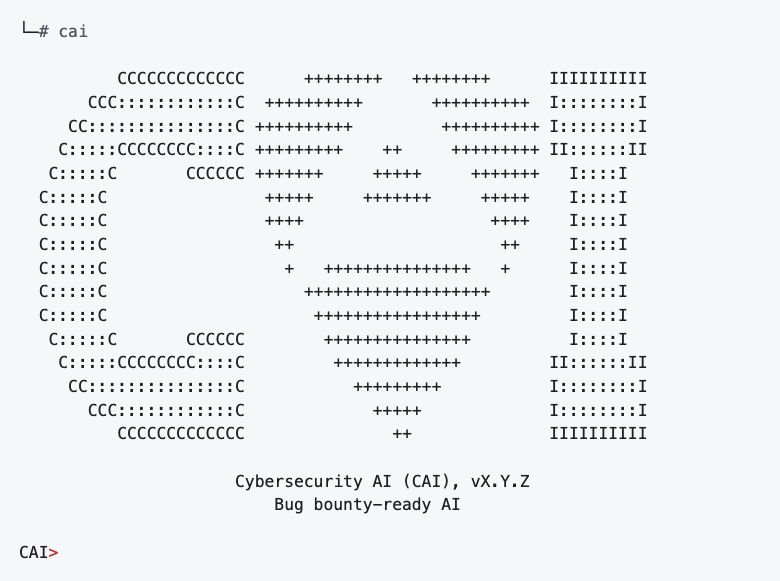}
  };

  \fill[graph_gray!15, rounded corners=6pt] (-7.8, -2.2) rectangle (-3.2, 2.2);
  \node[font=\scriptsize\itshape, text=graph_navy!60] at (-5.5, 2.5) {\textsc{Phase 1: Game-Theoretic AI Analysis (G-CTR)}};

  \fill[graph_teal!8, rounded corners=6pt] (-2.8, -2.2) rectangle (3.2, 2.2);
  \node[font=\scriptsize\itshape, text=graph_navy!60] at (0.2, 2.5) {\textsc{Phase 2: Game-Theoretic AI Guidance (Digest Generation)}};

  \fill[graph_lightcyan!15, rounded corners=6pt] (3.6, -2.2) rectangle (7.8, 2.2);
  \node[font=\scriptsize\itshape, text=graph_navy!60] at (5.7, 2.5) {\textsc{Phase 3: Agent Execution (ReAct)}};
  \node[keynode] (graph) at (-6.6, -1.3) {\small Attack\\Graph\\Generation};
  \node[datanode] (nash) at (-5.2, 0) {\scriptsize Nash\\Equilibrium};
  \node[altnode] (data) at (-4.0, 1.3) {\scriptsize G-CTR\\Results};
  
  \node[datanode] (algo) at (-1.2, 1.1) {\scriptsize Algorithmic\\ digest};
  \node[keynode] (llm) at (-1.2, -1.1) {\small LLM \\digest};
  \node[altnode, minimum size=1.0cm] (digest) at (1.5, 0) {\scriptsize Strategic\\\scriptsize Interpretation};
  
  \node[keynode] (agent) at (4.5, 0) {\small Plan\\ \scriptsize(LLM)};
  \node[datanode, minimum width=2.0cm] (actions) at (6.6, 1.0) {\small Act\\ \scriptsize(Tools)};
  \node[datanode] (observe) at (6.6, -1.0) {\scriptsize Scan \&\\Update\\ (Context)};
  
  \draw[-{Stealth[scale=1.2]}, line width=1.1pt, graph_navy]
      (graph.south east) to[bend left=-100] (nash.east);
  \draw[-{Stealth[scale=1.2]}, line width=1.1pt, graph_navy, rounded corners=4pt]
      (nash.west) to[bend left=100] (data.north west);
  
  \draw[-{Stealth[scale=1.2]}, line width=1.1pt, graph_navy, rounded corners=4pt]
      (data.east) .. controls +(1.5,0) and +(-1.2,0.6) .. (algo.west);
  \draw[-{Stealth[scale=1.2]}, line width=1.1pt, graph_navy, rounded corners=4pt]
      (data.east) .. controls +(1.5,0) and +(-1.2,-0.6) .. (llm.west);
  
  \draw[-{Stealth[scale=1.2]}, line width=1.1pt, graph_navy, rounded corners=4pt]
      (algo.east) .. controls +(0.8,0) and +(-0.8,0.5) .. (digest.north west);
  \draw[-{Stealth[scale=1.2]}, line width=1.1pt, graph_navy, rounded corners=4pt]
      (llm.east) .. controls +(0.8,0) and +(-0.8,-0.5) .. (digest.south west);
  
  \draw[-{Stealth[scale=1.2]}, line width=1.1pt, graph_navy]
      (digest.east) -- (agent.west);
  \draw[-{Stealth[scale=1.2]}, line width=1.1pt, graph_navy, rounded corners=4pt]
      (agent.north east) .. controls +(0.6,0.4) and +(-0.6,0) .. (actions.west);
  \draw[-{Stealth[scale=1.2]}, line width=1.1pt, graph_navy, rounded corners=4pt]
      (observe.west) .. controls +(-0.6,0) and +(0.6,-0.4) .. (agent.south east);
  
  \draw[-{Stealth[scale=1.2]}, line width=1.1pt, graph_navy, rounded corners=6pt]
      (actions.south) .. controls +(0,-0.4) and +(0,0.4) .. (observe.north);
  
  \draw[-{Stealth[scale=1.2]}, line width=1.1pt, dashed, graph_navy, rounded corners=8pt]
      (observe.south) .. controls +(0,-1.2) and +(0,-1.2) .. (graph.south)
      node[pos=0.5, below, font=\scriptsize, inner sep=10pt, rounded corners=1pt] {every 5 interactions ($\sim$80 tools)};
  
  \node[font=\tiny, text=black!80!black, fill=white, inner sep=1.5pt, rounded corners=1pt, draw=graph_teal!30, line width=0.3pt]
      at (-5.2, 0.5) {$<$5ms};
  \node[font=\tiny, text=graph_teal!80!black, fill=white, inner sep=1.5pt, rounded corners=1pt, draw=graph_teal!30, line width=0.3pt]
      at (-6.6, -0.5) {$\approx$20s};

  \node[font=\tiny, text=black!80!black, fill=white, inner sep=1.5pt, rounded corners=1pt, draw=graph_navy!20, line width=0.3pt]
      at (-1.2, 1.6) {$<$10ms};
  \node[font=\tiny, text=graph_teal!80!black, fill=white, inner sep=1.5pt, rounded corners=1pt, draw=graph_navy!20, line width=0.3pt]
      at (-1.2, -0.6) {$\approx$28.3s};
  \node[font=\tiny, text=graph_teal!80!black, fill=white, inner sep=1.5pt, rounded corners=1pt, draw=graph_navy!20, line width=0.3pt]
      at (4.5, 0.5) {$\approx$10s};

  \node[font=\tiny, text=black!80!black, fill=white, inner sep=1.5pt, rounded corners=1pt, draw=graph_navy!20, line width=0.3pt]
      at (6.6, 1.5) {$\approx$60s};

  \draw[line width=0.8pt, graph_navy!40] (3.3, -1.8) -- (3.3, 1.8);
  \draw[line width=0.8pt, graph_navy!40] (3.4, -1.8) -- (3.4, 1.8);



  \draw[line width=1.2pt, graph_teal!70, rounded corners=2pt] (-7.8, -3.6) -- (-7.8, -3.9) -- (3.3, -3.9) -- (3.3, -3.6);
  \node[font=\scriptsize\bfseries, text=white, fill=graph_teal, inner sep=3pt, rounded corners=2pt, draw=graph_teal!80!black, line width=0.6pt]
      at (-2.25, -3.9) {Time Budget: $\approx$50s};

  \draw[line width=1.2pt, graph_lightcyan!90!black, rounded corners=2pt] (3.4, -3.6) -- (3.4, -3.9) -- (7.8, -3.9) -- (7.8, -3.6);
  \node[font=\scriptsize\bfseries, text=white, fill=graph_lightcyan!90!black, inner sep=3pt, rounded corners=2pt, draw=graph_lightcyan!80!black, line width=0.6pt]
      at (5.6, -3.9) {Time Budget: $\approx$70s};

  \end{tikzpicture}

  \begin{minipage}[t]{0.48\textwidth}
    \begin{algorithm}[H]
      \tiny
      \caption{\textcolor{cai_primary}{\small \texttt{GenerateDigest()}: Game-Theoretic Digest for Guidance}}
      \label{alg:digest_generation}
      \begin{algorithmic}[1]
      \Require Attack graph $G$ \tikz[remember picture, overlay] \node (alg1start) {};
      \Ensure Strategic digest $D$ (markdown)
      \Statex
      \Statex \colorbox{gray!10}{\textcolor{cai_primary}{\textbf{\hspace{0.15cm}Phase 1: Game-Theoretic AI Analysis\hspace{0.15cm}}}}
      \State $N \gets \text{ComputeNashEquilibrium}(G)$ \Comment{\textcolor{cai_primary}{Nash equilibrium from attack graph}}
      \State $P \gets \text{IdentifyAttackPaths}(G, N)$ \Comment{\textcolor{cai_primary}{Probability-weighted paths}}
      \State $S \gets \text{InterpretStrategicPosition}(N)$ \Comment{\textcolor{cai_primary}{Attacker success probability}}
      \Statex
      \Statex \colorbox{gray!10}{\textcolor{cai_primary}{\textbf{\hspace{0.15cm}Phase 2: Strategic Interpretation\hspace{0.15cm}}}}
      \State $C \gets \text{IdentifyCriticalNodes}(P, G)$ \Comment{\textcolor{cai_primary}{High-variance decision points}}
      \State $B \gets \{(e, p) \in P \mid p < 0.95\}$ \Comment{\textcolor{cai_primary}{Bottlenecks: $p < 95\%$}}
      \State $H \gets \{(e, p) \in P \mid p > 0.90\}$ \Comment{\textcolor{cai_primary}{High-risk: $p > 90\%$}}
      \Statex
      \If{mode = \textcolor{cai_accent}{\textbf{algorithmic}}}
          \State $D \gets \text{FormatTemplate}(P, C, B, H)$ \Comment{\textcolor{cai_primary}{Rule-based digest}}
      \Else \Comment{\textcolor{cai_primary}{mode = \textcolor{cai_secondary}{\textbf{llm}}}}
          \State $\text{prompt} \gets \text{ConstructPrompt}(G, N, P, S)$
          \State $D \gets \text{LLMInference}(\texttt{alias1}, \text{prompt}, T{=}0.3)$
          \If{$D = \text{NULL}$} \Comment{\textcolor{cai_primary}{API failure $\rightarrow$ fallback}}
              \State $D \gets \text{FormatTemplate}(P, C, B, H)$
          \EndIf
      \EndIf
      \State \Return $D$ \tikz[remember picture, overlay] \node (alg1end) {};
      \end{algorithmic}
    \end{algorithm}
  \end{minipage}%
  \hfill
  \begin{minipage}[t]{0.48\textwidth}
    \begin{algorithm}[H]
      \tiny
      \caption{\textcolor{cai_primary}{\small Game-theoretic AI closed-loop feedback algorithm}}
      \label{alg:game_theoretic_ai}
      \begin{algorithmic}[1]
      \State $G \gets \text{InitializeGraph}()$, $D \gets \text{NULL}$ \Comment{\textcolor{cai_primary}{Initial attack graph, no digest}}
      \State $\pi \gets 0.0$, $i \gets 0$ \Comment{\textcolor{cai_primary}{Strategic position, agent interaction number}}
      \Statex
      \While{$\pi < 1.0 \land i < I_{\text{max}}$} \Comment{\textcolor{cai_primary}{Until success or max interactions}}
          \Statex \colorbox{gray!10}{\textcolor{cai_primary}{\textbf{\hspace{0.15cm}Phase 3: Agent Execution\hspace{0.15cm}}}}
          \State $a \gets \text{SelectAction}(G, D)$ \Comment{\textcolor{cai_primary}{Agent chooses action guided by digest}}
          \State $(s, o) \gets \text{Execute}(a)$ \Comment{\textcolor{cai_primary}{Tool execution, observe outcome}}
          \State $G \gets \text{UpdateGraph}(G, a, s, o)$ \Comment{\textcolor{cai_primary}{Refine graph with observations}}
          \State $i \gets i + 1$
          \Statex
          \If{$ i \bmod 5 = 0$} \Comment{\textcolor{cai_primary}{Trigger every 5 interactions ($\sim$80 tools)}}
              \Statex \colorbox{gray!10}{\textcolor{cai_primary}{\textbf{\hspace{0.15cm}Phase 1: Game-Theoretic AI Analysis\hspace{0.15cm}}}}
              \State $N \gets \text{ComputeNashEquilibrium}(G)$ \Comment{\textcolor{cai_primary}{Recompute equilibrium}}
              \State $\pi \gets \text{ExtractAttackerValue}(N)$ \Comment{\textcolor{cai_primary}{Update strategic position}}
              \Statex
              \Statex \colorbox{gray!10}{\textcolor{cai_primary}{\textbf{\hspace{0.15cm}Phase 2: Strategic Interpretation\hspace{0.15cm}}}}
              \State \tikz[remember picture, baseline=(alg2call.base)] \node[inner sep=0pt] (alg2call) {$D \gets \text{GenerateDigest}(G, N)$}; \Comment{\textcolor{cai_primary}{Update strategic guidance}}
          \EndIf
      \EndWhile
      \Statex
      \State \Return $G, \pi$ \Comment{Final attack graph and strategic outcome}
      \end{algorithmic}
    \end{algorithm}
  \end{minipage}

  \caption{Game-Theoretic AI closed-loop feedback architecture with three operational phases. The example attack graph (top-left) shows probability-weighted paths computed by G-CTR analysis. Strategic guidance using algorithm \ref{alg:digest_generation} produces the digest (top-center) that gets injected into the CAI agent's system prompt. The agent then executes security testing (top-right) with graph refinement every 5 interactions ($\sim$80 tool calls) by following algorithm \ref{alg:game_theoretic_ai}. Phases 1-2 operate with $\approx$50s budget, running parallel to Phase 3's $\approx$70s cycles for real-time strategic adaptation.}
  \label{fig:system_architecture}
\end{figure}

The integration between G-CTR and CAI establishes a closed-loop architecture where game-theoretic reasoning and AI-driven penetration testing mutually reinforce each other. This bidirectional strategic feedback mechanism transforms unstructured security narratives (CAI logs) into structured game-theoretic models (attack graphs and Nash equilibria), which in turn guide subsequent penetration tests toward more focused and effective security assessments. G-CTR's defender strategies identify critical nodes requiring additional security testing---the high-value defensive chokepoints where resource allocation yields maximum risk reduction. Conversely, attacker strategies highlight optimal exploitation paths---probability-weighted attack sequences that maximize success likelihood given observed system vulnerabilities.

To operationalize this closed-loop architecture, our game-theoretic AI architecture transforms raw Nash equilibrium computations (generated by G-CTR) into actionable strategic guidance through a \emph{digest generation} pipeline. Resulting digest is then integrated into the Agent's system prompt for further guiding the security exercise\footnote{We discovered empirically that guiding Agents via system prompt modifications was preferred. Alternatives explored included user and assistant prompts added to the LLM context. The latter showed worse results and hampered reasoning in mid- to long security exercises}. The system implements two interpretation modes---\textbf{algorithmic} (rule-based template processing) and \textbf{LLM} (linguistic reasoning)---both consuming identical game-theoretic inputs but producing strategically distinct outputs. This methodological framework enables rigorous empirical evaluation of the feedback loop's effectiveness, with detailed results presented in Section~\ref{subsection:strategic_feedback_loop_results} based on 44 independent penetration testing exercises.

Figure~\ref{fig:system_architecture} illustrates the operational feedback loop architecture. The system continuously cycles through four phases: (1) game-theoretic analysis computes Nash equilibria from the current attack graph ($<$5ms computational overhead), (2) digest generation translates equilibrium statistics into strategic guidance (algorithmic mode: $<$10ms; LLM mode: 10-46s mean), (3) agents execute tools guided by the injected digest, and (4) observations update the attack graph, triggering the next analysis cycle approximately every 5 interactions. This continuous refinement enables real-time strategic adaptation with negligible computational overhead ($<$2\% of total test duration for digest generation).

The two digest generation modes---\texttt{algorithmic} and \texttt{llm}---enable controlled empirical evaluation of strategic interpretation. Both modes consume identical Nash equilibrium inputs (game values, defender resource allocations, probability-weighted attack paths) but employ fundamentally different translation mechanisms as formalized in Algorithm~\ref{alg:digest_generation}. \emph{Bottlenecks} (p < 95\%) and \emph{high-risk transitions} (p > 90\%) are opposite ends of the attack path spectrum: bottlenecks are weak attack transitions where the defender has advantage andthe attacker struggles to succeed, while high-risk transitions are strong attack transitions where the attacker  succeeds easily and defenses are minimal. From a game-theoretic perspective, bottlenecks show where the attacker needs to improve capabilities or find alternatives, while high-risk transitions show where the defender should prioritize hardening efforts---together they reveal the strategic balance of power across the attack graph. The \texttt{algorithmic} mode applies fixed thresholds ($p > 0.9$ for high-risk transitions, $p < 0.5$ for bottlenecks) and assembles markdown templates via deterministic string concatenation. The \texttt{llm} mode sends all CTR data to the \texttt{alias1} model with a 350-word structured prompt (temperature=0.3) requesting interpretation across five sections: attack paths, bottlenecks, critical nodes, high-risk transitions, and tactical guidance.

Algorithm~\ref{alg:digest_generation} details the complete digest generation process, showing how both modes share a common preprocessing pipeline before diverging at the interpretation stage. The algorithmic branch assembles markdown sections via template formatting functions, while the LLM branch constructs a structured prompt and invokes external model inference. The fallback mechanism (line 12) ensures robustness: if LLM inference fails, the system automatically reverts to algorithmic mode. Across 44 tests (Section~\ref{subsection:strategic_feedback_loop_results}), zero digest generation failures occurred: when LLM mode encounters API errors or timeouts, the system automatically falls back to algorithmic mode, guaranteeing agents always receive strategic guidance. This operational robustness---zero test failures attributable to digest unavailability---validates the architectural decision to maintain both modes. The fallback mechanism proved essential for production deployment: while LLM mode provides superior performance (42.9\% success rate vs 20.0\%, as seen in Section~\ref{subsection:strategic_feedback_loop_results}), the algorithmic mode ensures system resilience against external service dependencies at negligible additional implementation cost. Algorithm~\ref{alg:game_theoretic_ai} presents the ReAct agent execution methodology augmented with the game-theoretic AI guidance every 5 interactions.

\FloatBarrier
\section{Results} \label{section:Results}

This section presents a comprehensive evaluation of the G-CTR framework across multiple dimensions, validating both its game-theoretic analysis capabilities and its strategic feedback mechanisms for guiding cyber attack and defense operations. Our evaluation is structured in four complementary parts. First, we present a \textbf{qualitative comparison} of attack graph quality (Section~\ref{subsection:qualitative_evaluation}, Table~\ref{tab:qualitative_comparison_attack_graphs}), contrasting LLM-generated outputs with human expert annotations across five real-world cybersecurity exercise domains to assess structural accuracy, vulnerability identification, and logical coherence. Second, we provide a \textbf{quantitative assessment} of LLM automation performance (Section~\ref{subsection:quantitative_evaluation}), measuring time efficiency, cost reduction, and computational overhead relative to manual expert analysis. Third, we evaluate the \textbf{efficiency of the strategic feedback loop} for guiding enhanced penetration testing (Section~\ref{subsection:strategic_feedback_loop_results}), demonstrating for the \texttt{shockwave-report} cyber range challenge \cite{sanzgomez2025cybersecurityaibenchmarkcaibench} how game-theoretic digest generation (Algorithm~\ref{alg:digest_generation}) translates Nash equilibrium computations into actionable guidance that improves CAI agent effectiveness across 44 independent attempts of this challenge. Finally, we present \textbf{ablation studies} (Section~\ref{subsection:ablations}) examining the impact of individual architectural components, including digest generation modes (algorithmic vs. LLM-based interpretation), feedback frequency, and strategic interpretation mechanisms. Together, these evaluations demonstrate that G-CTR not only automates attack graph generation with high fidelity but also closes the strategic feedback loop, enabling game-theoretic reasoning to guide operational security assessments in real-time.

\subsection{Qualitative Comparison: Graph Quality of Human Experts vs. LLMs} \label{subsection:qualitative_evaluation} 

This subsection presents a qualitative comparison of attack graphs from LLMs and human experts, summarized in Table \ref{tab:qualitative_comparison_attack_graphs}. The analysis focuses on structural accuracy (e.g., node count, vulnerability nodes) and logical coherence in representing attack scenarios, evaluating both the correctness and completeness of vulnerability identification and scenario flow.

\renewcommand{\arraystretch}{1.15}
\begin{longtable}{>{\raggedright\arraybackslash}p{2.5cm}>{\raggedleft\arraybackslash}p{1.8cm}>{\centering\arraybackslash}p{0.6cm}>{\centering\arraybackslash}p{0.6cm}>{\centering\arraybackslash}p{0.6cm}p{7.7cm}}
    \toprule
    \textbf{Domain} & \textbf{Graph Source} 
    & \textbf{\#N.} 
    & \textbf{\#V.} 
    & \textbf{Id.} 
    & \textbf{Scenario Representation} \\
    \midrule
    \endfirsthead
    \multicolumn{6}{c}%
    {\tablename\ \thetable\ -- \textit{Continued from previous page}} \\
    \toprule
    \textbf{Domain} & \textbf{Graph Source} 
    & \textbf{\#N.} 
    & \textbf{\#V.} 
    & \textbf{Id.} 
    & \textbf{Scenario Representation} \\
    \midrule
    \endhead

    \midrule \multicolumn{6}{r}{\textit{Continued on next page}} \\
    \endfoot

    \bottomrule
    \endlastfoot

        \multirow{6}{*}{\textbf{\textcolor{cai_primary}{kolesa.kz}}$^{\ref{appendix_kolesa}}$}
          & \texttt{gpt--4o}             & 4 & 1 & \textcolor{cai_primary}{\ding{51}} & Overly simplistic; it fails to convey the procedural complexity of the exercise (e.g., omission of key elements like the IP address).\\
          & \texttt{o3}                  & 6 & 1 & \textcolor{cai_primary}{\ding{51}} & It includes all essential information. \\
          & \texttt{grok--4}             & 5 & 1 & \textcolor{cai_primary}{\ding{51}} & It includes all essential information. \\
          & \texttt{claude sonnet 4}     & 8 & 1 & \textcolor{cai_primary}{\ding{51}} & Comprehensive. Includes all essential information along with relevant details, such as trial--and--error steps and final report generation. \\
          & \texttt{alias0}     & 7 & 1 & \textcolor{cai_primary}{\ding{51}} & Correct overall; includes trial--and--error, but it does not mention the final report. Node selection and information are appropriate. \\
          & \cellcolor{cai_light!30}\textcolor{cai_dark}{\textbf{Human}} & \cellcolor{cai_light!30}9 & \cellcolor{cai_light!30}1 & \cellcolor{cai_light!30}-- & \cellcolor{cai_light!30}-- \\
        \midrule

        \multirow{6}{*}{\textbf{\textcolor{cai_primary}{mercadolibre.com}}$^{\ref{appendix_mercadolibre}}$}
          & \texttt{gpt--4o}             & 6 & 1 & \textcolor{red}{\ding{55}} & Simplistic; it notes the API is vulnerable but fails to pinpoint the Insecure Direct Object
          Reference (IDOR).\\
          & \texttt{o3}                  & 6 & 1 & \textcolor{cai_primary}{\ding{51}} & Accurate and complete, correctly identifying the IDOR and scenario complexity, but assumes immediate exploitability, omitting trial--and--error. \\
          & \texttt{grok--4}             & 7 & 1 & \textcolor{cai_primary}{\ding{51}} & It includes all essential information. \\
          & \texttt{claude sonnet 4}     & 7 & 1 & \textcolor{cai_primary}{\ding{51}} & Comprehensive. Includes all essential information along with relevant details, such as trial--and--error steps and final report generation.  \\
          & \texttt{alias0}     & 7 & 1 & \textcolor{cai_primary}{\ding{51}} & Accurate; appropriate node selection. Demonstrates flow through trial-and-error and final report. \\
          & \cellcolor{cai_light!30}\textcolor{cai_dark}{\textbf{Human}} & \cellcolor{cai_light!30}7 & \cellcolor{cai_light!30}2 & \cellcolor{cai_light!30}-- & \cellcolor{cai_light!30}-- \\
        \midrule

        \multirow{6}{*}{\textbf{\textcolor{cai_primary}{pornbox.com}}$^{\ref{appendix_pornbox}}$}
          & \texttt{gpt--4o}             & 8 & 3 & \textcolor{cai_primary}{\ding{51}} & Comprehensive. Includes all essential information along with relevant details, such as trial--and--error steps and final report generation. \\
          & \texttt{o3}                  & 9 & 2 & \textcolor{cai_primary}{\ding{51}} & Accurate and complete, correctly identifying the vulnerabilities and scenario complexity, but assumes immediate exploitability, omitting trial--and--error. \\
          & \texttt{grok--4}             & 9 & 4 & \textcolor{cai_primary}{\ding{51}} & Comprehensive. Includes all essential information, including trial-and-error steps and final PoC generation. Effectively connects discovered vulnerabilities, notably through node ID:3 redirection. \\
          & \texttt{claude sonnet 4}     & 9 & 4 & \textcolor{cai_primary}{\ding{51}} & Comprehensive, with key details covered. Node selection is strong, though some labeled vulnerabilities are questionable.\\
          & \texttt{alias0}     & 9 & 3 & \textcolor{cai_primary}{\ding{51}} & It includes all essential information and effectively selects node information and vulnerable nodes. \\
          & \cellcolor{cai_light!30}\textcolor{cai_dark}{\textbf{Human}} & \cellcolor{cai_light!30}8 & \cellcolor{cai_light!30}3 & \cellcolor{cai_light!30}-- & \cellcolor{cai_light!30}-- \\
        \midrule

        \multirow{6}{*}{\textbf{\textcolor{cai_primary}{hm.com}}$^{\ref{appendix_hm}}$}
          & \texttt{gpt--4o}             & 12 & 1 & \textcolor{cai_primary}{\ding{51}} & Good overall; covers all key details, including trial--and--error. Effectively find the SSL Certificate Mismatch vulnerability. Relevant omissions: subdomain missing in node ID (though noted in extra info document) and hallucinated node numbering using \texttt{message\_id} instead of consistent IDs. \\
          & \texttt{o3}                  & 9 & 1 & \textcolor{cai_primary}{\ding{51}} & Comprehensive. Includes trial--and--error and final PoC. Correct nodes identified, but path 1–9 omits node 6; should be 1–6–9. Whether this omission is an issue is debatable. \\
          & \texttt{grok--4}             & 13 & 1 & \textcolor{cai_primary}{\ding{51}} & Comprehensive. Includes trial-and-error and final PoC. While nodes are correctly identified, the representation could be more concise, with potential for node merging. \\
          & \texttt{claude sonnet 4}     & 13 & 1 & \textcolor{cai_primary}{\ding{51}} & Comprehensive. Covers all key details, including trial--and--error and final PoC. Maintains coherent flow with well--connected steps throughout the scenario. \\
          & \texttt{alias0}     & 13 & 1 & \textcolor{cai_primary}{\ding{51}} & Comprehensive. It includes all essential information.  \\
          & \cellcolor{cai_light!30}\textcolor{cai_dark}{\textbf{Human}} & \cellcolor{cai_light!30}10 & \cellcolor{cai_light!30}2 & \cellcolor{cai_light!30}-- & \cellcolor{cai_light!30}-- \\
        \midrule

        \multirow{6}{*}{\textbf{\textcolor{cai_primary}{media.guilded.gg}}$^{\ref{appendix_mediaguildedgg}}$}
          & \texttt{gpt--4o}             & 14 & 2 & \textcolor{cai_primary}{\ding{51}} & Comprehensive. Covers all key details, including trial--and--error and final PoC. Maintains coherent flow with well--connected steps.\\
          & \texttt{o3}                  & 10 & 2 & \textcolor{cai_primary}{\ding{51}}  & Comprehensive. Covers all key details, though some trial-and-error steps are missing. Vulnerability is well represented. \\
          & \texttt{grok--4}             & 13 & 2 & \textcolor{cai_primary}{\ding{51}} & Comprehensive. Includes trial-and-error and final PoC. Maintains clear, sequential flow through a single coherent path. Vulnerability is well represented. Minor aspect: Attack graphs are meant to represent the exercise conceptually, not replicate every step in a single linear path; thus, whether a single path accurately reflects this abstraction is debatable. \\
          & \texttt{claude sonnet 4}     & 13 & 4 & \textcolor{cai_primary}{\ding{51}} & Comprehensive. Includes trial--and--error and final report. Maintains clear flow. Node selection and information are appropriate. However, some nodes labeled as vulnerable (e.g., node ID: 13) do not represent actual vulnerabilities but rather report elements from the exercise. \\
          & \texttt{alias0}     & 13 & 4 & \textcolor{cai_primary}{\ding{51}} & Representative of the overall exercise, though some nodes could be consolidated to reduce redundancy.  \\
          & \cellcolor{cai_light!30}\textcolor{cai_dark}{\textbf{Human}} & \cellcolor{cai_light!30}14 & \cellcolor{cai_light!30}2 & \cellcolor{cai_light!30}-- & \cellcolor{cai_light!30}-- \\
        \bottomrule

\caption{Qualitative comparison of attack-graph content generated by large-language models (LLMs) and human experts across five exercise domains. \texttt{Graph Source} refers to the origin of the graph analyzed in the evaluation. This includes the specific model or the human expert. \texttt{\#N.} refers to the total number of nodes present in the graph, while \texttt{\#V.} Nodes indicates how many of those nodes are considered vulnerable within the scenario. \texttt{Id.} denotes whether the model (or human) correctly identified the presence of vulnerabilities (\textcolor{cai_primary}{\ding{51}}) or failed to do so (\ding{55}). \texttt{Scenario Representation} is a qualitative evaluation describing issues observed in the graph, such as hallucinations, overly simple or complex structures, unnecessary data, missing exploration attempts, or missing logical connections. See Appendices \ref{appendix_kolesa}, \ref{appendix_mercadolibre}, \ref{appendix_pornbox}, \ref{appendix_hm}, \ref{appendix_mediaguildedgg} for full attack graphs examples and evaluation details.}
\label{tab:qualitative_comparison_attack_graphs}
\end{longtable}
\renewcommand{\arraystretch}{1.0}

In the \textbf{kolesa.kz} scenario, \texttt{claude-sonnet-4} stood out for producing the most complete and coherent attack graph. It included 8 nodes in total, identifying 1 as vulnerable--closely matching the human--generated reference, which had 9 nodes and 1 vulnerability. Claude’s graph effectively captured the procedural flow, including intermediate steps such as trial--and--error and the generation of a final report. In contrast, \texttt{gpt-4o}'s graph was overly simplistic, containing only 4 nodes and omitting key elements. While both \texttt{o3} and \texttt{grok-4} generated correct and complete representations with 6 and 5 nodes respectively, their graphs lacked the depth of procedural granularity shown by Claude. \texttt{alias0} produced a structurally sound Attack Graph, only missing some information related to the final report. Notably, all models correctly identified the vulnerability in this domain, but the richness and interpretability of the graphs varied significantly.

In the \textbf{mercadolibre.com} exercise, a scenario centered around an IDOR vulnerability, \texttt{claude-sonnet} \texttt{--4} and \texttt{grok-4} again delivered robust results. Each generated 7-node graphs identifying 1 vulnerable node, reflecting a strong understanding of the attack path. The human expert's graph (involving 1 human expert)
was slightly more granular, with 7 nodes and 2 marked as vulnerable, reflecting a higher sensitivity to nuanced security elements.  \texttt{o3} produced a 6-node graph with a correct vulnerability identification, although it assumed immediate exploitability and omitted trial-and-error details. \texttt{gpt-4o}, while producing a structurally simple 6--node graph, failed to identify the key vulnerability, significantly limiting the usefulness of its representation. \texttt{alias0} also generated a 7--node graph and correctly retrieved the IDOR vulnerability. This case highlights how vulnerability detection accuracy, combined with appropriate procedural modeling, is essential to generating actionable graphs.

The \textbf{pornbox.com} scenario was particularly effective in revealing differences in model performance. The human expert constructed an 8-node graph with 3 vulnerabilities, as correctly modeled by \texttt{alias0}, while both \texttt{claude-sonnet-4} and \texttt{grok-4} exceeded this level of detail with 9 nodes and 4 vulnerabilities. Grok and Claude also demonstrated an advanced modeling capability by integrating redirection logic--specifically, using node ID: 3 to connect different paths in the graph. \texttt{o3} generated a similarly detailed graph (9 nodes, 2 vulnerabilities), capturing core aspects but omitting some of the more subtle connections. \texttt{gpt-4o} produced a graph with 8 nodes and 3 vulnerabilities, matching the human graph in numbers but lacking some of the deeper logical links.

In the \textbf{hm.com} exercise, which focused on identifying an SSL certificate mismatch, all models correctly detected the vulnerability. However, differences emerged in how they structured the graphs. \texttt{claude-sonnet-4} and \texttt{grok-4} produced highly detailed graphs with 13 nodes each, while \texttt{gpt-4o} and \texttt{o3} had 12 and 9 nodes, respectively. The human graph contained 10 nodes, 2 labeled as vulnerable. \texttt{gpt-4o}'s graph, although mostly correct, introduced inconsistencies by using message identifiers instead of coherent node labels, which compromised clarity. \texttt{o3} omitted an intermediate node in the attack path, slightly reducing the traceability of the exploit sequence. \texttt{Claude-sonnet-4}’s and \texttt{alias0} output maintained a logical progression and included all necessary steps, making it one of the strongest performers in this domain.

Finally, \textbf{media.guilded.gg} represented the most complex scenario in the evaluation. The human expert generated a 14-node graph with 2 vulnerabilities. All models approached or matched this level of detail: \texttt{gpt-4o} also created 14 nodes and 2 vulnerabilities, while \texttt{claude-sonnet-4} and \texttt{grok-4} produced 13 nodes each, though \texttt{claude-sonnet-4} labeled 4 as vulnerable, potentially overestimating due to misclassification of informational elements. \texttt{claude-sonnet-4} and \texttt{grok-4} maintained a clear and coherent path through the graph, effectively mirroring the logical flow of the attack.  \texttt{o3} delivered a slightly leaner graph with 10 nodes and 2 vulnerabilities, still accurate but missing some trial-and-error steps. \texttt{alias0}'s Attack Graph was consistent, though some nodes could be merged without losing meaning.

In summary, across five cybersecurity exercise domains, LLMs generally demonstrated strong capabilities to identify relevant nodes that represent key attack stages. Most models accurately flagged vulnerable nodes. Compared to human-generated graphs, which were more comprehensive and detailed, LLMs identified between 70\% to 90\% of the key nodes found by human experts but often missed finer details, particularly stages like reconnaissance and final reports.

\subsection{Quantitative Evaluation: Performance of LLM Automation vs. Gold Standard/Human Experts} \label{subsection:quantitative_evaluation}

\begin{table}[!h]
    \centering
    \small
    \renewcommand{\arraystretch}{1.3}
    \setlength{\tabcolsep}{5pt}
    \begin{tabular}{llrrrrrr}
        \toprule
        \textbf{Domain} & \textbf{LLM Model} 
        & \textcolor{cai_primary}{\textbf{$t_{\text{Attack Graph}}$} (s)} 
        & \textcolor{cai_primary}{\textbf{$t_{\text{CTR}}$} (s)} 
        & \textcolor{cai_primary}{\textbf{$t_{\text{total}}$} (s)}
        & \textcolor{cai_primary}{\textbf{Cost (\$)}} 
        & \textbf{$t_{ratio}$} 
        & \textcolor{cai_primary}{\textbf{\timecolor{$c_{ratio}$}}} \\
        \midrule

        \multirow{5}{*}{\textbf{kolesa.kz}$^{\ref{appendix_kolesa}}$}
        & \texttt{gpt--4o}      & \textbf{14} & 0.004 & \textbf{15.0} & 0.12 & $\textbf{128} \times$ & 187$\times$ \\
        & \texttt{o3}         & 16 & 0.003 & 17 & \textbf{0.10}s  & 112 $\times$ & \textbf{225} $\times$  \\
        & \texttt{grok--4}     & 30 & 0.003 & 31 & 0.15 & 60 $\times$  & 150 $\times$  \\
        & \texttt{claude-sonnet-4} & 16 & 0.003 & 17 & 0.16 & 112$\times$  & 140 $\times$  \\
        & \texttt{alias0} & 20 & 0.003 & 20 & 0.35 & 88 $\times$  & 62 $\times$  \\
          & \textcolor{human_color}{\textbf{Human}} 
          & \textcolor{human_color}{\textbf{1800} (30 m)} 
         & \textcolor{human_color}{\textbf{---}}
         & \textcolor{human_color}{\textbf{---}}
         & \textcolor{human_color}{\textbf{22.5}} 
         & \textcolor{human_color}{\textbf{---}} 
         & \textcolor{human_color}{\textbf{---}} \\
        \midrule

        {\textbf{mercadolibre.com}$^{\ref{appendix_mercadolibre}}$}
        & \texttt{gpt--4o}               & \textbf{9.4} & 0.003 & \textbf{10} & \textbf{0.05} & \textbf{191} $\times$  & \textbf{450} $\times$  \\
        & \texttt{o3}         & 27 & 0.004 & 27 & 0.05 & 66 $\times$  & \textbf{450} $\times$  \\
        & \texttt{grok--4}           & 27 & 0.004 & 24 & 0.07 & 66 $\times$  & 321 $\times$  \\
        & \texttt{claude--sonnet--4}           & 12 & 0.003 & 18 & 0.09 & 163 $\times$ & 250 $\times$ \\
        & \texttt{alias0} & 15 & 0.003 & 15 & 0.16 & 116 $\times$  & 145 $\times$  \\
        & \textcolor{human_color}{\textbf{Human}} 
          & \textcolor{human_color}{\textbf{1800} (30 m)} 
         & \textcolor{human_color}{\textbf{---}}
         & \textcolor{human_color}{\textbf{---}}
         & \textcolor{human_color}{\textbf{22.5}}
         & \textcolor{human_color}{\textbf{---}}
         & \textcolor{human_color}{\textbf{---}} \\
        \midrule

        {\textbf{pornbox.com}$^{\ref{appendix_pornbox}}$}
        & \texttt{gpt--4o}          & \textbf{11} & 0.005 & \textbf{12} & 0.12 & \textbf{245} $\times$ & 281 $\times$ \\
        & \texttt{o3}               & 22 & 0.005 & 23 & \textbf{0.11} & 122 $\times$ & \textbf{306} $\times$ \\
        & \texttt{grok--4}           & 27 & 0.004 & 28 & 0.15 & 100 $\times$ & 225 $\times$ \\
        & \texttt{claude-sonnet-4}  & 13 & 0.004 & 14 & 0.19 & 207 $\times$ & 177 $\times$ \\
        & \texttt{alias0} & 23 & 0.004 & 24 & 0.36 & 113 $\times$ & 94 $\times$ \\
        & \textcolor{human_color}{\textbf{Human}} 
          & \textcolor{human_color}{\textbf{2700} (45 m)} 
         & \textcolor{human_color}{\textbf{---}}
         & \textcolor{human_color}{\textbf{---}}
         & \textcolor{human_color}{\textbf{33.75}} 
         & \textcolor{human_color}{\textbf{---}} 
         & \textcolor{human_color}{\textbf{---}} \\
        \midrule

        {\textbf{hm.com}$^{\ref{appendix_hm}}$} 
        & \texttt{gpt--4o}            & \textbf{20} & 0.004 & \textbf{21} & 0.1 & \textbf{135}  $\times$ & 373  $\times$ \\
        & \texttt{o3}               & 26 & 0.004 & 27 & \textbf{0.09} & 103  $\times$ & \textbf{375}  $\times$ \\
        & \texttt{grok--4}           & 36 & 0.004 & 37 & 0.13 & 75  $\times$ & 259  $\times$ \\
        & \texttt{claude-sonnet-4}   & 30 & 0.003 & 31 & 0.13 & 90  $\times$ & 259  $\times$ \\
        & \texttt{alias0} & 26 & 0.004 & 26 & 0.28 & 102 $\times$  & 120 $\times$  \\
        & \textcolor{human_color}{\textbf{Human}} 
          & \textcolor{human_color}{\textbf{2700} (45 m)} 
         & \textcolor{human_color}{\textbf{---}}
         & \textcolor{human_color}{\textbf{---}}
         & \textcolor{human_color}{\textbf{33.75}} 
         & \textcolor{human_color}{\textbf{---}} 
         & \textcolor{human_color}{\textbf{---}} \\
        \midrule

        {\textbf{media.guilded.gg}$^{\ref{appendix_mediaguildedgg}}$}
        & \texttt{gpt--4o}           & 56 & 0.004 & 58 & 0.19 & 86 $\times$  & 355 $\times$ \\
        & \texttt{o3}               & \textbf{32} & 0.004 & \textbf{35} & \textbf{0.18} & \textbf{168} $\times$  & \textbf{375} $\times$  \\
        & \texttt{grok--4}           & 41 & 0.005 & 44 & 0.27 & 131 $\times$  & 250 $\times$  \\
        & \texttt{claude-sonnet-4}   & 38 & 0.005 & 41 & 0.28 & 142 $\times$  & 241 $\times$ \\
        & \texttt{alias0} & 46 & 0.006 & 47 & 0.64 & 115 $\times$  & 104 $\times$  \\
          & \textcolor{human_color}{\textbf{Human}} 
          & \textcolor{human_color}{\textbf{5400} (90 m)} 
         & \textcolor{human_color}{\textbf{---}}
         & \textcolor{human_color}{\textbf{---}} 
         & \textcolor{human_color}{\textbf{67.5}} 
         & \textcolor{human_color}{\textbf{---}}
         & \textcolor{human_color}{\textbf{---}} \\
        \bottomrule
    \end{tabular}
     \caption{Comparison of Attack Graph construction time and cost between LLMs and humans across five domains. For each domain, five LLMs generate Attack Graphs, which are compared against a human expert annotation Gold Standard. \textbf{$t_{\text{Attack Graph}}$} denotes the time required by an LLM to generate an attack graph; for humans, it reflects expert annotation time. \textbf{$t_{\text{CTR}}$} represents the time taken to compute the CTR pipeline. \textbf{$t_{\text{total}}$} is the end--to--end execution time of the full \texttt{G-CTR} pipeline. \textbf{Cost} reflects the LLM inference cost or equivalent human labor cost. In the latter case, we assume an hourly rate of \$45. \textbf{$t_{ratio}$} and \textbf{$c_{ratio}$} express relative time and cost efficiency, normalized against the human baseline. For numbers above 10 we do not show decimals; for values below 10, we keep one decimal and round accordingly. For \textbf{$t_{\text{CTR}}$}, we preserve three decimal places, rounding the last digit.}
    
\label{tab:llm_vs_human_efficiency_all_domains}
\end{table}

To quantify the efficiency and scalability gains of LLM-based automation, we benchmarked five state-of-the-art language models against gold-standard annotations produced by human security experts across five real-world cybersecurity exercises. Table~\ref{tab:llm_vs_human_efficiency_all_domains} presents comprehensive performance metrics including execution time, inference cost, computational overhead, and efficiency ratios relative to manual analysis.

\paragraph{Time Efficiency.}
LLM-based attack graph generation demonstrated substantial temporal advantages over manual annotation. Across all exercises and models, automated generation required 10--46 seconds per exercise, exhibiting task-dependent variation correlated with log complexity (message count and scenario depth). In contrast, human experts required 30--90 minutes per exercise for equivalent graph construction, yielding speedup factors ranging from \textbf{60$\times$ to 245$\times$} relative to manual workflows. Model-specific performance varied systematically: \texttt{gpt-4o} achieved the lowest median latency across four of five domains, while \texttt{o3} delivered optimal performance (32 seconds) on the most complex exercise (\texttt{media.guilded.gg}, 358 messages). The slowest models, \texttt{grok-4} and \texttt{alias0}, required 27--41 seconds per execution, remaining two orders of magnitude faster than human annotation.

\paragraph{Cost Efficiency.}
Economic analysis reveals similarly dramatic cost advantages for automated generation. LLM inference costs ranged from \$0.05 to \$0.64 per exercise, depending on model architecture and input complexity. Human expert annotation, estimated at a conservative labor rate of \$45/hour, cost \$22.50--\$67.50 per exercise, resulting in \textbf{cost reduction factors of 62$\times$ to 450$\times$}. Cost efficiency varied substantially across models: \texttt{o3} and \texttt{gpt-4o} consistently achieved the lowest inference costs (\$0.05--\$0.18), while \texttt{claude-sonnet-4} and \texttt{alias0} were more expensive (\$0.09--\$0.64). Notably, the higher-cost models produced qualitatively superior outputs (Section~\ref{subsection:qualitative_evaluation}), suggesting a quality-cost tradeoff wherein premium models deliver enhanced structural completeness and logical coherence at 2--4$\times$ higher inference cost while maintaining 100$\times$+ cost advantages over human annotation.

\paragraph{Computational Overhead of Game-Theoretic Analysis.}
The Cut-the-Rope (CTR) game-theoretic computation layer, responsible for Nash equilibrium calculation and strategic analysis, introduced negligible computational overhead: all CTR operations completed in under 5 milliseconds across all exercises and graph sizes. This empirical result confirms that the computational bottleneck in G-CTR lies exclusively in LLM inference rather than game-theoretic reasoning, validating the architectural decision to integrate CTR analysis without compromising real-time performance requirements.

\paragraph{Model Selection and Performance Tradeoffs.}
Cross-model comparison reveals systematic tradeoffs between efficiency and output quality. \texttt{gpt-4o} and \texttt{o3} optimize for speed and cost, delivering adequate structural accuracy for high-throughput or resource-constrained deployments. Conversely, \texttt{claude-sonnet-4} and \texttt{alias0} prioritize graph completeness and interpretability, exhibiting superior performance on complex multi-stage exercises such as \texttt{hm.com} (118 messages) and \texttt{media.guilded.gg} (358 messages). These empirical performance characteristics enable context-dependent model selection: rapid triage and large-scale assessments benefit from lightweight models, while high-stakes investigations requiring maximal fidelity justify premium model deployment.



\subsection{Game-Theoretic AI Digest Results for Guiding Attack and Defense} \label{subsection:strategic_feedback_loop_results}

This subsection presents empirical validation of this feedback loop architecture through 44 independent penetration testing exercises targeting the \textcolor{cai_primary}{\texttt{shockwave\_report}} Cyber Range Challenge (Shellshock CVE-2014-6271 vulnerability). We demonstrate how Nash equilibrium analysis operationally guides offensive security testing in real-time scenarios, quantifying improvements in vulnerability discovery rates, time-to-exploitation and economic efficiency. The empirical evidence establishes that strategic game-theoretic guidance enables autonomous security systems to not merely find vulnerabilities, but to reason strategically about their exploitation in a more efficient and effective manner.

The evaluation employed a randomized comparative design across three configurations, all using the state-of-the-art~\cite{sanzgomez2025cybersecurityaibenchmarkcaibench} \texttt{alias1} Cybersecurity LLM\footnote{Refer to \url{https://aliasrobotics.com/alias1.php} for more details and to \cite{sanzgomez2025cybersecurityaibenchmarkcaibench} for benchmark results.} and using the \texttt{red\_teamer} Agent in CAI. The configurations were: No G-CTR baseline ($n=15$), G-CTR with LLM-generated strategic digests ($n=14$), and G-CTR with algorithmic rule-based digests ($n=15$). All tests targeted identical infrastructure (\texttt{shockwave\_report} Cyber Range CTF) with maximum duration 2410 seconds (40 minutes). Success was defined as flag extraction within time limit. 
Table~\ref{tab:mode_performance} summarizes performance results.

\begin{table}[h]
    \centering
    \caption{Empirical performance comparison: algorithmic vs. LLM digest modes ($n=44$ tests).$^\dagger$ The cost per success is calculated based on the pricing of the \texttt{alias1} model (5 EUR per 1M tokens in + out) and the total cost of the test and the number of successful tests. The reason why the cost per success is higher for the baseline is because despite having a shorter duration, it had more tool calls which saturated faster the cost per LLM interaction.}
    \label{tab:mode_performance}
    \small
    \begin{tabular}{@{}lcccccc@{}}
    \toprule
    \textcolor{cai_color}{\textbf{Mode}} & \textcolor{cai_color}{\textbf{Tests}} & \textcolor{cai_color}{\textbf{Successes}} & \textcolor{cai_color}{\textbf{Success Rate}} & \textcolor{cai_color}{\textbf{Avg Duration}} & \textcolor{cai_color}{\textbf{Tool Variance}} & \textcolor{cai_color}{\textbf{Cost/Success}} \\
    \midrule
    \rowcolor{gray!5}
    \textbf{No G-CTR} & 15 & 2 & 13.3\% & 16.7 min & 1.6$\times$ & \$2.71$^\dagger$ \\
    \textcolor{cai_color}{(baseline)} & & & & & & \\
    \addlinespace
    \textbf{G-CTR \texttt{algorithmic}} & 15 & 3 & 20.0\% & 22.5 min & 6.2$\times$ & \$0.32 \\
    \textcolor{cai_color}{$\Delta$ over baseline} & & \textcolor{cai_color}{+1} & \textcolor{cai_color}{+6.7 pp} & \textcolor{cai_color}{-5.8 min} & \textcolor{cai_color}{-3.9$\times$} & \textcolor{cai_color}{8.5$\times$ better} \\
    \addlinespace
    \textbf{G-CTR \texttt{llm}} & 14 & 6 & 42.9\% & 20.2 min & 1.2$\times$ & \$0.12 \\
    \textcolor{cai_color}{$\Delta$ over baseline} & & \textcolor{cai_color}{+4} & \textcolor{cai_color}{+29.6 pp} & \textcolor{cai_color}{-3.5 min} & \textcolor{cai_color}{1.3$\times$ better} & \textcolor{cai_color}{23$\times$ better} \\
    \addlinespace
    \rowcolor{gray!10}
    \textbf{Digest comparison} & & \textbf{+3} & \textbf{+22.9 pp} & \textbf{+2.3 min} & \textbf{5.2$\times$ lower} & \textbf{2.7$\times$ better} \\
    ($\Delta$ \texttt{llm} vs. \texttt{algorithmic}) & & & & & & \\
    \bottomrule
    \end{tabular}
    \end{table}


The 44-test evaluation dataset enables direct comparison of digest generation modes under controlled conditions.  The \texttt{llm} digest generation achieves a 29.6\% higher success rate than \texttt{algorithmic} mode (42.9\% vs 20.0\%) despite similar Nash equilibrium inputs\footnote{Despite using the same game-theoretic method, we can't guarantee they are identical but similar due to the stochastic nature of the LLMs used to produce the corresponding attack graph as depicted in Figure \ref{fig:system_architecture}.}. This isolates the contribution of LLM strategic interpretation: both modes receive similar game-theoretic analysis, but LLM digests provide contextual guidance that agents operationalize more effectively. The 5.2$\times$ lower tool usage variance (1.2$\times$ vs 6.2$\times$ range) indicates more consistent agent behavior under LLM guidance, suggesting the LLM interpretation reduces behavioral uncertainty.

\begin{figure}[h!]
\centering
\begin{tikzpicture}
\begin{axis}[
    width=0.95\textwidth,
    height=7cm,
    ylabel={Time (minutes)},
    ylabel style={font=\large\bfseries},
    symbolic x coords={No G-CTR, G-CTR algorithmic, G-CTR llm},
    xtick=data,
    xticklabels={\text{No G-CTR}, \text{G-CTR \textcolor{cai_color}{\texttt{algorithmic}}}, \text{G-CTR \textcolor{cai_color}{\texttt{llm}}}},
    xticklabel style={font=\normalsize\bfseries,anchor=north,yshift=-2pt},
    yticklabel style={font=\normalsize},
    tick label style={font=\normalsize},
    nodes near coords,
    every node near coord/.style={font=\bfseries\normalsize,black,yshift=3pt},
    legend style={at={(0.5,-0.18)},anchor=north,legend columns=2,font=\normalsize,
                 cells={anchor=west},
                 draw=black,line width=1pt,fill=white,fill opacity=0.95},
    ymin=-3,
    ymax=135,
    bar width=35pt,
    enlarge x limits=0.25,
    axis x line*=bottom,
    axis y line*=left,
    ymajorgrids=true,
    xmajorgrids=false,
    grid style={draw=gray!10,line width=0.3pt},
    every axis plot/.append style={mark=none},
    clip=false
]
\addplot+[sharpbar,mark=*,mark size=3pt,mark options={fill=cai_dark,draw=cai_dark},fill=cai_dark!40,draw=black,line width=0.3pt] coordinates {(No G-CTR,16.7) (G-CTR algorithmic,22.5) (G-CTR llm,20.2)};
\addplot+[sharpbar,mark=*,mark size=3pt,mark options={fill=cai_primary!40,draw=cai_dark},fill=cai_primary!85,draw=black,line width=0.3pt] coordinates {(No G-CTR,126) (G-CTR algorithmic,112.5) (G-CTR llm,47)};

\legend{Average Time Duration ($T_{\text{avg}}$), Expected Time $\mathbb{E}[T_{\text{success}}]$ (Eq.~\ref{eq:expected_time})}
\end{axis}
\end{tikzpicture}
\caption{Time-to-vulnerability comparison across methods. \textcolor{cai_dark}{$\bullet$} Average Time Duration $(T_{\text{avg}})$. \textcolor{cai_primary!40}{$\bullet$} Expected time $\mathbb{E}[T_{\text{success}}] = T_{\text{avg}} / P_{\text{success}}$ from Eq.~\ref{eq:expected_time}, which accounts for failed attempts. Despite 21\% longer raw duration (20.2 vs 16.7 min), G-CTR \texttt{llm} delivers a 2.67$\times$ reduction in expected time (126 $\rightarrow$ 47 min) relative to No G-CTR because of its 3.21$\times$ higher success probability.}
\label{fig:time_analysis}
\end{figure}
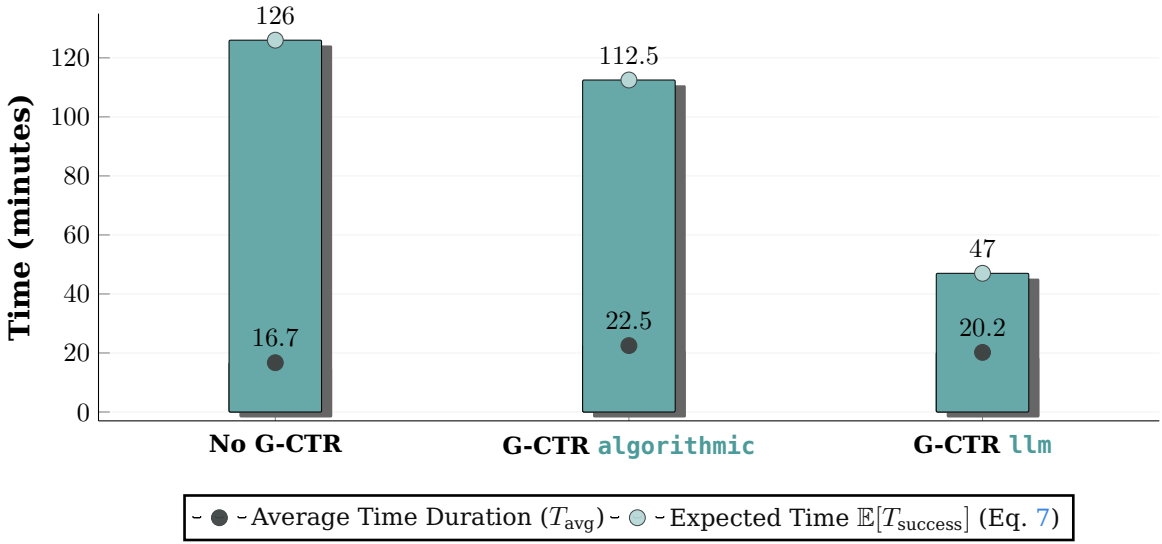

Figure~\ref{fig:time_analysis} visualizes the temporal efficiency analysis, accounting for failure probability in expected time calculations. While G-CTR \texttt{llm}-guided successful tests exhibit 21\% longer individual duration (20.2 vs 16.7 min), the 3.21$\times$ improvement in success probability dominates the performance metric, yielding 2.67$\times$ reduction in expected time to vulnerability discovery. The expected time metric ($\mathbb{E}[T_{\text{success}}]$) properly accounts for the cost of failed attempts as shown in Equation~\ref{eq:expected_time}:

\begin{equation}
\mathbb{E}[T_{\text{success}}] = \frac{T_{\text{avg}}}{P_{\text{success}}} = \begin{cases}
126 \text{ min} & \text{\textbf{No G-CTR}: } \frac{16.7}{0.133} \\
112.5 \text{ min} & \text{\textbf{G-CTR \texttt{algorithmic}}: } \frac{22.5}{0.200} \\
47 \text{ min} & \text{\textbf{G-CTR \texttt{llm}}: } \frac{20.2}{0.429}
\end{cases}
\label{eq:expected_time}
\end{equation}

Taken together, these results confirm that \texttt{llm} digest-based strategic guidance delivers \textbf{substantial improvements}
to offensive cybersecurity exercises in Cyber Range CTFs, achieving a remarkable 29.6 percentage point increase in success rate (from 13.3\% to 42.9\%—a 222\% relative improvement), reducing the expected time to success by 2.67$\times$ (126 $\rightarrow$ 47 min) and the cost per success by 23$\times$ (from \$2.71 to \$0.12).

\subsection{Ablations} \label{subsection:ablations}

To substantiate the aggregate performance trends reported in Section~\ref{subsection:strategic_feedback_loop_results}, we conduct ablation experiments across two complementary evaluation regimes. The first subsubsection extends the analysis of the \textcolor{cai_primary}{\texttt{shockwave\_report}} Cyber Range Challenge, examining in detail how the two digest generation strategies (\texttt{algorithmic} vs. \texttt{llm}) impact system performance. The second subsubsection transitions to Attack and Defense (A\&D) CTF exercises, evaluating how the same architectural feedback loop performs under competitive, score-driven conditions that better reflect real-world cybersecurity operations.

\subsubsection{LLM Digest vs. Algorithmic Digest: Isolating the Interpretation}

We isolate the role of LLM interpretation by further dissecting the results from Section~\ref{subsection:strategic_feedback_loop_results}. Table~\ref{tab:digest_modes} summarizes the architectural characteristics and empirical observables recorded across the comparison cohort. Table~\ref{tab:llm_vs_algo} aggregates the primary outcome metrics, enabling direct quantification of success probability, runtime, tool-use variance, and economic efficiency for each digest mode.

\begin{table}[h]
    \centering
    \caption{Architectural and empirical comparison of algorithmic and LLM digest modes derived from 44 cyber-range penetration-testing exercises (additional metrics in Section~\ref{subsection:strategic_feedback_loop_results}).}
    \label{tab:digest_modes}
    \small
    \begin{tabular*}{\textwidth}{@{\extracolsep{\fill}}>{\raggedright\arraybackslash}p{3.8cm}>{\raggedright\arraybackslash}p{5.2cm}>{\raggedright\arraybackslash}p{5.2cm}@{}}
    \toprule
    \textcolor{cai_color}{\textbf{Property}} & \textcolor{cai_color}{\textbf{algorithmic Mode}} & \textcolor{cai_color}{\textbf{llm Mode}} \\
    \midrule
  
    \textbf{\textcolor{cai_color}{Success Rate}} & 20.0\% (3/15 tests) & \textbf{42.9\%} (6/14 tests) \\
    \addlinespace[0.15cm]


    \textbf{\textcolor{cai_color}{Implementation}} & Rule-based template processing with fixed thresholds & LLM inference (\texttt{alias1}) with 350-word structured prompt \\
    \addlinespace[0.15cm]

    \textbf{\textcolor{cai_color}{Strategic Interpretation}} & Deterministic thresholds: $p > 0.9$ (high-risk), $p < 0.5$ (bottleneck) & Reasoning with contextual probability interpretation and tactical recommendations \\
    \addlinespace[0.3cm]

    \textbf{\textcolor{cai_color}{Latency}} & Instant ($<$10ms) & 10--46s (mean: 28.3s $\pm$ 11.2s) \\
    \addlinespace[0.15cm]


    \textbf{\textcolor{cai_color}{Output Consistency}} & 100\% deterministic & Stochastic (temp=0.3); 5.2$\times$ lower behavioral variance \\
    \addlinespace[0.15cm]

    \textbf{\textcolor{cai_color}{Tool Usage Variance}} & 6.2$\times$ range (512--3186 tools) & \textbf{1.2$\times$} range (860--1034 tools) \\
    \addlinespace[0.3cm]

    \textbf{\textcolor{cai_color}{Cost per Success}} & \$0.32 & \textbf{\$0.12} (2.7$\times$ more efficient) \\
    \addlinespace[0.15cm]

    \textbf{\textcolor{cai_color}{Failure Mode}} & Guaranteed output if CTR data exists & API errors $\rightarrow$ fallback to algorithmic mode \\
  
    \bottomrule
    \end{tabular*}
\end{table}

\begin{table}[h]
\centering
\caption{Aggregate performance metrics for algorithmic and LLM digest generation modes.}
\label{tab:llm_vs_algo}
\small
\begin{tabular}{@{}lccccc@{}}
\toprule
\textcolor{cai_color}{\textbf{Configuration}} & \textcolor{cai_color}{\textbf{$n$}} & \textcolor{cai_color}{\textbf{Success Rate}} & \textcolor{cai_color}{\textbf{Avg Duration}} & \textcolor{cai_color}{\textbf{Tool Variance}} & \textcolor{cai_color}{\textbf{Cost/Success}} \\
\midrule
\rowcolor{gray!5}
G-CTR \texttt{algorithmic} & 15 & 20.0\% (3/15) & 22.5 min & 6.2$\times$ & \$0.32 \\
G-CTR \texttt{llm} & 14 & 42.9\% (6/14) & 20.2 min & 1.2$\times$ & \$0.12 \\
\midrule
\rowcolor{gray!10}
\textbf{$\Delta$ (\texttt{llm} - \texttt{algorithmic})} & & \textbf{+22.9 pp} & \textbf{-2.3 min} & \textbf{-83\%} & \textbf{-\$0.20} \\
\bottomrule
\end{tabular}
\end{table}

Across the 29 matched digest runs (15 \texttt{algorithmic}, 14 \texttt{llm}), swapping the interpreter (\texttt{algorithmic} vs. \texttt{llm}) more than doubles observed task success (20.0\%$\rightarrow$42.9\%, +22.9 percentage points). This ablation demonstrates that the \texttt{llm} digest mode provides a superior interpretation of the game-theoretic signals compared to the \texttt{algorithmic} mode.

The same intervention tightens behavioral dispersion. Tool-use range contracts from 6.2$\times$ (512--3186 invocations) to 1.2$\times$ (860--1034), implying that the \texttt{llm} digest acts as a variance-reducing controller even though the model samples with temperature~0.3. Results suggest that the sampling stochasticity is dominated by the constraints (the context) supplied in the \texttt{llm} digest processing, not by randomness in the underlying LLM model. Further experiments with different temperature values would help to confirm this hypothesis.

Efficiency markers shift in concert. Average episode duration decreases by 2.3 minutes despite the 28.3~s $\pm$ 11.2~s digest generation latency, demonstrating that LLM guidance more than compensates for its inference overhead. Cost per successful breach falls from \$0.32 to \$0.12 (2.7$\times$ improvement), indicating that higher success probability and lower resource dispersion jointly reduce marginal compute expenditure.\\
\newline
With the cyber-range effects established, we next test whether the same control advantage (via the \texttt{llm} digest mode) persists under live Attack and Defense CTF scoring, which better reflects real-world cybersecurity conditions.


\subsubsection{Studying Offensive and Defensive Enhancements in Realistic Cybersecurity Exercises}

We transition to evaluating the full game-theoretic AI architecture and G-CTR implementation in Attack and Defense (A\&D) CTF scenarios. Unlike the single-agent cyber-range exercises above, A\&D CTFs introduce competitive scoring dynamics and simultaneous offensive/defensive operations that better reflect real-world security engagements. This realism stems from several key factors: (1) \textbf{Adversarial pressure}, where defenders must actively counter live attacks rather than merely hardening systems against theoretical threats, (2) \textbf{Resource competition}, as both sides operate under time and computational constraints that mirror operational security budgets, (3) \textbf{Dynamic threat landscapes}, where attack vectors evolve in real-time as defenders implement countermeasures, and (4) \textbf{Multi-objective optimization}, requiring teams to balance offensive success against defensive resilience—a fundamental tension present in actual cybersecurity operations where organizations must simultaneously protect assets while maintaining business functionality. For a more detailed description of the A\&D CTFs, please refer to \cite{balassone2025cybersecurity,sanzgomez2025cybersecurityaibenchmarkcaibench}.

\newcommand{\configswatch}[1]{
  \tikz[baseline=-0.65ex]{\draw[draw=black!55, fill=#1, line width=0.2pt] (0,0) rectangle (0.55,0.25);}%
}

\newcommand{\configswatchpattern}[2]{
  \tikz[baseline=-0.65ex]{
    \filldraw[fill=#1, draw=black!55, line width=0.2pt] (0,0) rectangle (0.55,0.25);
    \path[pattern=north east lines, pattern color=#2] (0,0) rectangle (0.55,0.25);
    \draw[draw=black!55, line width=0.2pt] (0,0) rectangle (0.55,0.25);
  }%
}

\begin{table}[h!]
    \centering
    \caption{Team configurations and legend used throughout the Attack and Defense experiments. Colors and patterns match Figure~\ref{fig:ad_cowsay_results}.}
    \label{tab:gctr_configs}
    \small
    \begin{tabular*}{\textwidth}{@{\extracolsep{\fill}}p{0.5cm}p{3.4cm}>{\raggedright\arraybackslash}p{12cm}@{}}
        \toprule
          & \textcolor{cai_color}{\textbf{Configuration}} & \textcolor{cai_color}{\textbf{Operational Description}} \\
        \midrule
        \configswatch{NoGCTR} & \textbf{No G-CTR} (baseline) & Red and blue agents act without strategic overlays, relying solely on the \texttt{alias1} LLM's Phase~3 (ReAct) behaviors from Figure~\ref{fig:system_architecture}. \\
        \configswatch{RedGCTR} & \textcolor{RedGCTR}{\textbf{Red G-CTR}} & Only the attacker receives G-CTR guidance; the defending agent remains without guidance. Both agents run in parallel, with context separation. \\
        \configswatch{BlueGCTR} & \textcolor{BlueGCTR}{\textbf{Blue G-CTR}} & Only the defender receives G-CTR guidance while the attacker operates without it. Both agents run in parallel, with context separation. \\
        \configswatch{PurplePlain} & \textcolor{PurplePlain}{\textbf{Purple G-CTR}} & Both agents, attacker and defender, receive G-CTR guidance with independent attack graphs constructed for each agent. Agents run in parallel and with no context sharing, exposing how dual but uncoupled strategies perform.\\
        \configswatch{PurpleMerged} & \textcolor{PurpleMerged}{\textbf{Purple G-CTR$_{\mathbf{merged}}$}} & Similar to \textcolor{PurpleCoop}{\textbf{Purple G-CTR}} but both agents share the same G-CTR attack graph and context, yielding the highest degree of synchronized strategic planning.\\
        \bottomrule
    \end{tabular*}
\end{table}

\begin{figure}[h]
    \centering
    \begin{subfigure}[t]{0.51\textwidth}
        \centering
        \includegraphics[width=\textwidth]{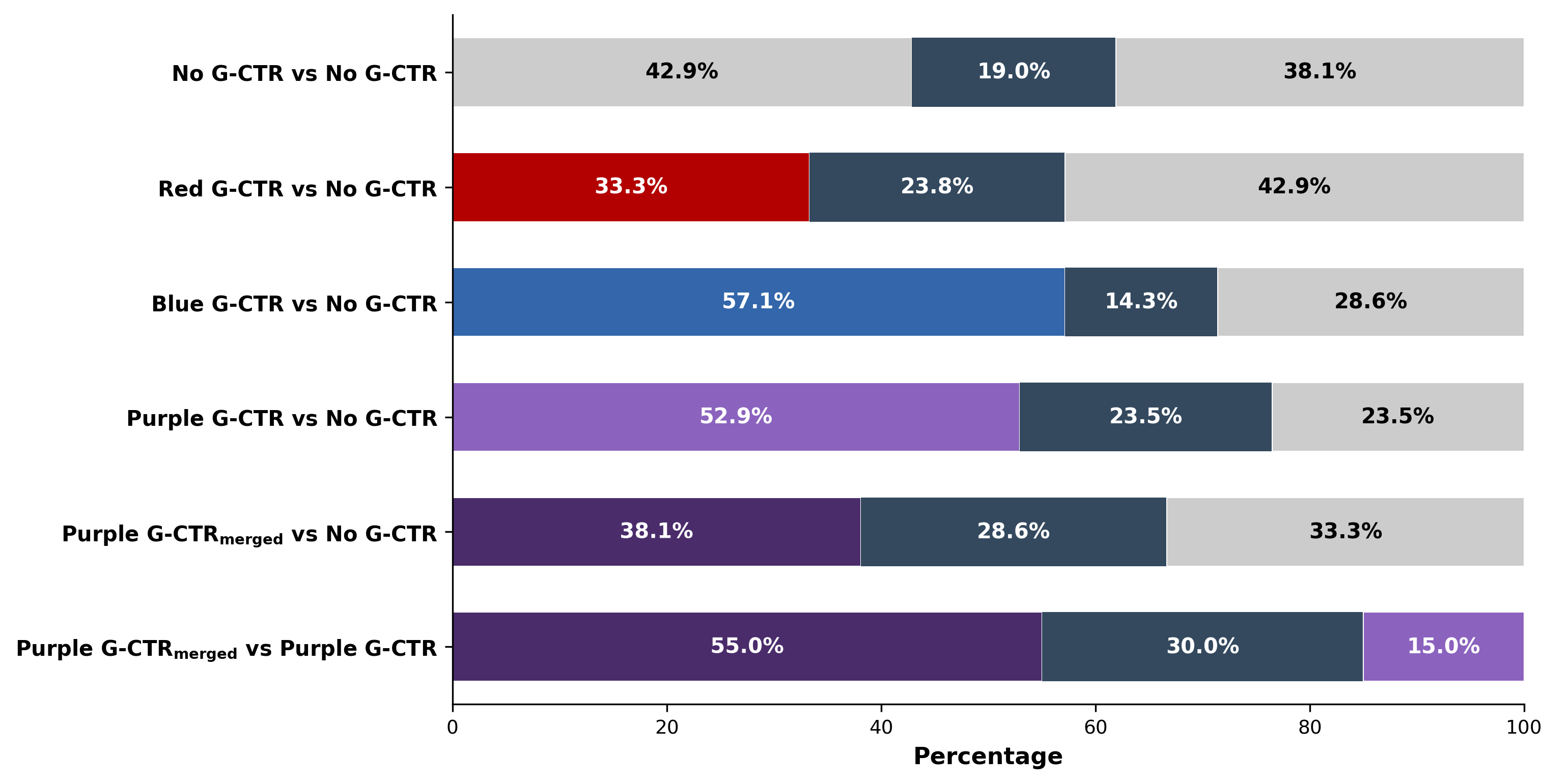}
        \caption{Cowsay Challenge Results.}
        \label{fig:ad_cowsay_results}
    \end{subfigure}\hfill%
    \begin{subfigure}[t]{0.45\textwidth}
        \centering
        \includegraphics[width=\textwidth]{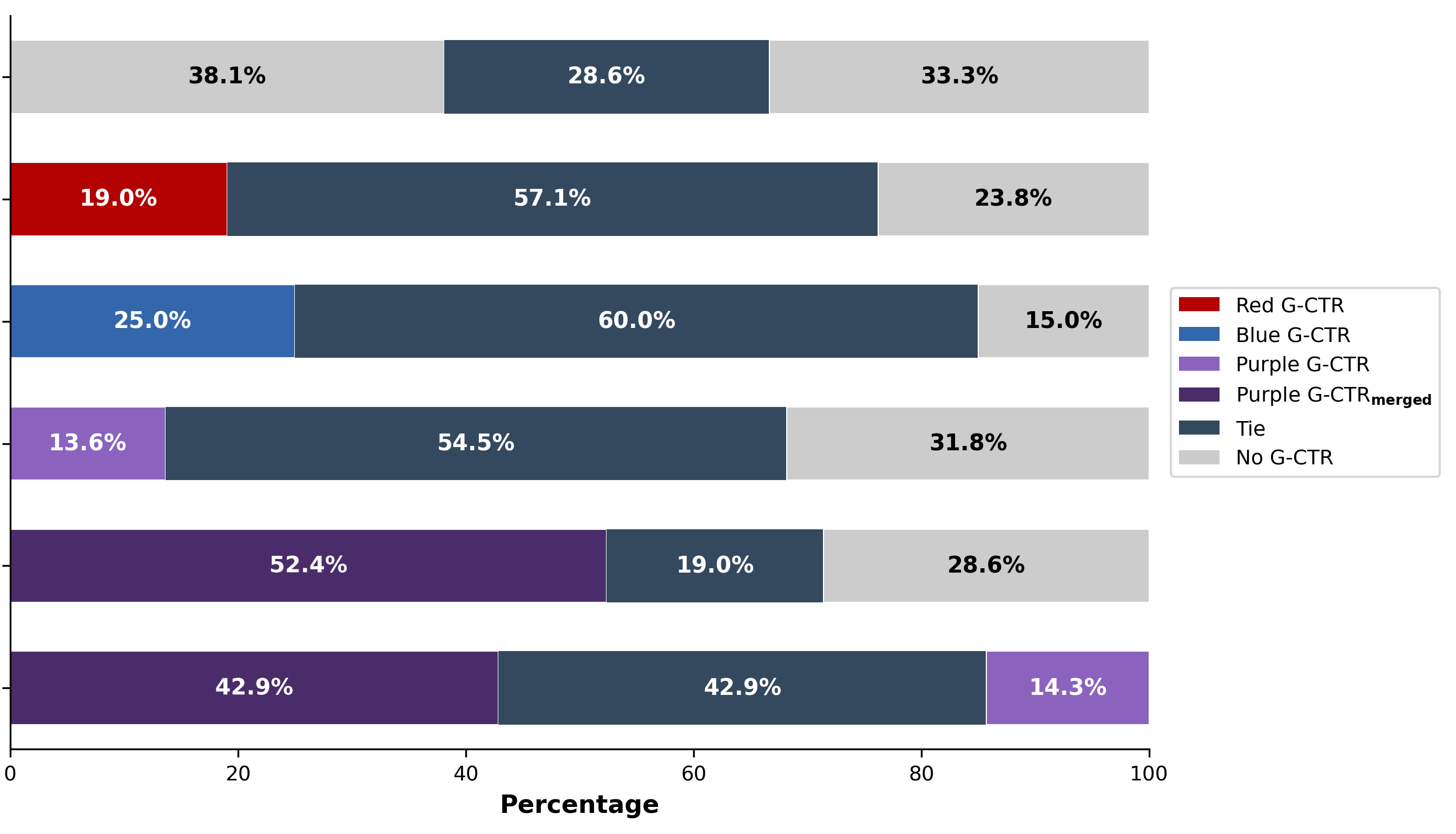}
        \caption{Pingpong Challenge Results.}
        \label{fig:ad_pingpong_results}
    \end{subfigure}
    \caption{Attack and Defense (A\&D) challenge results comparing different G-CTR team configurations across two scenarios. Outcome distributions are measured across 25 best-of-one matches per team pairings. Stacked bars report the percentage of team1 wins (left), ties (center), and team2 wins (right) for each team configuration. Colors and hatch patterns follow Table~\ref{tab:gctr_configs}; all agents use the \texttt{llm} digest mode, the \texttt{alias1} execution model, and the same CTR timing parameters ($\lambda_a=2$ attacker rate, $\lambda_d=1$ defender rate).}
    \label{fig:ad_results}
\end{figure}

To ablate the impact of different strategic configurations, we used the best performing digest from the cyber-range exercises (the \texttt{llm} digest mode) and systematically varied team compositions by controlling how G-CTR guidance gets integrated into the agents of each team. For that purpose, we created various configurations that varied the amount of G-CTR guidance that apply either exclusively to the team's attacker, exclusively to the team's defender or to both attacker and defender. We then confront each configuration against the baseline configuration (\textbf{No G-CTR}). Across all teams and evaluations, agents use the same \texttt{alias1} LLM model. Each team runs all its agents with parallel capabilities (refer to \cite{mayoralvilches2025caiopenbugbountyready} for details on CAI's parallel capabilities). Table~\ref{tab:gctr_configs} formalizes the team compositions used in the Attack and Defense (A\&D) experiments. Each entry displays the same glyph that appears in Figures~\ref{fig:ad_cowsay_results} and~\ref{fig:ad_pingpong_results}, ensuring the prose, the table legend, and the stacked-bar visualization share a common color and pattern vocabulary.


To evaluate G-CTR's effectiveness in A\&D scenarios, we conducted experiments on two ``very easy'' difficulty challenges: \texttt{cowsay} and \texttt{pingpong}. Both scenarios feature command injection vulnerabilities and privilege escalation vectors, making them accessible entry points for evaluating strategic guidance in adversarial settings. Each experiment consisted of multiple competitive rounds \cite{balassone2025cybersecurity} using the combinations of configurations described above.

Figure \ref{fig:ad_cowsay_results} presents the \texttt{cowsay} challenge which introduces shallow exploitation depth but rapid tactical churn, making coordination effects measurable. Figure \ref{fig:ad_pingpong_results} presents the \texttt{pingpong} scenario which requires sustained service uptime in addition to exploit delivery, amplifying the value of defensive foresight. 

Against our baseline \textbf{No G-CTR}, \textcolor{RedGCTR}{\textbf{Red G-CTR}} achieves a 33.3\%/\textcolor{cai_primary}{\textbf{42.9}}\% win/loss ratio in \texttt{cowsay} and a 19\%/\textcolor{cai_primary}{\textbf{23.8}}\% in \texttt{pingpong}, indicating that strategic guidance provides limited benefit for attacker-only guided configurations. We hypothesize that this performance degradation occurs because the guided attacker, despite receiving strategic direction, fails to leverage it effectively in dynamic scenarios. The underlying mechanism appears to be that attackers become fixated on specific exploitation paths that may no longer be viable, and because the system prompt carries significant weight in the guidance mechanism, the guided attacker exhibits reduced responsiveness to contextual updates (the need to pivot to alternative attack vectors). This performance decline supports our conclusion that \emph{attacker-only guidance cannot compensate for an unguided defense}. 

In contrast, \textcolor{BlueGCTR}{\textbf{Blue G-CTR}} achieves a high victory fraction (\textcolor{cai_primary}{\textbf{57.1}}\%/28.6\% win/loss ratio in \texttt{cowsay} and \textcolor{cai_primary}{\textbf{25}}\%/15\% in \texttt{pingpong}) confirming that defensive guidance provided by G-CTR supplies a large marginal gain. 

\textcolor{PurplePlain}{\textbf{Purple G-CTR}} also exceeded baseline performance with \textcolor{cai_primary}{\textbf{52.9}}\%/23.5\% win/loss ratios in \texttt{cowsay}, hinting that activating guidance on both sides stabilizes play without requiring further coordination. This observation is however contradicted by the fact that it struggles in \texttt{pingpong} with a 13.6\%/\textcolor{cai_primary}{\textbf{31.8}}\% win/loss ratio, suggesting that scenario complexity may interact with guidance effectiveness in ways that warrant further investigation. We argue that this contradiction hints at the need for deeper exploration of how teams with shared context and merged guidance (utilizing a single attack graph for all team members) might achieve better alignment and communication, potentially resolving the performance inconsistencies observed across different challenge types. Correspondingly, we propose \textcolor{PurpleMerged}{\textbf{Purple G-CTR$_{\mathbf{merged}}$}}, which implements a team reasoning stack with shared context and a single attack graph for all team members. Results of \textcolor{PurpleMerged}{\textbf{Purple G-CTR$_{\mathbf{merged}}$}} against our baseline delivers a \textcolor{cai_primary}{\textbf{38.1}}\%/33.3\% win/loss ratio in \texttt{cowsay} and \textcolor{cai_primary}{\textbf{52.4}}\%/28.6\% in \texttt{pingpong}, wins in both scenarios and suggesting that this last configuration with \emph{merged} guidance (shared context and a single attack graph) delivers the best performance in both scenarios.

We challenge this last hypothesis by evaluating the performance of \textcolor{PurpleMerged}{\textbf{Purple G-CTR$_{\mathbf{merged}}$}} against \textcolor{PurplePlain}{\textbf{Purple G-CTR}}. Results show that \textcolor{PurpleMerged}{\textbf{Purple G-CTR$_{\mathbf{merged}}$}} achieves a \textcolor{cai_primary}{\textbf{55}}\%/15\% win/loss ratio in \texttt{cowsay} and \textcolor{cai_primary}{\textbf{42.9}}\%/14.3\% in \texttt{pingpong}, confirming our conjecture: \emph{merged} guidance delivers the best performance.

Synthesizing across both scenarios, we observe a consistent hierarchy. Purely offensive guidance tops out at a losing record 33.3\% vs \textcolor{cai_primary}{\textbf{42.9}}\% (0.78:\textcolor{cai_primary}{\textbf{1}}), showing that attacker-only prompting cannot keep pace with unguided defenses. Defensive-only guidance improves the contest to roughly \textcolor{cai_primary}{\textbf{2}}:1 (\textcolor{cai_primary}{\textbf{57.1}}\% vs 28.6\%), while dual guidance without agents sharing context or attack graphs contributes another step change relative to independent dual guidance. In absolute terms, \textcolor{PurpleMerged}{\textbf{Purple G-CTR$_{\mathbf{merged}}$}} converts \texttt{pingpong} series \textcolor{cai_primary}{\textbf{52.4}}\% to 28.6\% ($\approx$\textcolor{cai_primary}{\textbf{1.8}}:1) when facing \textbf{No G-CTR}, and it defeats \textcolor{PurplePlain}{\textbf{Purple G-CTR}} in \texttt{cowsay} \textcolor{cai_primary}{\textbf{55}}\% to 15\% ($\approx$\textcolor{cai_primary}{\textbf{3.7}}:1), implying that opponents would need nearly four coordinated wins to offset every merged victory. Because all teams run the same \texttt{alias1} executor and \texttt{llm} digest, these ratios isolate the effect of marrying generative interpretation with Nash-equilibrium control. The resulting game-theoretic AI architecture (LLM+G-CTR) halves opponent success, compresses attacker variance, and delivers a reproducible breakthrough: under identical tooling and budgets, game-theoretic guidance lets AI teams execute with the decisiveness of coordinated human red/blue cells, establishing a new breakthrough performance for realistic cybersecurity exercises.

\section{Discussion} \label{section:Discussion}

In this study, we have presented and evaluated G-CTR, a game-theoretic AI framework for guiding cyber attack and defense using LLMs. Through quantitative evaluation across five real--world cybersecurity exercises, we demonstrated that G-CTR offers substantial gains in \textbf{scalability}, \textbf{automation}, and \textbf{cost-efficiency}. LLMs generated complete attack graphs in under 40 seconds, compared to up to 90 minutes for human experts, achieving a speedup between \textbf{60$\times$} and \textbf{245$\times$}. Inference costs ranged from \$0.05 to \$0.28, \textbf{reducing} analysis \textbf{costs by over 140$\times$} relative to expert labor. The G-CTR layer introduced negligible overhead (<5ms), confirming that the primary computational cost lies in LLM inference. While gpt-4o and o3 delivered the best balance of speed and cost, Claude Sonnet 4 consistently produced the most structurally complete and logically coherent graphs, especially in complex scenarios.

Qualitatively, LLMs showed strong capabilities in identifying key attack nodes--particularly in vulnerable stages--with an average node correspondence of 70–90\% compared to expert-generated graphs. Simpler logs led to highly accurate outputs, while more complex scenarios revealed subtle model weaknesses, including hallucinated nodes, semantic misclassifications, and occasional omissions of steps. \texttt{gpt-4o} was more prone to hallucinations, whereas \texttt{claude-sonnet-4} provided more context and detailed outputs, although at a higher cost. Overall, while LLMs tend to emphasize direct attack paths and occasionally oversimplify nuanced stages, G-CTR shows promise as a scalable, automated complement to human-driven security analysis, particularly when rapid or repeated assessments are needed.

While G-CTR has shown strong performance in both efficiency and output quality, several areas offer promising directions for future enhancement. One such factor is the configuration of graph complexity, particularly the minimum and maximum number of nodes. Across five domains (Table~\ref{tab:qualitative_comparison_attack_graphs}), instructing LLMs to emit 6--15 nodes with at most four vulnerable stages yielded the most faithful graphs; however, adaptive tuning based on scenario complexity could further improve completeness and interpretability in edge cases.

Another area of opportunity lies in prompt engineering. We observed that different LLMs respond better to tailored prompt structures, suggesting that customized system prompts per model could unlock further performance gains. This opens the door to more fine-grained control over LLM behavior, especially as new models and capabilities emerge.

Our analysis reinforces that model choice should be aligned with operational needs. Lighter models like \texttt{gpt-4o} and \texttt{o3} offer speed and cost--efficiency ideal for triage or scalable analysis, while more detailed outputs from Claude Sonnet 4 may be preferable in high-risk or high-stakes investigations. These tradeoffs highlight the flexibility of G-CTR in adapting to different deployment contexts, from automated threat assessments to analyst-in-the-loop workflows.

Extending the framework to Attack and Defense CTFs (Figures~\ref{fig:ad_cowsay_results} and~\ref{fig:ad_pingpong_results}) reveals a similar hierarchy of guidance strategies. Purely offensive prompting remains net-negative (33.3\% wins vs 42.9\% losses, 0.78:1), defensive-only guidance pushes results to roughly 2:1 (57.1\% vs 28.6\%), and dual guidance with shared graphs adds another step change. The \textcolor{PurpleMerged}{\textbf{Purple G-CTR$_{\mathbf{merged}}$}} team---where red and blue agents share both context and G-CTR attack graph---wins \texttt{pingpong} matches 52.4\% to 28.6\% (\textasciitilde1.8:1) against \textbf{No G-CTR} and defeats \textcolor{PurplePlain}{\textbf{Purple G-CTR}} in \texttt{cowsay} 55\% to 15\% (\textasciitilde3.7:1). Because every configuration reuses the same \texttt{alias1} executor and \texttt{llm} digest, these ratios isolate the benefit of combining generative semantics with Nash-equilibrium coordination, establishing \textcolor{PurpleMerged}{\textbf{Purple G-CTR$_{\mathbf{merged}}$}} as a reproducible, best-in-class agent for live adversarial play.

%
In summary, G--CTR demonstrates that combining LLM-based automation with lightweight game-theoretic reasoning enables fast, cost-effective, and contextually rich attack graph generation. While human expertise remains unmatched in capturing subtle and indirect attack behaviors, LLMs are rapidly closing the gap and can already support large-scale or time-sensitive threat assessments.


\section{Conclusion}

This paper presented a game-theoretic AI guidance architecture that unifies automated penetration testing with strategic defense planning. At its core lies Generative Cut-the-Rope (G-CTR), the component that extracts attack graphs, computes Nash equilibria, and feeds guidance back into CAI agents. Our key findings demonstrate both the feasibility and practical impact of this architecture:

\textbf{Automated Graph Generation Works.} LLMs successfully extract structured attack graphs from unstructured CAI logs with 70-90\% node correspondence to expert annotations, achieving 60-245$\times$ speedup. This automation eliminates the primary bottleneck in applying game-theoretic analysis to real-world security data.

\textbf{Strategic Feedback Enhances Pentesting.} By computing optimal attack/defense strategies and feeding them back to CAI, G-CTR enables focused security assessments that prioritize high-value targets. Algorithm~\ref{alg:digest_generation} serves as the critical translation layer, converting Nash equilibrium computations into actionable strategic guidance that CAI agents consume during penetration testing. Our empirical evaluation across 44 independent exercises demonstrates that this game-theoretic guidance substantially improves pentesting effectiveness: LLM-mode digests achieved 42.9\% success rates compared to 20.0\% for algorithmic templates, with the strategic feedback loop enabling agents to identify high-risk attack paths (transitions with $p > 0.9$) and recognize defensive bottlenecks (transitions with $p < 0.95$) that would otherwise remain obscured in unguided exploration.

\textbf{Breakthrough Attack and Defense Guidance.} When applied to simultaneous red/blue competitions, the same LLM digest plus G-CTR coordination unlocks a new best-in-class agent. In two realistic Attack and Defense cybersecurity exercises (Figures~\ref{fig:ad_cowsay_results} and~\ref{fig:ad_pingpong_results}), the \textcolor{PurpleMerged}{\textbf{Purple G-CTR$_{\mathbf{merged}}$}} configuration---where both teams share a single strategic context and attack graph---wins matches roughly 1.8:1 against the LLM-only baseline and 3.7:1 against independently guided teams. Purely offensive prompting, by contrast, remains sub-baseline (33.3\% to 42.9\%). This establishes GenAI+game-theory coupling as the differentiator for multi-agent cyber exercises.

\textbf{Reducing Ambiguity and Suppressing Hallucinations.} Hallucinations are mitigated by constraining the LLM's reasoning with an external, continuously updated game-theoretic control signal: G-CTR transforms the AI's own context (or logs) into an attack graph, computes equilibria, and injects a digest that anchors the model to the statistically strongest paths and chokepoints. This keeps the LLM focused on what is actually happening in the environment rather than drifting into speculative or irrelevant actions. The experimental results provide measurable evidence of this effect: in a 44-run cyber-range benchmark, adding the G-CTR digest more than doubled success rates (13.3\% $\rightarrow$ 42.9\%), reduced tool-use variance by 5.2$\times$, and cut cost-per-success by 23$\times$, all indicators of reduced hallucination and tighter behavioral coherence. This effectiveness stems from G-CTR's bidirectional translation capability: it converts narrative AI security outputs into formal game-theoretic inputs (Nash equilibria, probability-weighted paths, critical nodes) and then produces strategic guidance that integrates seamlessly back into agent system prompts via Algorithm~\ref{alg:digest_generation}. This closed-loop architecture---where attack graphs inform strategic analysis, which in turn guides subsequent penetration testing through digest injection---enables continuous refinement and represents a concrete step toward cybersecurity superintelligence by collapsing the LLM's search space and keeping the model tightly anchored to the most strategically relevant parts of the problem.

Future work must now optimize the balance between hallucination mitigation and creative attack discovery—exploring whether controlled randomness (e.g., temperature schedules) can unlock novel strategies without sacrificing fidelity, while refining LLM interpretation to handle even richer telemetry. Building on that foundation, we will consider run adversarial-robustness exercises against best human red teams.

Taken together, these advances position the proposed game-theoretic AI architecture—and its G-CTR core—as a decisive step toward cybersecurity superintelligence that not only find vulnerabilities but strategically orchestrate how AI attackers and defenders respond in realistic, high-pressure environments.

\section{Acknowledgements}

This research has been partly funded by the European Innovation Council (EIC) as part of the accelerator project ``RIS'' (GA 101161136) - HORIZON-EIC-2023-ACCELERATOR-01 call. Thanks to the Alias Robotics team members for their various reviews. Thanks also to Alfonso Muñoz Muñoz for his strategic game-theoretic insights and critical review of the scientific value of our work.

\bibliography{bibliography}

\clearpage
\cleardoublepage
\begingroup
\begin{center}
\vspace*{4cm}
{\fontsize{24}{28}\selectfont\color{cai_primary}\textbf{APPENDICES}}
\vspace{1cm}

\begin{tikzpicture}
\fill[cai_primary!30] (0,0) rectangle (\textwidth,0.5pt);
\end{tikzpicture}



\end{center}
\endgroup
\appendix

\section{Appendix 1: A Refresher on Cut-The-Rope (CTR) and Game Theory \cite{ctr_paper2}} 
\label{appendix_game_theory_ctr}
The Cut-the-Rope (CTR) model introduces a formal game-theoretic framework for modeling stealthy cyberattacks in probabilistic attack graphs under asynchronous timing. It assumes the attacker is already present in the system and advances continuously in time along a directed path toward a critical target node \( v_0 \), following a stochastic process, typically a Poisson process, with rate \( \lambda \) that determines the average number of steps the attacker takes between defender actions. In contrast, the defender operates in discrete time, inspecting one node per round from an admissible set \( AS_1 \subseteq V \setminus \{v_0\} \). If the defender inspects a node that lies on the attacker’s current path before the attacker reaches \( v_0 \), the path is ``cut'' and the attacker is forced to restart. This asynchronous, probabilistic setup models the temporal and informational asymmetry inherent in advanced persistent threat (APT) scenarios, capturing both the stealth and adaptiveness of real-world attackers.

CTR models the system as a directed acyclic attack graph \( G = (V, E) \), where:
\begin{itemize}
  \item \( V \) is the set of nodes representing system components or privileges;
  \item \( E \subseteq V \times V \) is the set of directed edges representing potential exploits.
\end{itemize}

The attacker's goal is to reach a designated critical asset \( v_0 \in V \), while the defender performs \textit{spot checks} to prevent this. The attacker is assumed to already be in the system, at an unknown node \( \theta \in V \setminus \{v_0\} \), and proceeds along an attack path \( \pi \), which is a directed path from \( \theta \) to \( v_0 \).

The defender can inspect nodes from a set \( AS_1 \subseteq V \setminus \{v_0\} \). The attacker selects a strategy over pairs \( (\theta, \pi) \), where \( \theta \in V(\pi) \subseteq V \setminus \{v_0\} \) is the starting point, and \( \pi \in AS_2 \) is a feasible attack path ending at \( v_0 \). The game is defined by the quintuple:
\begin{equation}
(G, v_0, AS_1, AS_2, f_N)
\end{equation}

where \( f_N \) is a probability distribution over the number of steps the attacker can take during a single round (e.g., following a Poisson-, geometric or other distribution).\\\\

\noindent\textbf{Defender's expected success}


The probability that the attacker has reached a particular node \( v \) on path \( \pi \), given the rate \( \lambda \), is:

\begin{equation}
P_\pi(v \mid \theta, \lambda) = 
\frac{f_{\mathrm{Pois}}(d_\pi(\theta, v); \lambda)}
{\sum_{x \in V(\pi)} f_{\mathrm{Pois}}(d_\pi(\theta, x); \lambda)}
\end{equation}

This expression simply says: "Among all possible positions the attacker could reach on path \( \pi \), how likely is it that they reached node \( v \)?"

Here:
\begin{itemize}
    \item \( \theta \in V \setminus \{v_0\} \): the attacker’s starting node,
    \item \( \pi \): the attack path from \( \theta \) to the target \( v_0 \),
    \item \( d_\pi(\theta, v) \): the number of steps from \( \theta \) to node \( v \) along \( \pi \),
    \item \( V(\pi) \subseteq V \): the set of nodes on path \( \pi \),
    \item \( \lambda \): the attack rate (average steps per round),
    \item \( f_{\mathrm{Pois}}(k; \lambda) \): the Poisson mass function:
\end{itemize}

\[
f_{\mathrm{Pois}}(k; \lambda) = \frac{\lambda^k e^{-\lambda}}{k!}
\]

This function gives the probability of making exactly \( k \) steps between two defender actions. \\\\

\noindent\textbf{Defender's utility}

Let \( \sigma_d \in \Delta(AS_1) \) be the defender's mixed strategy: a probability distribution over the set \( AS_1 \subseteq V \setminus \{v_0\} \) of nodes that the defender is allowed to inspect. Each round, the defender picks a node \( c \in AS_1 \) to inspect, with probability \( \sigma_d(c) \). The defender’s expected probability of catching the attacker (e.g., cutting the attack path) is:

\begin{equation}
U_D(\sigma_d, \pi, \theta) = 
\sum_{c \in AS_1} \sigma_d(c) \cdot P_\pi(c \mid \theta, \lambda)
\end{equation}

Here:
\begin{itemize}
    \item \( c \in AS_1 \): a candidate node for inspection (excluding the target \( v_0 \)),
    \item \( \sigma_d(c) \): the probability the defender chooses to inspect node \( c \),
    \item \( P_\pi(c \mid \theta, \lambda) \): the likelihood the attacker is at node \( c \), given path \( \pi \) and attack rate \( \lambda \).
\end{itemize}

\vspace{0.5em}
\noindent\textbf{Attacker’s success probability}

The attacker succeeds if none of the nodes they use are inspected during the round. This gives the attacker's success utility:

\begin{equation}
U_A = 1 - U_D(\sigma_d, \pi, \theta)
\end{equation} \\\\

\vspace{0.5em}
\noindent\textbf{Game objective: optimal defender strategy}

This is a zero-sum game: the attacker's gain is the defender's loss. The defender seeks the strategy \( \sigma_d^\star \) that minimizes the attacker's best-case chance of success across all possible attack paths and starting points. Formally, the optimal defense is:

\begin{equation}
\sigma_d^\star = \arg \min_{\sigma_d} 
\max_{\pi \in AS_2,~ \theta \in V(\pi)} 
\left( 1 - \sum_{c \in AS_1} \sigma_d(c) \cdot P_\pi(c \mid \theta, \lambda) \right)
\end{equation}

Where:
\begin{itemize}
    \item \( AS_2 \): the set of all possible attack paths ending at \( v_0 \),
    \item \( V(\pi) \): the set of nodes on path \( \pi \),
    \item \( \theta \in V(\pi) \): the possible (unknown) starting location of the attacker.
\end{itemize}

This defines a \textit{minimax} game: the defender prepares for the worst-case attack, and ensures that even under maximum uncertainty, the probability of a successful breach is as low as possible. The resulting \( \sigma_d^\star \) is the optimal randomized defense policy.\\\\

\vspace{0.5em}
\noindent\textbf{Game solution: Nash equilibrium}

The interaction between the attacker and defender is modeled as a zero-sum game, in which the attacker selects a hidden entry point \( \theta \in V \setminus \{v_0\} \) and a path \( \pi \in AS_2 \), while the defender chooses a randomized inspection strategy \( \sigma_d \in \Delta(AS_1) \). The defender aims to minimize the probability of a successful undetected attack, while the attacker aims to maximize it.

A central solution concept in this game is the \textit{Nash equilibrium}, which is a pair of strategies such that neither player can improve their expected utility by unilaterally deviating. In this zero-sum setting, the defender's optimal strategy \( \sigma_d^\star \) corresponds to the minimax solution, and the attacker's best response is to select the pair \( (\pi^\star, \theta^\star) \) that maximizes their success probability under \( \sigma_d^\star \).

Formally, the equilibrium condition is defined as:

\begin{equation}
U_D(\sigma_d^\star, \pi, \theta) \geq U_D(\sigma_d, \pi, \theta) \quad \forall \sigma_d, \quad \text{and} \quad U_A(\sigma_d^\star, \pi^\star, \theta^\star) \geq U_A(\sigma_d^\star, \pi, \theta) \quad \forall (\pi, \theta)
\end{equation}

Here:
\begin{itemize}
    \item \( U_D(\sigma_d, \pi, \theta) \): the defender’s expected probability of detection,
    \item \( U_A = 1 - U_D \): the attacker’s probability of success,
    \item \( \sigma_d^\star \): the defender's optimal mixed strategy,
    \item \( (\pi^\star, \theta^\star) \): the attacker's best-response path and entry point under \( \sigma_d^\star \).
\end{itemize}

This equilibrium represents a stable solution in which the defender has maximally reduced the attacker's chance of success, even under worst-case uncertainty. No player can unilaterally improve their utility, i.e., the strategy pair is mutually optimal.

\section{Appendix 2: Attack Graph Gold Standard Dataset}
\label{appendix_annotation}




In order to evaluate the performance of large language models (LLMs) in comparison to human analysts, a manually annotated dataset of attack graphs was constructed using real-world vulnerability reports. This dataset serves as a gold standard for assessing the quality of generated graphs, enabling direct comparison between machine-generated and human-constructed outputs. Two annotators were involved in the process, and annotation of each log typically required between 30 to 60 minutes depending on its complexity.

Five logs were selected from cybersecurity exercises derived from bug bounty evaluations as described in \cite{aliasrobotics2025cai}. These exercises originated from HackerOne, a platform where real vulnerabilities have been discovered and validated. These logs were chosen specifically because they are known to contain confirmed vulnerabilities, providing a reliable basis for evaluating attack graph reconstructions.

For consistency, the same list of messages used as input for the LLMs was extracted from each report. This ensured alignment between the human and model interpretations of the data.

Each log was replayed with corresponding tool and analyzed manually. Messages were reviewed one by one, and annotations were recorded using a structured format that captures each attack step, its preconditions, and its effects. This format is designed to support subsequent evaluation tasks. An example of the annotation schema is:

\begin{verbatim}
{
  "nodes": [
    {
      "id": "A",
      "name": "Start Node",
      "info": "Initial point of the attack",
      "vulnerability": false,
      "message_id": 101
    },
    {
      "id": "B",
      "name": "Exploit Attempt",
      "info": "Attempt to exploit a weakness",
      "vulnerability": true,
      "message_id": 102
    }
  ],
  "edges": [
    {
      "source": "A",
      "target": "B"
    }
  ]
}
\end{verbatim}

A key contribution of this work is the introduction of a standardized annotation format for manually constructed attack graphs derived from natural language reports. Although existing formats support general graph structures, few are tailored to the specific semantics of attack chains. The proposed format aims to promote reproducibility, support benchmarking efforts, and encourage future work in evaluating LLMs for cybersecurity applications.

\begin{table}[!h]
    \centering
    \small
    \renewcommand{\arraystretch}{1.3}
    \begin{tabular}{llll}
        \toprule
        \textbf{Domain} & \textbf{Messages} & \textbf{Tokens} & \textbf{Vulnerabilities} \\
        \midrule
        \textbf{kolesa.kz} & 72 & 44,959 & Hardcoded API key \\
        \textbf{mercadolibre.com} & 47 & 424,838 & User enumeration \\
        \textbf{pornbox.com} & 80 & 100,139 & No rate limit \\
        \textbf{hm.com} & 118 & 40,616 & SSL certificate mismatch \\
        \textbf{media.guilded.gg} & 358 & 8,959,492 & URL Confusion Attack, SSL cert mismatch \\
        \bottomrule
    \end{tabular}
    \caption{Summary of the exercises included in the Attack Graph Gold Standard Dataset across five domains. For each domain, the table aggregates the total number of messages exchanged, total tokens processed (Input + Output tokens), and the vulnerabilities identified during execution. The model for the vulnerability search was \texttt{claude-3-7-sonnet-20250219}.}
    \label{tab:llm_domain_summary_grouped}
\end{table}

\section{Appendix 3: kolesa.kz}
\label{appendix_kolesa}

%

%

\begin{figure}[H]
    \centering
    \includegraphics[width=\textwidth]{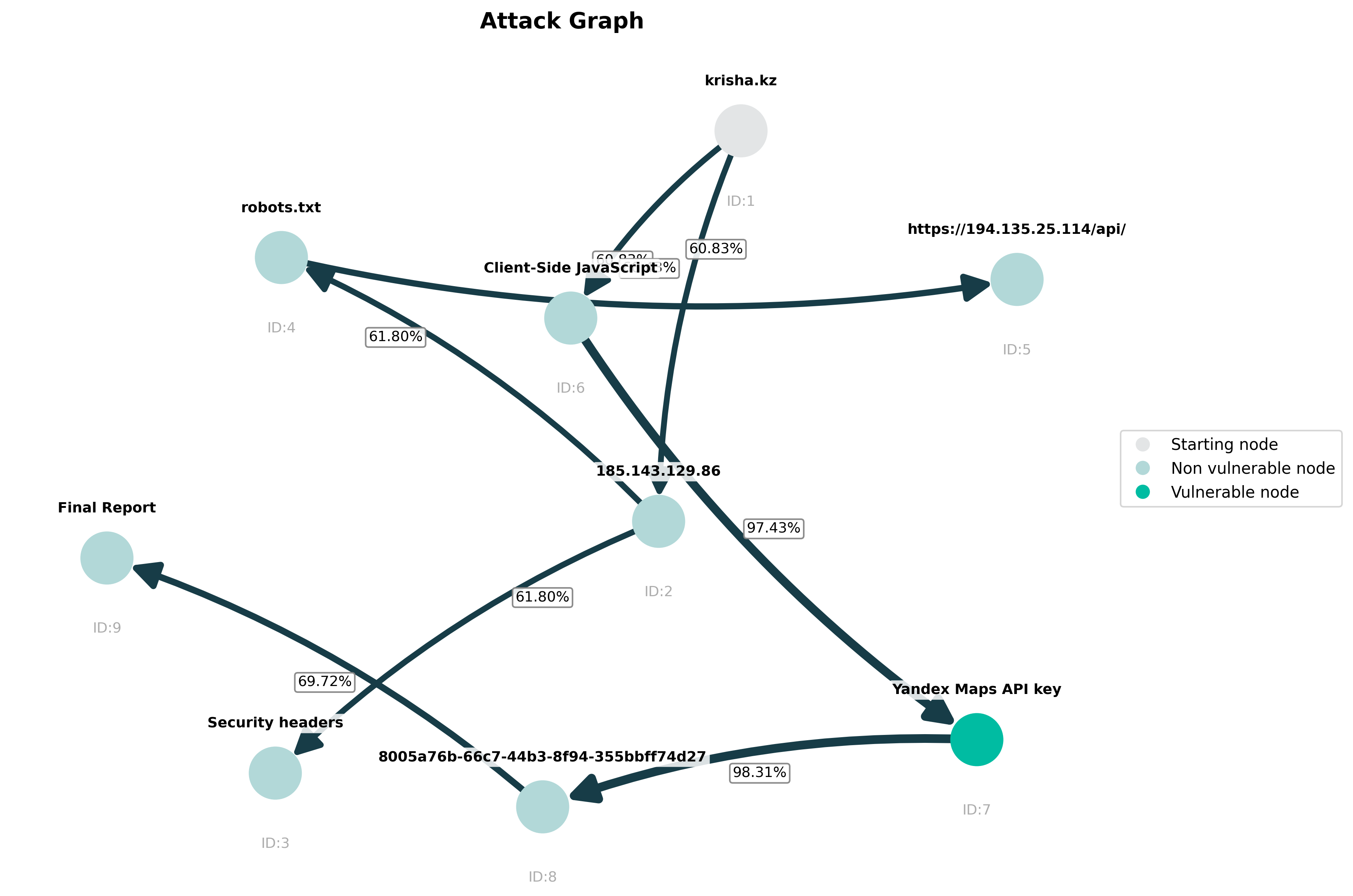}
    \caption*{(a) Human Attack Graph}

    \vspace{0.5cm}

\begin{minipage}{0.4\textwidth}
    \centering
    \begin{tabular}{cc}
    \hline
     \textcolor{cai_primary}{\textbf{Node ID}} & \textcolor{cai_primary}{\textbf{Probability}} \\
    \hline
    6 & 1.000000 \\
    \hline
    \end{tabular}
    \caption*{(b) Optimal Defense Strategy}
\end{minipage}
\hfill
\begin{minipage}{0.45\textwidth}
    \centering
    \begin{tabular}{ccc}
    \hline
\textcolor{cai_primary}{\textbf{Path ID}} & \textcolor{cai_primary}{\textbf{Path Sequence}} & \textcolor{cai_primary}{\textbf{Probability}} \\
    \hline
    1 & 1 $\textcolor{cai_primary}{\rightarrow}$ 6 $\textcolor{cai_primary}{\rightarrow}$ 7 & 1.000000 \\
    \hline
    \end{tabular}
    \caption*{(c) Attacker Strategy}
\end{minipage}

\vspace{0.5cm}

\begin{minipage}{\textwidth}
    \centering
    \colorbox{cai_light}{%
        \parbox{0.85\textwidth}{%
            \centering
            \vspace{1.2ex}
            \textcolor{cai_dark}{\textbf{Game Equilibrium}}\\[0.5ex]
            \textcolor{cai_primary}{Defender can keep attacker success below:} \textbf{0.000000}\\[0.3ex]
            \textcolor{cai_primary}{Attacker can guarantee success probability of:} \textbf{0.000000}
            \vspace{1.2ex}
        }%
    }
    \caption*{(d) Game Equilibrium}
\end{minipage}

    \caption{\textbf{Ground-Truth attack graph annotated by Humans in the kolesa.kz domain}: This manually constructed graph illustrates the real-world steps taken in a bug bounty exercies of the kolesa.kz domain, starting with the subdomain krisha.kz. The process followed a natural reconnaissance path, including inspection of server IPs, analysis of security headers and robots.txt, and exploration of API endpoints. Attention then shifted to client-side JavaScript, where there was identified a hardcoded Yandex Maps API key (8005a76b-66c7-44b3-8f94-355bbff74d27) embedded in the window.data object. The exposed key represents a potential vulnerability due to improper handling of sensitive credentials. The flow concludes with the creation of a final report. All connections and insights were derived and annotated manually by the researcher, without automated tooling.}
    \label{fig:kolesa_human}
\end{figure}
%

\begin{figure}[H]
    \centering
    \includegraphics[width=\textwidth]{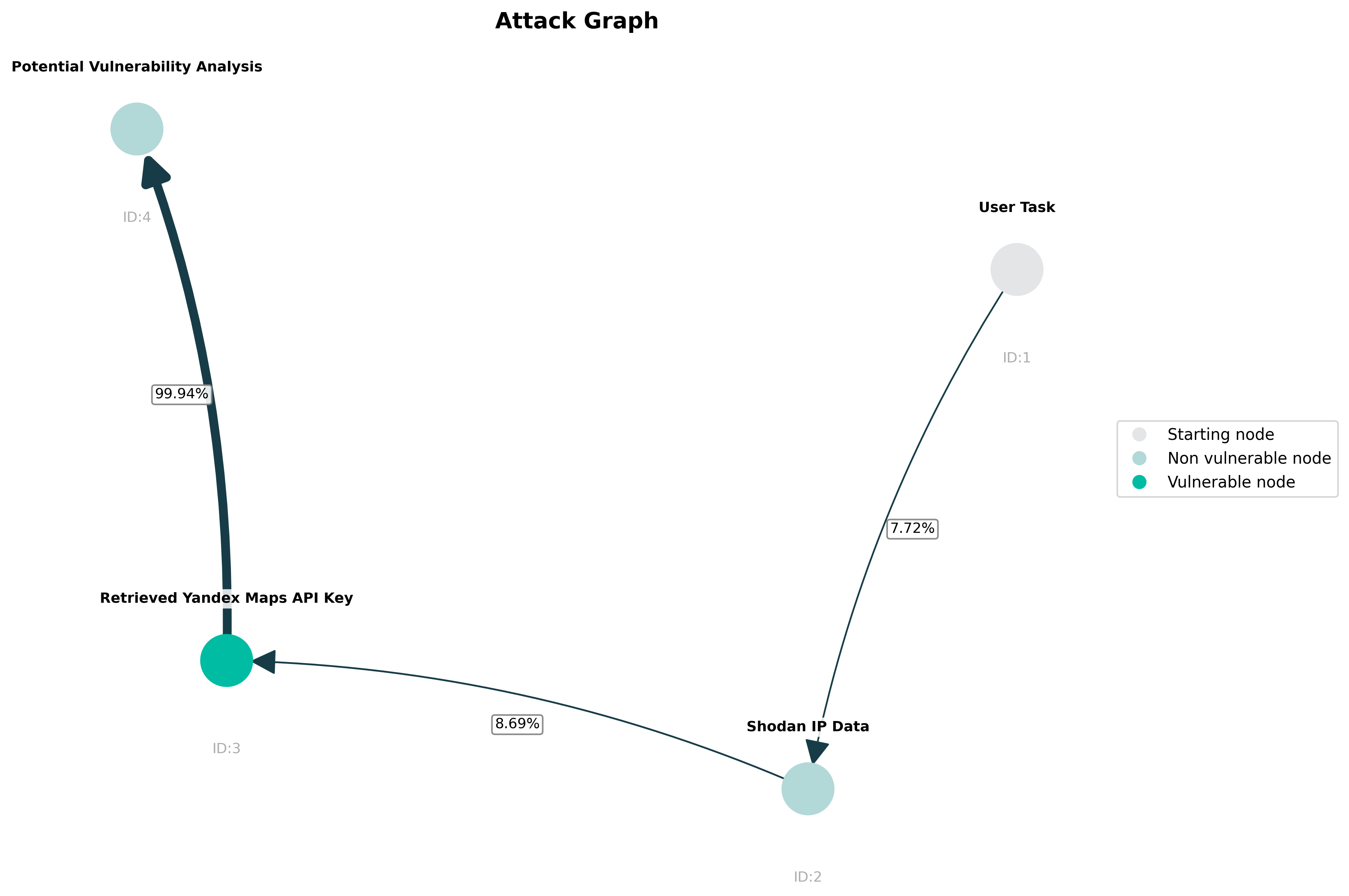}
    \caption*{(a) Attack Graph Example}

    \vspace{0.5cm}
    
    \begin{minipage}{0.4\textwidth}
        \centering
        \renewcommand{\arraystretch}{1.4}
        \begin{tabular}{cc}
        \hline
         \textcolor{cai_primary}{\textbf{Node ID}} & \textcolor{cai_primary}{\textbf{Probability}} \\
        \hline
        2 & 1.000000 \\
        \hline
        \end{tabular}
        \caption*{(b) Optimal Defense Strategy}
    \end{minipage}
    \hfill
    \begin{minipage}{0.45\textwidth}
        \centering
        \renewcommand{\arraystretch}{1.4}
        \begin{tabular}{ccc}
        \hline
\textcolor{cai_primary}{\textbf{Path ID}} & \textcolor{cai_primary}{\textbf{Path Sequence}} & \textcolor{cai_primary}{\textbf{Probability}} \\
        \hline
        1 & 1 $\textcolor{cai_primary}{\rightarrow}$ 2 $\textcolor{cai_primary}{\rightarrow}$ 3  & 1.000000 \\
        \hline
        \end{tabular}
        \caption*{(c) Attacker Strategy}
    \end{minipage}
    
    \vspace{0.5cm}
    
    \begin{minipage}{\textwidth}
        \centering
        \colorbox{cai_light}{%
            \parbox{0.85\textwidth}{%
                \centering
                \vspace{1.2ex}
                \textcolor{cai_dark}{\textbf{Game Equilibrium}}\\[0.5ex]
                \textcolor{cai_primary}{Defender can keep attacker success below:} \textbf{0.000000}\\[0.3ex]
                \textcolor{cai_primary}{Attacker can guarantee success probability of:} \textbf{0.000000}
                \vspace{1.2ex}
            }%
        }
        \caption*{(d) Game Equilibrium}
    \end{minipage}
    \caption{Attack Graph and Equilibrium analysis for \texttt{gpt-4o}}
    \label{fig:kolesa_gpt}
\end{figure}

\begin{figure}[H]
    \centering
    \includegraphics[width=\textwidth]{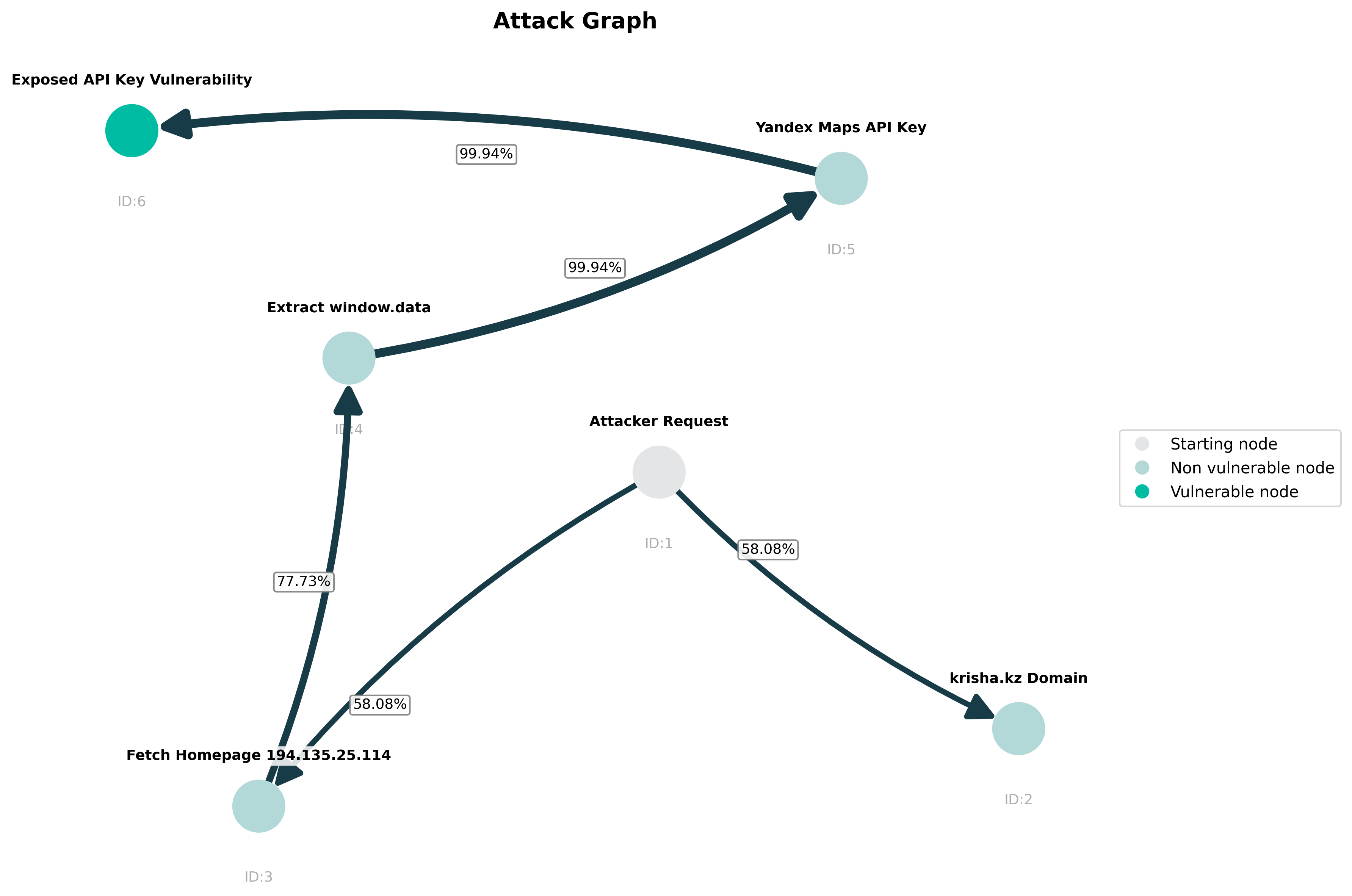}
    \caption*{(a) Attack Graph Example}

    \vspace{0.5cm}

    \begin{minipage}{0.45\textwidth}
        \centering
        \renewcommand{\arraystretch}{1.4}
        \begin{tabular}{cc}
        \hline
         \textcolor{cai_primary}{\textbf{Node ID}} & \textcolor{cai_primary}{\textbf{Probability}} \\
        \hline
        5 & 1.000000 \\
        3 & 0.000000 \\
        4 & 0.000000 \\
        \hline
        \end{tabular}
        \caption*{(b) Optimal Defense Strategy}
    \end{minipage}
    \hspace{0.02\textwidth}
    \begin{minipage}{0.50\textwidth}
        \centering
        \renewcommand{\arraystretch}{1.4}
        \begin{tabular}{ccc}
        \hline
\textcolor{cai_primary}{\textbf{Path ID}} & \textcolor{cai_primary}{\textbf{Path Sequence}} & \textcolor{cai_primary}{\textbf{Probability}} \\
        \hline
        1 & 1 $\textcolor{cai_primary}{\rightarrow}$ 3 $\textcolor{cai_primary}{\rightarrow}$ 4 $\textcolor{cai_primary}{\rightarrow}$ 5 $\textcolor{cai_primary}{\rightarrow}$ 6  & 1.000000 \\
        \hline
        \end{tabular}
        \caption*{(c) Attacker Strategy}
    \end{minipage}
    
    \vspace{0.5cm}
    
    \begin{minipage}{\textwidth}
        \centering
        \colorbox{cai_light}{%
            \parbox{0.85\textwidth}{%
                \centering
                \vspace{1.2ex}
                \textcolor{cai_dark}{\textbf{Game Equilibrium}}\\[0.5ex]
                \textcolor{cai_primary}{Defender can keep attacker success below:} \textbf{0.000000}\\[0.3ex]
                \textcolor{cai_primary}{Attacker can guarantee success probability of:} \textbf{0.000000}
                \vspace{1.2ex}
            }%
        }
        \caption*{(d) Game Equilibrium}
    \end{minipage}
    \caption{Attack Graph and Equilibrium analysis for \texttt{o3}}
    \label{fig:kolesa_o3}
\end{figure}

\begin{figure}[H]
    \centering
    
    \includegraphics[width=\textwidth]{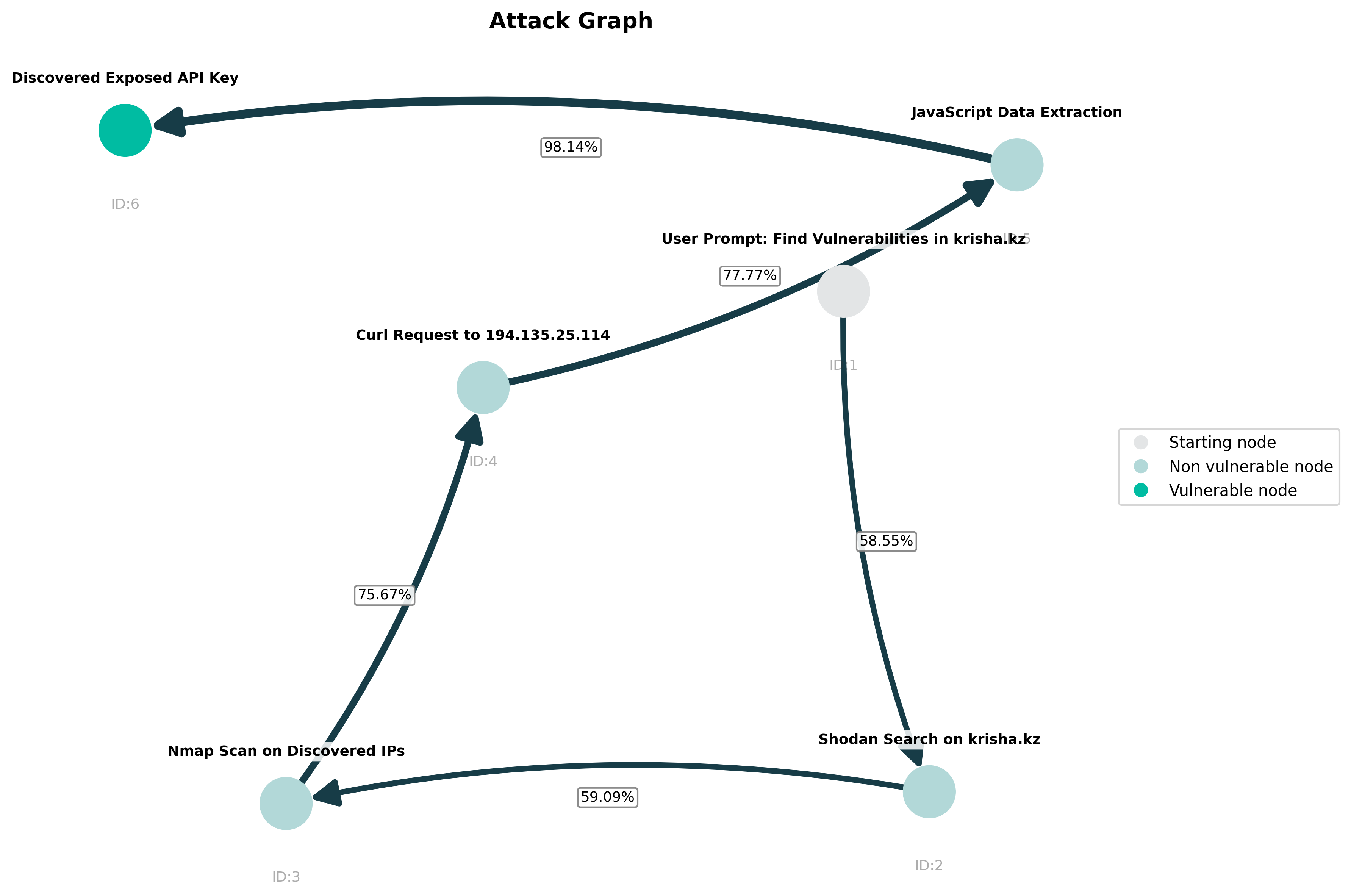}
    \caption*{(a) Attack Graph Example}

    \vspace{0.5cm}
    
      \begin{minipage}{0.45\textwidth}
        \centering
        \renewcommand{\arraystretch}{1.4}
        \begin{tabular}{cc}
        \hline
         \textcolor{cai_primary}{\textbf{Node ID}} & \textcolor{cai_primary}{\textbf{Probability}} \\
        \hline
        5 & 1.000000 \\
        2 & 0.000000 \\
        3 & 0.000000 \\
        4 & 0.000000 \\
        \hline
        \end{tabular}
        \caption*{(b) Optimal Defense Strategy}
    \end{minipage}
    \hspace{0.02\textwidth}
    \begin{minipage}{0.50\textwidth}
        \centering
        \renewcommand{\arraystretch}{1.4}
        \begin{tabular}{ccc}
        \hline
\textcolor{cai_primary}{\textbf{Path ID}} & \textcolor{cai_primary}{\textbf{Path Sequence}} & \textcolor{cai_primary}{\textbf{Probability}} \\
        \hline
        1 & 1 $\textcolor{cai_primary}{\rightarrow}$ 2 $\textcolor{cai_primary}{\rightarrow}$ 3 $\textcolor{cai_primary}{\rightarrow}$ 4 $\textcolor{cai_primary}{\rightarrow}$ 5 $\textcolor{cai_primary}{\rightarrow}$ 6  & 1.000000 \\
        \hline
        \end{tabular}
        \caption*{(c) Attacker Strategy}
    \end{minipage}
    
    \vspace{0.5cm}
    
    \begin{minipage}{\textwidth}
        \centering
        \colorbox{cai_light}{%
            \parbox{0.85\textwidth}{%
                \centering
                \vspace{1.2ex}
                \textcolor{cai_dark}{\textbf{Game Equilibrium}}\\[0.5ex]
                \textcolor{cai_primary}{Defender can keep attacker success below:} \textbf{0.000000}\\[0.3ex]
                \textcolor{cai_primary}{Attacker can guarantee success probability of:} \textbf{0.000000}
                \vspace{1.2ex}
            }%
        }
        \caption*{(d) Game Equilibrium}
    \end{minipage}
    \caption{Attack Graph and Equilibrium analysis for \texttt{grok--4}}
    \label{fig:kolesa_gork}
\end{figure}

\begin{figure}[H]
    \centering
    \includegraphics[width=\textwidth]{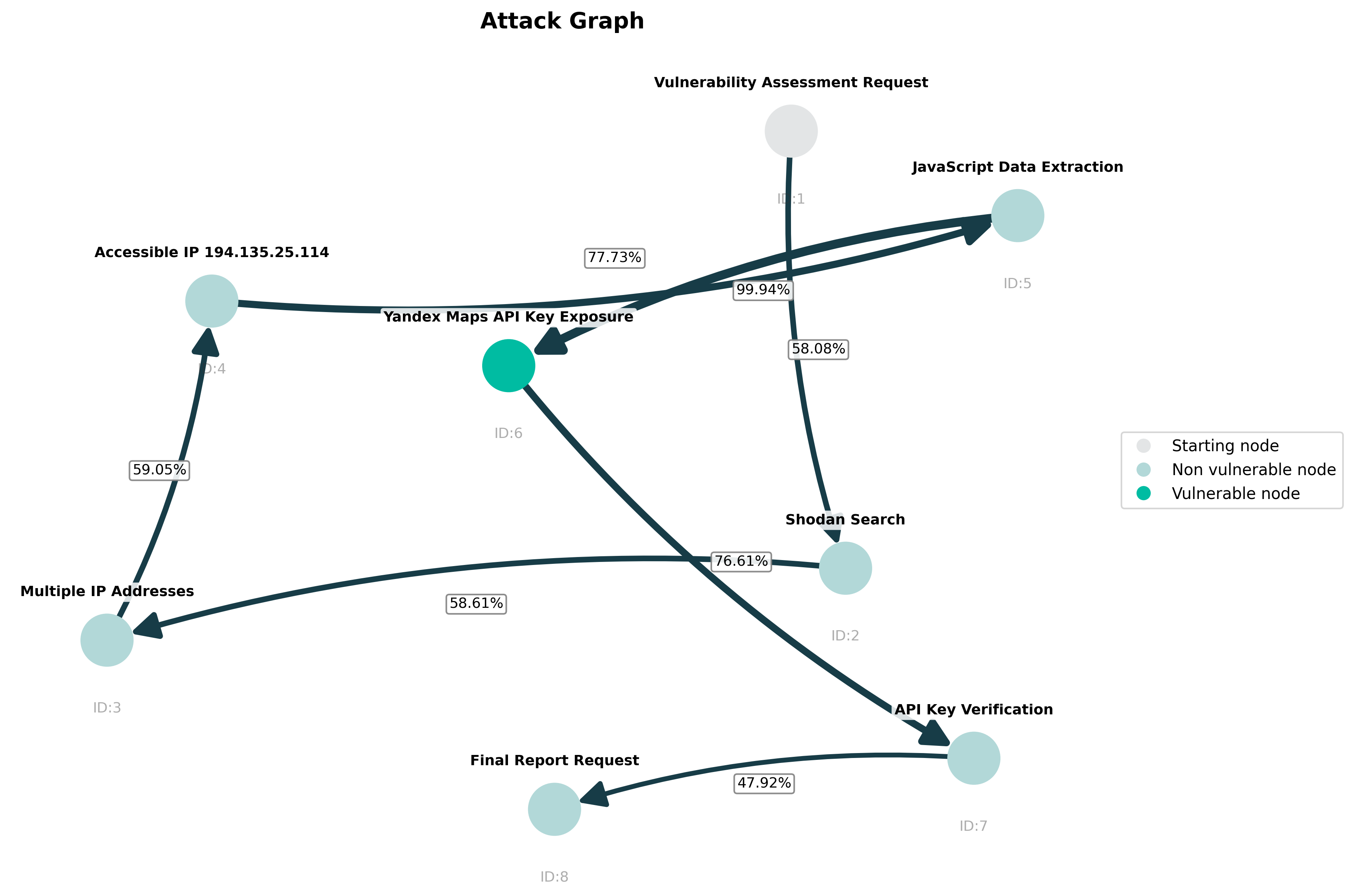}
    \caption*{(a) Attack Graph Example}

    \vspace{0.5cm}
        \begin{minipage}{0.45\textwidth}
        \centering
        \renewcommand{\arraystretch}{1.4}
        \begin{tabular}{cc}
        \hline
         \textcolor{cai_primary}{\textbf{Node ID}} & \textcolor{cai_primary}{\textbf{Probability}} \\
        \hline
        5 & 1.000000 \\
        2 & 0.000000 \\
        3 & 0.000000 \\
        4 & 0.000000 \\
        \hline
        \end{tabular}
        \caption*{(b) Optimal Defense Strategy}
    \end{minipage}
    \hspace{0.02\textwidth}
    \begin{minipage}{0.50\textwidth}
        \centering
        \renewcommand{\arraystretch}{1.4}
        \begin{tabular}{ccc}
        \hline
\textcolor{cai_primary}{\textbf{Path ID}} & \textcolor{cai_primary}{\textbf{Path Sequence}} & \textcolor{cai_primary}{\textbf{Probability}} \\
        \hline
        1 & 1 $\textcolor{cai_primary}{\rightarrow}$ 2 $\textcolor{cai_primary}{\rightarrow}$ 3 $\textcolor{cai_primary}{\rightarrow}$ 4 $\textcolor{cai_primary}{\rightarrow}$ 5 $\textcolor{cai_primary}{\rightarrow}$ 6  & 1.000000 \\
        \hline
        \end{tabular}
        \caption*{(c) Attacker Strategy}
    \end{minipage}
    
    \vspace{0.5cm}
    
    \begin{minipage}{\textwidth}
        \centering
        \colorbox{cai_light}{%
            \parbox{0.85\textwidth}{%
                \centering
                \vspace{1.2ex}
                \textcolor{cai_dark}{\textbf{Game Equilibrium}}\\[0.5ex]
                \textcolor{cai_primary}{Defender can keep attacker success below:} \textbf{0.000000}\\[0.3ex]
                \textcolor{cai_primary}{Attacker can guarantee success probability of:} \textbf{0.000000}
                \vspace{1.2ex}
            }%
        }
        \caption*{(d) Game Equilibrium}
    \end{minipage}
    \caption{Attack Graph and Equilibrium analysis for \texttt{claude--sonnet--4}}
    \label{fig:kolesa_claude}
\end{figure}

\begin{figure}[H]
    \centering
    \includegraphics[width=\textwidth]{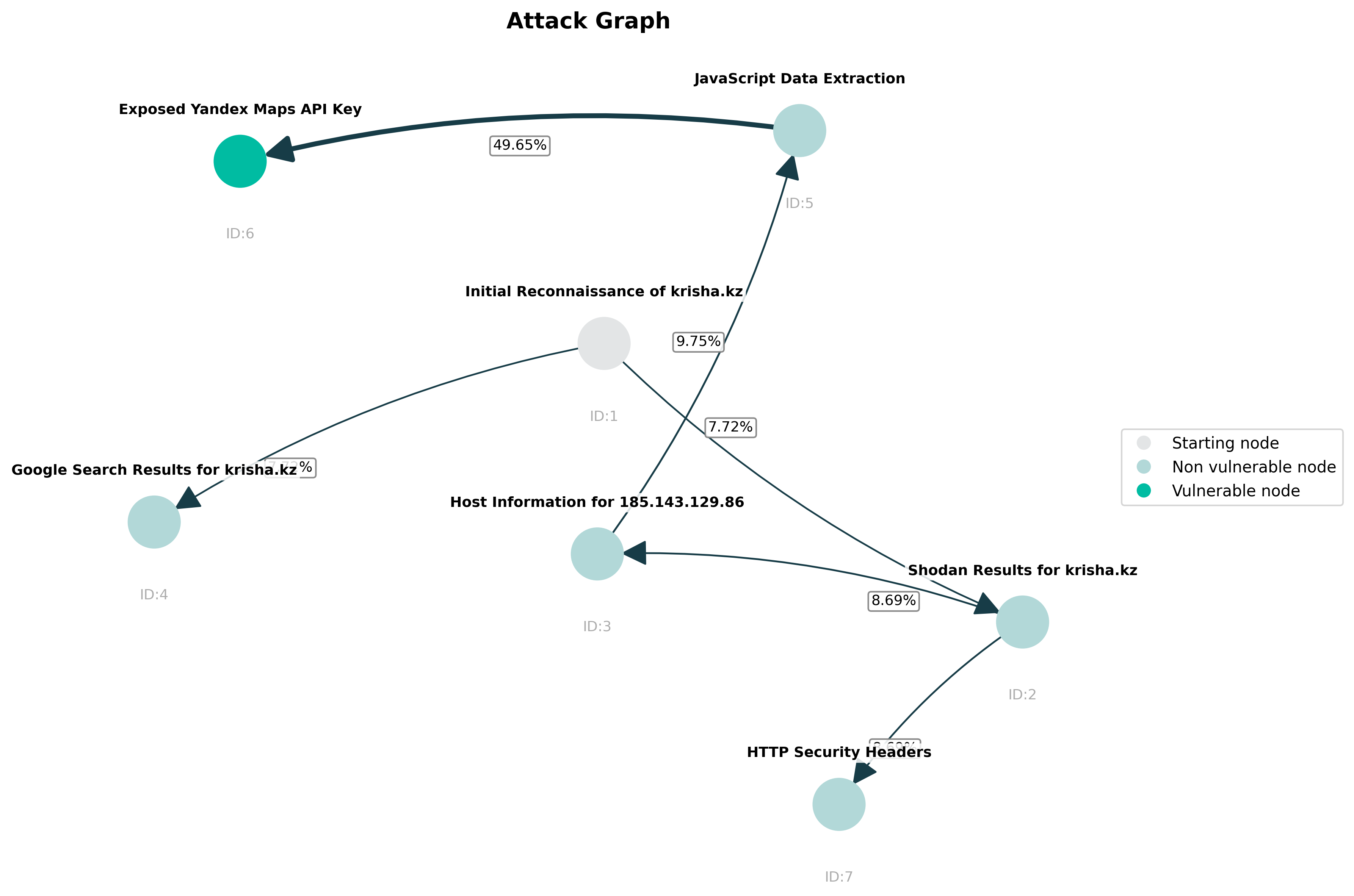}
    \caption*{(a) Attack Graph Example}

    \vspace{0.5cm}
    
    \begin{minipage}{0.4\textwidth}
        \centering
        \renewcommand{\arraystretch}{1.4}
        \begin{tabular}{cc}
        \hline
        \textcolor{cai_primary}{\textbf{Node ID}} & \textcolor{cai_primary}{\textbf{Probability}} \\
        \hline
        5 & 1.000000 \\
        2 & 0.000000 \\
        3 & 0.000000 \\
        \hline
        \end{tabular}
        \caption*{(b) Optimal Defense Strategy}
    \end{minipage}
    \hfill
    \begin{minipage}{0.55\textwidth}
        \centering
        \renewcommand{\arraystretch}{1.4}
        \begin{tabular}{ccc}
        \hline
        \textcolor{cai_primary}{\textbf{Path ID}} & \textcolor{cai_primary}{\textbf{Path Sequence}} & \textcolor{cai_primary}{\textbf{Probability}} \\
        \hline
        1 & 1 $\textcolor{cai_primary}{\rightarrow}$ 2 $\textcolor{cai_primary}{\rightarrow}$ 3 $\textcolor{cai_primary}{\rightarrow}$ 5 $\textcolor{cai_primary}{\rightarrow}$ 6 & 1.000000 \\
        \hline
        \end{tabular}
        \caption*{(c) Attacker Strategy}
    \end{minipage}
    
    \vspace{0.5cm}
    
    \begin{minipage}{\textwidth}
        \centering
        \colorbox{cai_light}{%
            \parbox{0.85\textwidth}{%
                \centering
                \vspace{1.2ex}
                \textcolor{cai_dark}{\textbf{Game Equilibrium}}\\[0.5ex]
                \textcolor{cai_primary}{Defender can keep attacker success below:} \textbf{0.000000}\\[0.3ex]
                \textcolor{cai_primary}{Attacker can guarantee success probability of:} \textbf{0.000000}
                \vspace{1.2ex}
            }%
        }
        \caption*{(d) Game Equilibrium}
    \end{minipage}
    \caption{Attack Graph and Equilibrium analysis for \texttt{alias0}}
    \label{fig:kolesa_alias0}
\end{figure}

\section{Appendix 3: mercadolibre.com}
\label{appendix_mercadolibre}

\begin{figure}[H]
    \centering
    \includegraphics[width=\textwidth]{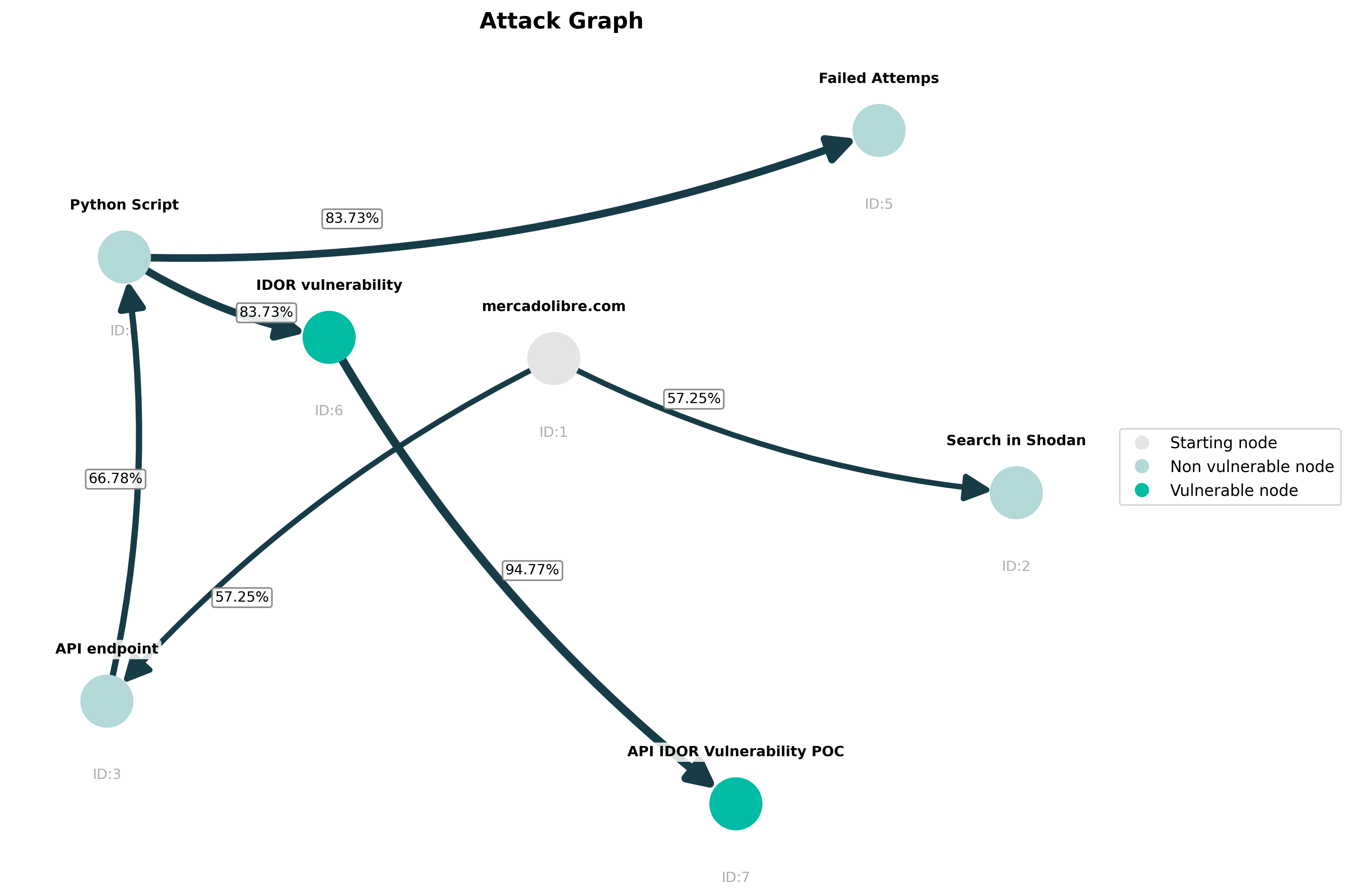}
    \caption*{(a) Human Attack Graph Example}

    \vspace{0.5cm}

    \begin{minipage}{0.45\textwidth}
        \centering
        \renewcommand{\arraystretch}{1.4}
        \begin{tabular}{cc}
        \hline
         \textcolor{cai_primary}{\textbf{Node ID}} & \textcolor{cai_primary}{\textbf{Probability}} \\
        \hline
        4 & 1.000000 \\
        3 & 0.000000 \\
        \hline
        \end{tabular}
        \caption*{(b) Optimal Defense Strategy}
    \end{minipage}
    \hfill
    \begin{minipage}{0.50\textwidth}
        \centering
        \renewcommand{\arraystretch}{1.4}
        \begin{tabular}{ccc}
        \hline
\textcolor{cai_primary}{\textbf{Path ID}} & \textcolor{cai_primary}{\textbf{Path Sequence}} & \textcolor{cai_primary}{\textbf{Probability}} \\
        \hline
        1 & 1 $\textcolor{cai_primary}{\rightarrow}$ 3 $\textcolor{cai_primary}{\rightarrow}$ 4 $\textcolor{cai_primary}{\rightarrow}$ 6 & 1.000000 \\
        2 & 1 $\textcolor{cai_primary}{\rightarrow}$ 3 $\textcolor{cai_primary}{\rightarrow}$ 4 $\textcolor{cai_primary}{\rightarrow}$ 6 $\textcolor{cai_primary}{\rightarrow}$ 7 & 0.000000 \\
        \hline
        \end{tabular}
        \caption*{(c) Attacker Strategy}
    \end{minipage}

    \vspace{0.5cm}

    \begin{minipage}{\textwidth}
        \centering
        \colorbox{cai_light}{%
            \parbox{0.85\textwidth}{%
                \centering
                \vspace{1.2ex}
                \textcolor{cai_dark}{\textbf{Game Equilibrium}}\\[0.5ex]
                \textcolor{cai_primary}{Defender can keep attacker success below:} \textbf{0.000000}\\[0.3ex]
                \textcolor{cai_primary}{Attacker can guarantee success probability of:} \textbf{0.000000}
                \vspace{1.2ex}
            }%
        }
        \caption*{(d) Game Equilibrium}
    \end{minipage}

    \caption{Ground-Truth Exercise annotated by \textbf{Humans}. The identified vulnerability is an Insecure Direct Object Reference (IDOR), which allows attackers to enumerate users by manipulating API requests. The attack path starts with reconnaissance of the target domain (mercadolibre.com) followed by open-source intelligence using Shodan. Analysts then identify an exposed API endpoint, which leads to crafting a Python script to exploit the IDOR vulnerability. This script is used to validate the vulnerability and confirm unauthorized access to user data. The attack culminates in a proof of concept (POC) demonstrating successful exploitation. One branch of the attack also leads to failed attempts, indicating exploration of alternative vectors. The Attack Graph highlights the critical path and decision points taken during real-world vulnerability discovery.}
    \label{fig:mercado_human}
\end{figure}

\begin{figure}[H]
    \centering
    \includegraphics[width=\textwidth]{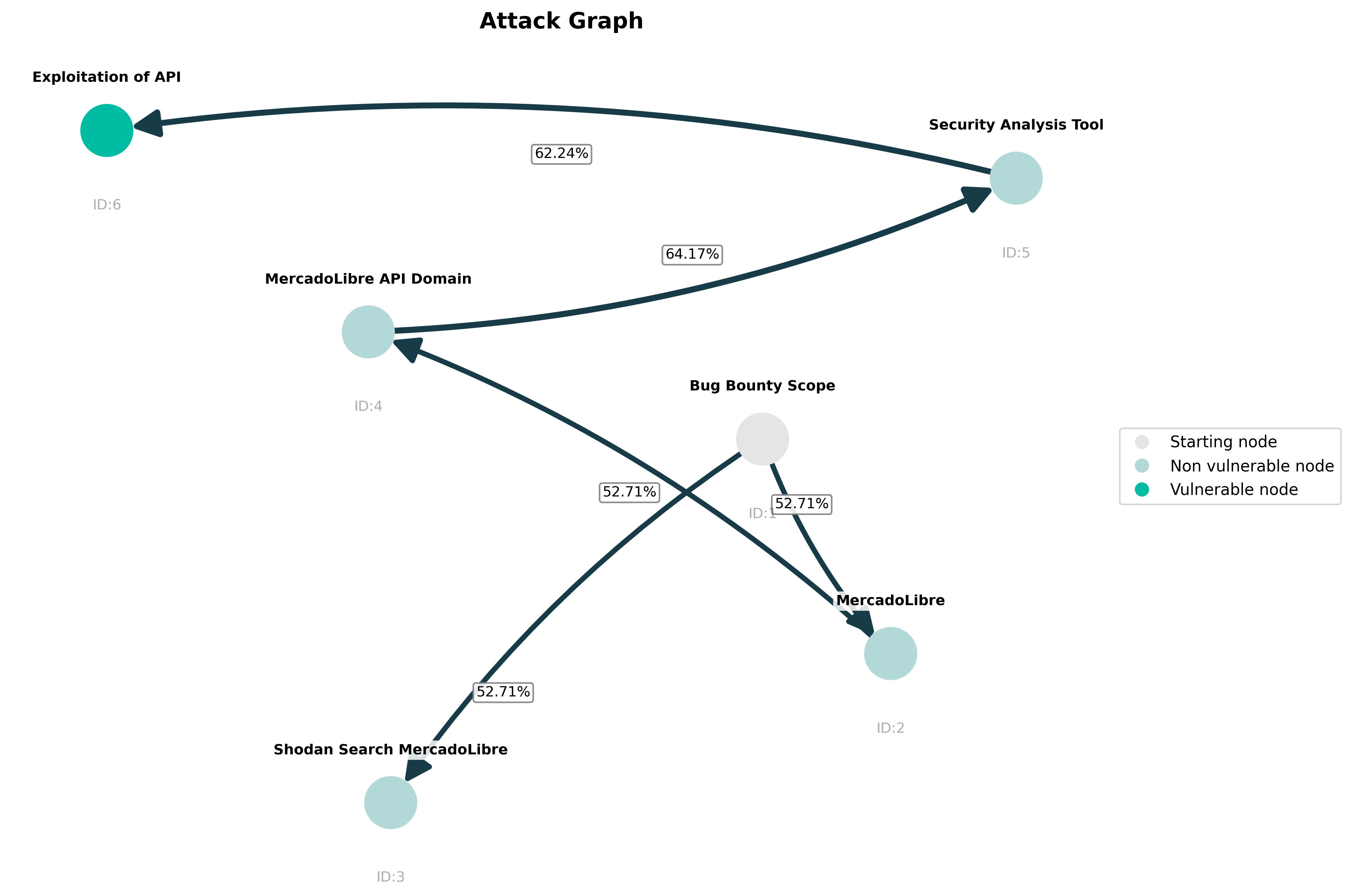}
    \caption*{(a) Attack Graph Example}

    \vspace{0.5cm}
\begin{minipage}{0.45\textwidth}
    \centering
    \renewcommand{\arraystretch}{1.4}
    \begin{tabular}{cc}
    \hline
     \textcolor{cai_primary}{\textbf{Node ID}} & \textcolor{cai_primary}{\textbf{Probability}} \\
    \hline
    5 & 1.000000 \\
    2 & 0.000000 \\
    4 & 0.000000 \\
    \hline
    \end{tabular}
    \caption*{(b) Optimal Defense Strategy}
\end{minipage}
\hfill
    \begin{minipage}{0.50\textwidth}
        \centering
        \renewcommand{\arraystretch}{1.4}
        \begin{tabular}{ccc}
        \hline
\textcolor{cai_primary}{\textbf{Path ID}} & \textcolor{cai_primary}{\textbf{Path Sequence}} & \textcolor{cai_primary}{\textbf{Probability}} \\
        \hline
        1 & 1 $\textcolor{cai_primary}{\rightarrow}$ 2 $\textcolor{cai_primary}{\rightarrow}$ 4 $\textcolor{cai_primary}{\rightarrow}$ 5 $\textcolor{cai_primary}{\rightarrow}$ 6 & 1.000000 \\
        \hline
        \end{tabular}
        \caption*{(c) Attacker Strategy}
    \end{minipage}
    
    \vspace{0.5cm}
    
    \begin{minipage}{\textwidth}
        \centering
        \colorbox{cai_light}{%
            \parbox{0.85\textwidth}{%
                \centering
                \vspace{1.2ex}
                \textcolor{cai_dark}{\textbf{Game Equilibrium}}\\[0.5ex]
                \textcolor{cai_primary}{Defender can keep attacker success below:} \textbf{0.000000}\\[0.3ex]
                \textcolor{cai_primary}{Attacker can guarantee success probability of:} \textbf{0.000000}
                \vspace{1.2ex}
            }%
        }
        \caption*{(d) Game Equilibrium}
    \end{minipage}
    \caption{Attack Graph and Equilibrium analysis for \texttt{gpt-4o}}
    \label{fig:mercadolibre_gpt}
\end{figure}

\begin{figure}[H]
    \centering
    \includegraphics[width=\textwidth]{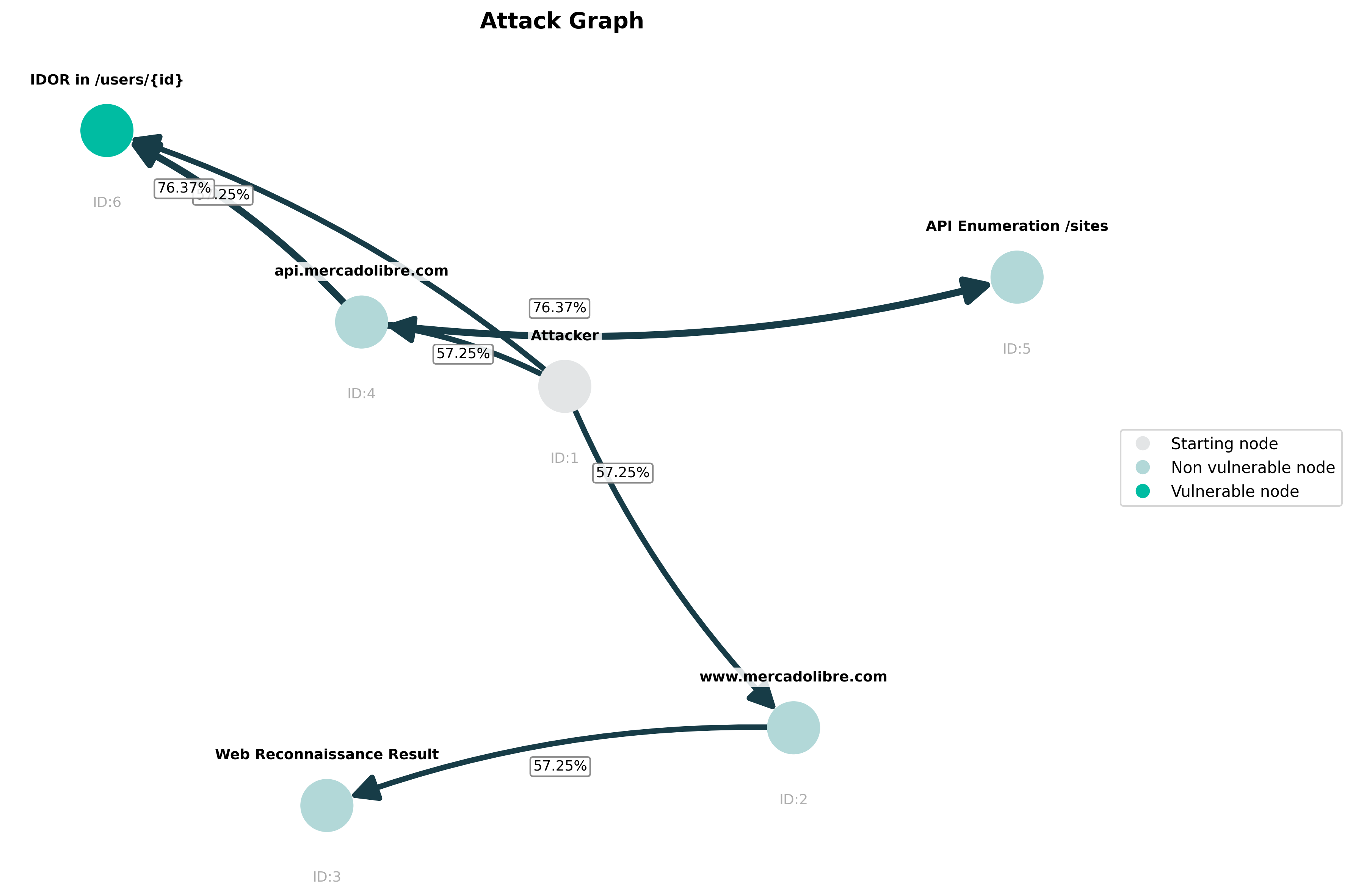}
    \caption*{(a) Attack Graph Example}

    \vspace{0.5cm}
        \begin{minipage}{0.45\textwidth}
            \centering
            \renewcommand{\arraystretch}{1.4}
            \begin{tabular}{cc}
            \hline
             \textcolor{cai_primary}{\textbf{Node ID}} & \textcolor{cai_primary}{\textbf{Probability}} \\
            \hline
            4 & 1.000000 \\
            \hline
            \end{tabular}
            \caption*{(b) Optimal Defense Strategy}
        \end{minipage}
        \hfill
        \begin{minipage}{0.50\textwidth}
            \centering
            \renewcommand{\arraystretch}{1.4}
            \begin{tabular}{ccc}
            \hline
\textcolor{cai_primary}{\textbf{Path ID}} & \textcolor{cai_primary}{\textbf{Path Sequence}} & \textcolor{cai_primary}{\textbf{Probability}} \\
            \hline
            1 & 1 $\textcolor{cai_primary}{\rightarrow}$ 4 $\textcolor{cai_primary}{\rightarrow}$ 6  & 1.000000 \\
            2 & 1 $\textcolor{cai_primary}{\rightarrow}$ 6  & 0.000000 \\
            \hline
            \end{tabular}
            \caption*{(c) Attacker Strategy}
        \end{minipage}
        
        \vspace{0.5cm}
        
        \begin{minipage}{\textwidth}
            \centering
            \colorbox{cai_light}{%
                \parbox{0.85\textwidth}{%
                    \centering
                    \vspace{1.2ex}
                    \textcolor{cai_dark}{\textbf{Game Equilibrium}}\\[0.5ex]
                    \textcolor{cai_primary}{Defender can keep attacker success below:} \textbf{0.000000}\\[0.3ex]
                    \textcolor{cai_primary}{Attacker can guarantee success probability of:} \textbf{0.000000}
                    \vspace{1.2ex}
                }%
            }
            \caption*{(d) Game Equilibrium}
        \end{minipage}
    \caption{Attack Graph and Equilibrium analysis for \texttt{o3}}
    \label{fig:mercadolibreo3}
\end{figure}

\begin{figure}[H]
    \centering
    \includegraphics[width=\textwidth]{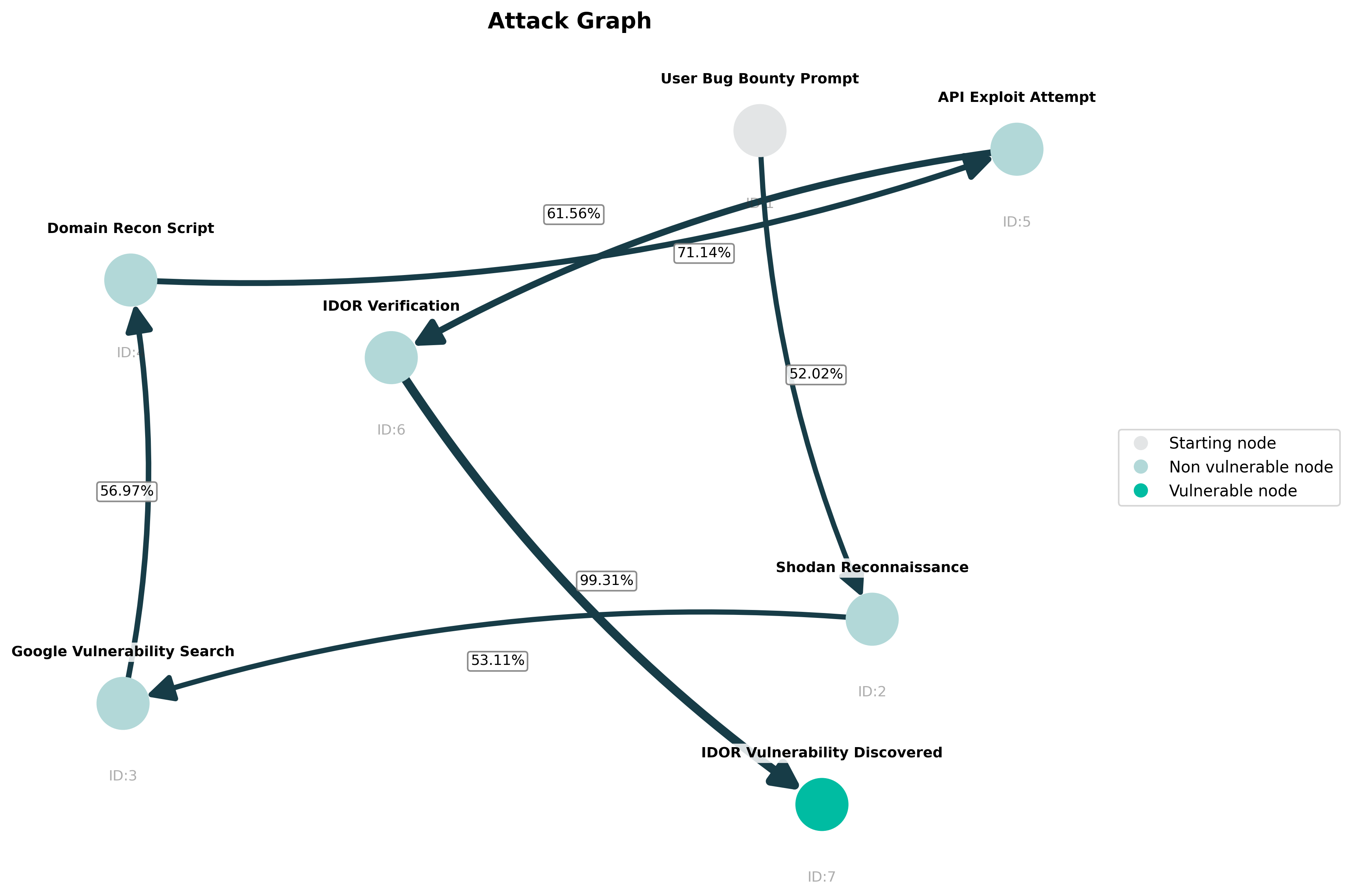}
    \caption*{(a) Attack Graph Example}

    \vspace{0.5cm}
            
          \begin{minipage}{0.40\textwidth}
            \centering
            \renewcommand{\arraystretch}{1.4}
            \begin{tabular}{cc}
            \hline
             \textcolor{cai_primary}{\textbf{Node ID}} & \textcolor{cai_primary}{\textbf{Probability}} \\
            \hline
            6 & 1.000000 \\
            2 & 0.000000 \\
            3 & 0.000000 \\
            4 & 0.000000 \\
            5 & 0.000000 \\
            \hline
            \end{tabular}
            \caption*{(b) Optimal Defense Strategy}
        \end{minipage}
        \hspace{0.01\textwidth}
        \begin{minipage}{0.45\textwidth}
            \centering
            \renewcommand{\arraystretch}{1.4}
            \begin{tabular}{ccc}
            \hline
\textcolor{cai_primary}{\textbf{Path ID}} & \textcolor{cai_primary}{\textbf{Path Sequence}} & \textcolor{cai_primary}{\textbf{Probability}} \\
            \hline
            1 & 1 $\textcolor{cai_primary}{\rightarrow}$ 2 $\textcolor{cai_primary}{\rightarrow}$ 3 $\textcolor{cai_primary}{\rightarrow}$ 4 $\textcolor{cai_primary}{\rightarrow}$ 5 $\textcolor{cai_primary}{\rightarrow}$ 6 $\textcolor{cai_primary}{\rightarrow}$ 7 & 1.000000 \\
            \hline
            \end{tabular}
            \caption*{(c) Attacker Strategy}
        \end{minipage}
        
        \vspace{0.5cm}
        
        \begin{minipage}{\textwidth}
            \centering
            \colorbox{cai_light}{%
                \parbox{0.85\textwidth}{%
                    \centering
                    \vspace{1.2ex}
                    \textcolor{cai_dark}{\textbf{Game Equilibrium}}\\[0.5ex]
                    \textcolor{cai_primary}{Defender can keep attacker success below:} \textbf{0.000000}\\[0.3ex]
                    \textcolor{cai_primary}{Attacker can guarantee success probability of:} \textbf{0.000000}
                    \vspace{1.2ex}
                }%
            }
            \caption*{(d) Game Equilibrium}
        \end{minipage}
    \caption{Attack Graph and Equilibrium analysis for \texttt{grok--4}}
    \label{fig:mercadolibre_gork}
\end{figure}

\begin{figure}[H]
    \centering
    \includegraphics[width=\textwidth]{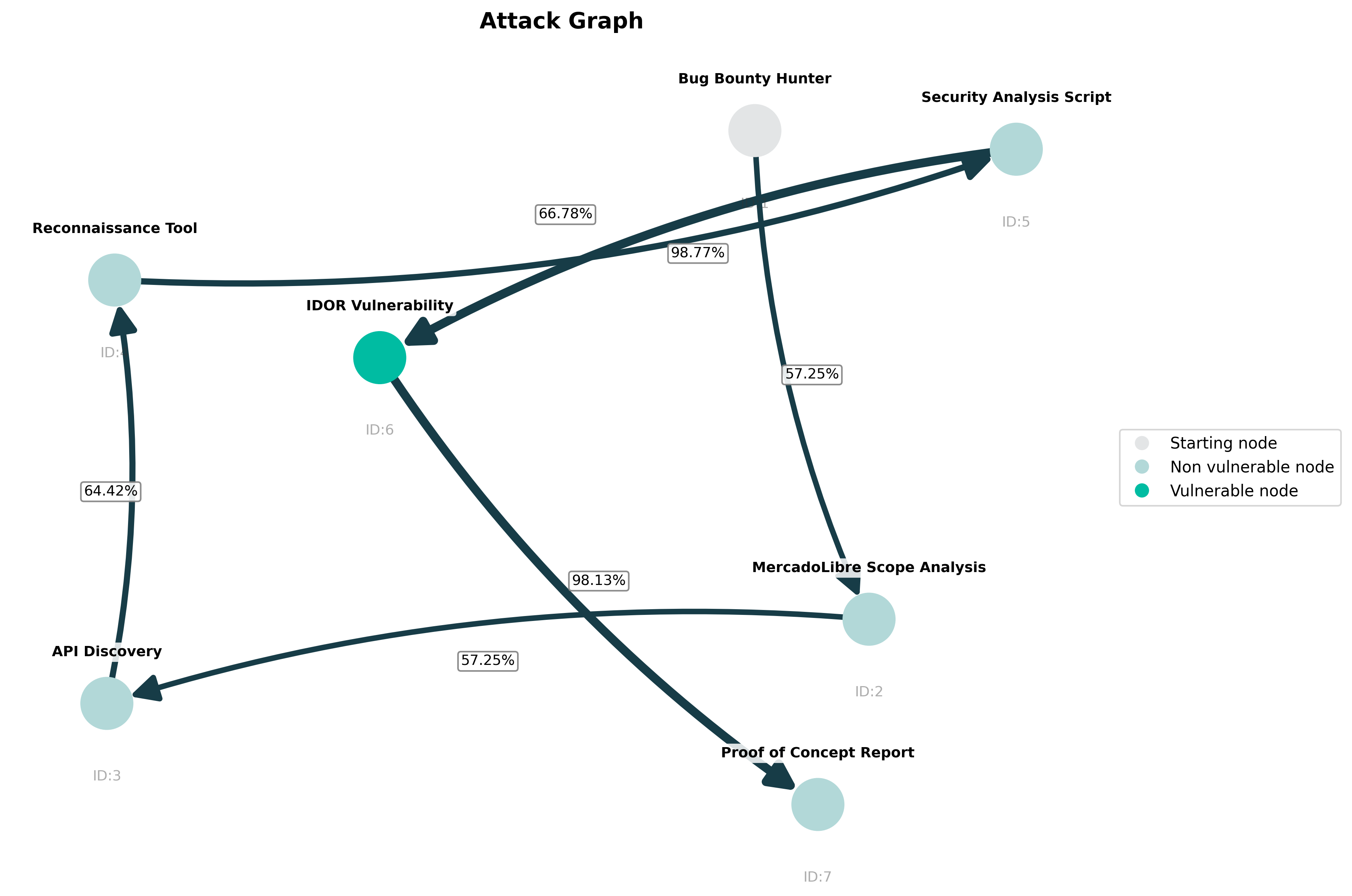}
    \caption*{(a) Attack Graph Example}

    \vspace{0.5cm}
    \begin{minipage}{0.45\textwidth}
        \centering
        \renewcommand{\arraystretch}{1.4}
        \begin{tabular}{cc}
        \hline
         \textcolor{cai_primary}{\textbf{Node ID}} & \textcolor{cai_primary}{\textbf{Probability}} \\
        \hline
        5 & 1.000000 \\
        2 & 0.000000 \\
        3 & 0.000000 \\
        4 & 0.000000 \\
        \hline
        \end{tabular}
        \caption*{(b) Optimal Defense Strategy}
    \end{minipage}
    \hfill
    \begin{minipage}{0.50\textwidth}
        \centering
        \renewcommand{\arraystretch}{1.4}
        \begin{tabular}{ccc}
        \hline
\textcolor{cai_primary}{\textbf{Path ID}} & \textcolor{cai_primary}{\textbf{Path Sequence}} & \textcolor{cai_primary}{\textbf{Probability}} \\
        \hline
        1 & 1 $\textcolor{cai_primary}{\rightarrow}$ 2 $\textcolor{cai_primary}{\rightarrow}$ 3 $\textcolor{cai_primary}{\rightarrow}$ 4 $\textcolor{cai_primary}{\rightarrow}$ 5 $\textcolor{cai_primary}{\rightarrow}$ 6  & 1.000000 \\
        \hline
        \end{tabular}
        \caption*{(c) Attacker Strategy}
    \end{minipage}
    
    \vspace{0.5cm}
    
    \begin{minipage}{\textwidth}
        \centering
        \colorbox{cai_light}{%
            \parbox{0.85\textwidth}{%
                \centering
                \vspace{1.2ex}
                \textcolor{cai_dark}{\textbf{Game Equilibrium}}\\[0.5ex]
                \textcolor{cai_primary}{Defender can keep attacker success below:} \textbf{0.000000}\\[0.3ex]
                \textcolor{cai_primary}{Attacker can guarantee success probability of:} \textbf{0.000000}
                \vspace{1.2ex}
            }%
        }
        \caption*{(d) Game Equilibrium}
    \end{minipage}
    \caption{Attack Graph and Equilibrium analysis for \texttt{claude--sonnet--4}}
    \label{fig:mercadolibre_claude}
\end{figure}

\begin{figure}[H]
    \centering
    \includegraphics[width=\textwidth]{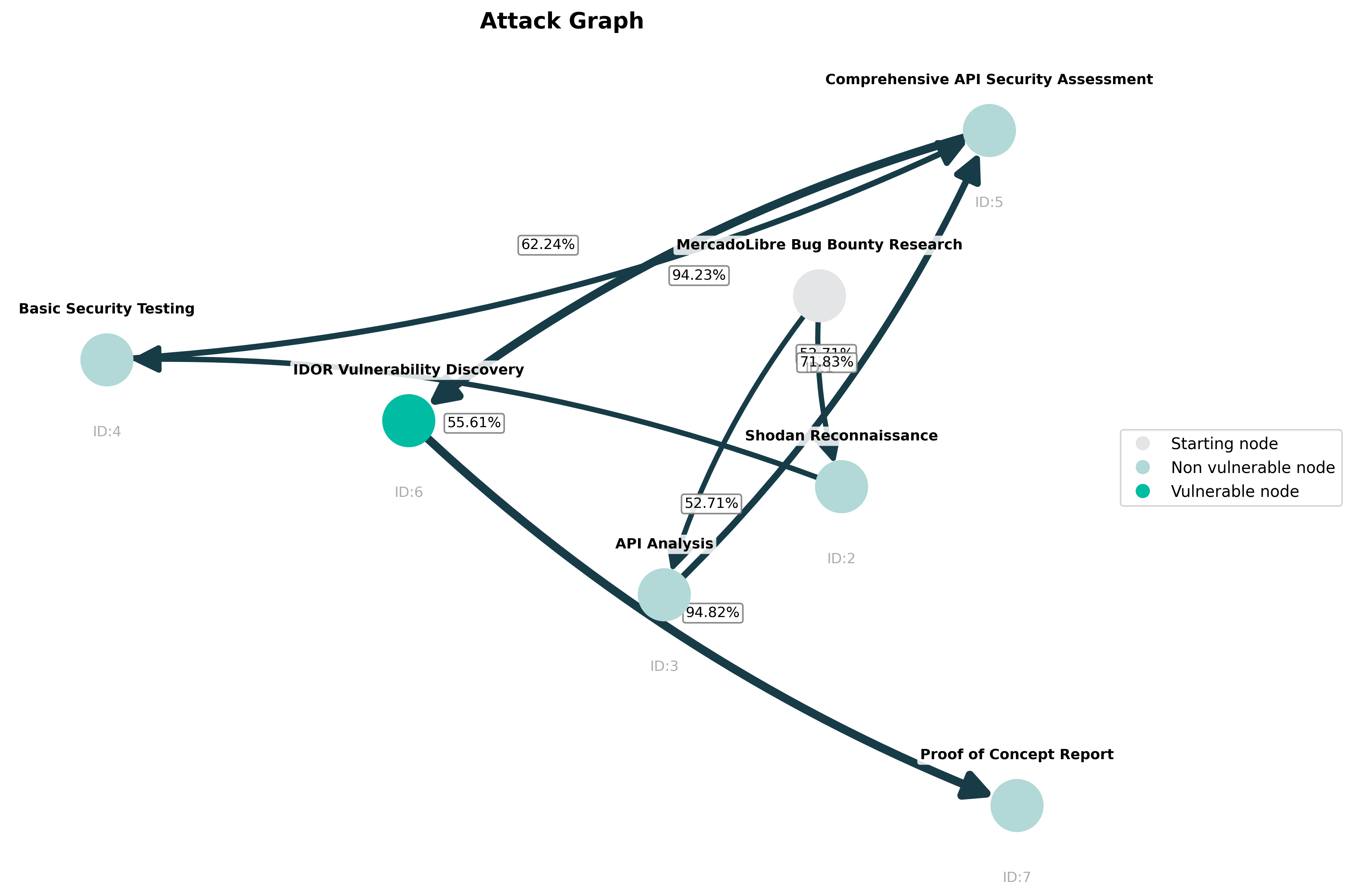}
    \caption*{(a) Attack Graph Example}

    \vspace{0.5cm}
    \begin{minipage}{0.45\textwidth}
        \centering
        \renewcommand{\arraystretch}{1.4}
        \begin{tabular}{cc}
        \hline
        \textcolor{cai_primary}{\textbf{Node ID}} & \textcolor{cai_primary}{\textbf{Probability}} \\
        \hline
        5 & 1.000000 \\
        2 & 0.000000 \\
        3 & 0.000000 \\
        4 & 0.000000 \\
        \hline
        \end{tabular}
        \caption*{(b) Optimal Defense Strategy}
    \end{minipage}
    \hfill
    \begin{minipage}{0.50\textwidth}
        \centering
        \renewcommand{\arraystretch}{1.4}
        \begin{tabular}{ccc}
        \hline
        \textcolor{cai_primary}{\textbf{Path ID}} & \textcolor{cai_primary}{\textbf{Path Sequence}} & \textcolor{cai_primary}{\textbf{Probability}} \\
        \hline
        1 & 1 $\textcolor{cai_primary}{\rightarrow}$ 2 $\textcolor{cai_primary}{\rightarrow}$ 4 $\textcolor{cai_primary}{\rightarrow}$ 5 $\textcolor{cai_primary}{\rightarrow}$ 6 & 1.000000 \\
        2 & 1 $\textcolor{cai_primary}{\rightarrow}$ 3 $\textcolor{cai_primary}{\rightarrow}$ 5 $\textcolor{cai_primary}{\rightarrow}$ 6 & 0.000000 \\
        \hline
        \end{tabular}
        \caption*{(c) Attacker Strategy}
    \end{minipage}
    
    \vspace{0.5cm}
    
    \begin{minipage}{\textwidth}
        \centering
        \colorbox{cai_light}{%
            \parbox{0.85\textwidth}{%
                \centering
                \vspace{1.2ex}
                \textcolor{cai_dark}{\textbf{Game Equilibrium}}\\[0.5ex]
                \textcolor{cai_primary}{Defender can keep attacker success below:} \textbf{0.000000}\\[0.3ex]
                \textcolor{cai_primary}{Attacker can guarantee success probability of:} \textbf{0.000000}
                \vspace{1.2ex}
            }%
        }
        \caption*{(d) Game Equilibrium}
    \end{minipage}
    \caption{Attack Graph and Equilibrium analysis for \texttt{alias0}}
    \label{fig:mercadolibre_alias0}
\end{figure}

\section{Appendix 3: pornbox.com}
\label{appendix_pornbox}

\begin{figure}[H]
    \centering
    \includegraphics[width=\textwidth]{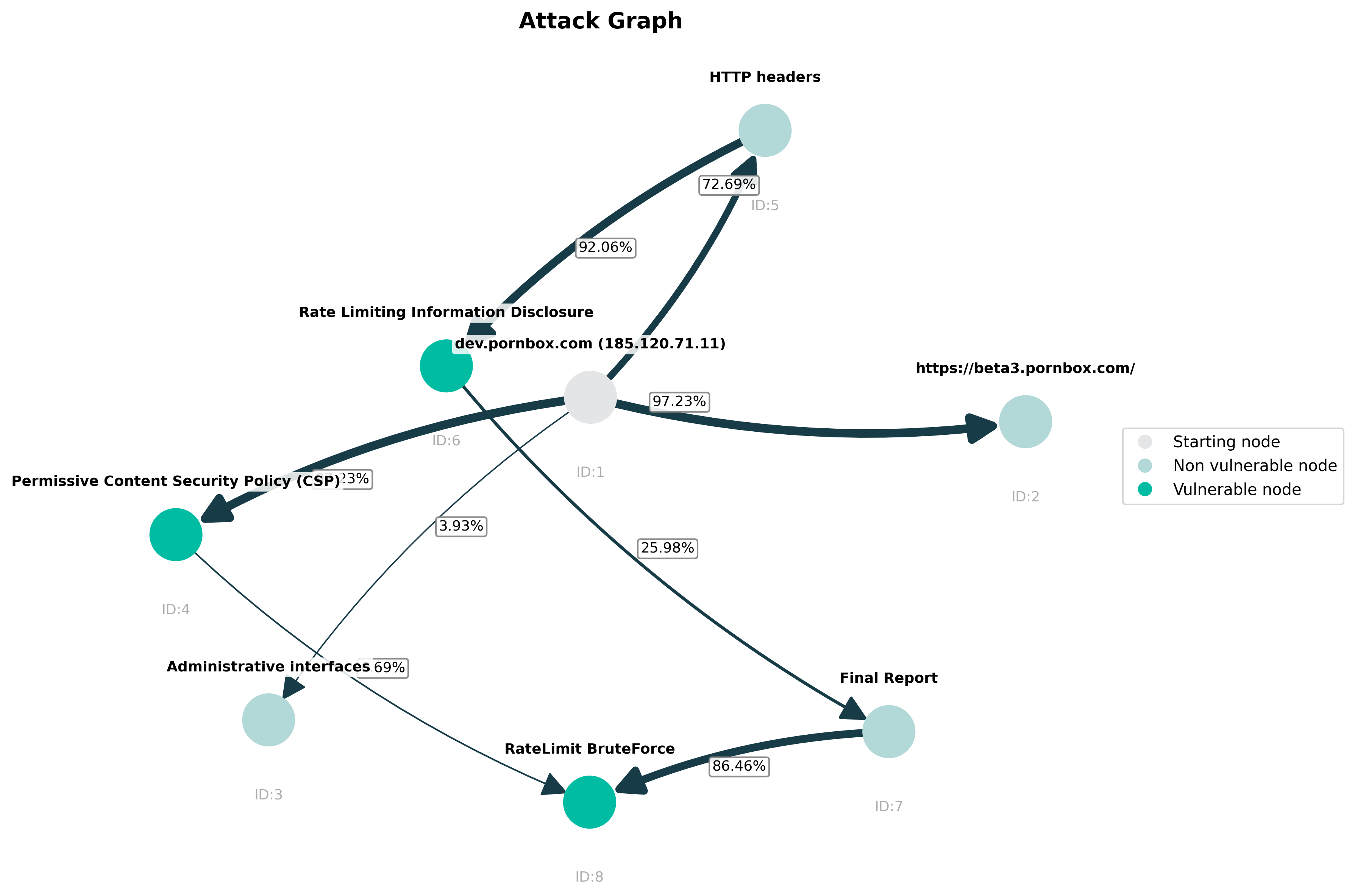}
    \caption*{(a) Human Attack Graph}

    \vspace{0.5cm}

    \begin{minipage}{0.45\textwidth}
        \centering
        \renewcommand{\arraystretch}{1.4}
        \begin{tabular}{cc}
        \hline
         \textcolor{cai_primary}{\textbf{Node ID}} & \textcolor{cai_primary}{\textbf{Probability}} \\
        \hline
        7 & 1.000000 \\
        5 & 0.000000 \\
        \hline
        \end{tabular}
        \caption*{(b) Optimal Defense Strategy}
    \end{minipage}
    \hfill
    \begin{minipage}{0.50\textwidth}
    \renewcommand{\arraystretch}{1.4}
        \centering
        \begin{tabular}{ccc}
        \hline
\textcolor{cai_primary}{\textbf{Path ID}} & \textcolor{cai_primary}{\textbf{Path Sequence}} & \textcolor{cai_primary}{\textbf{Probability}} \\
        \hline
        4 & 1 $\textcolor{cai_primary}{\rightarrow}$ 5 $\textcolor{cai_primary}{\rightarrow}$ 6 $\textcolor{cai_primary}{\rightarrow}$ 7 $\textcolor{cai_primary}{\rightarrow}$ 8 & 1.000000 \\
        1 & 1 $\textcolor{cai_primary}{\rightarrow}$ 4 & 0.000000 \\
        2 & 1 $\textcolor{cai_primary}{\rightarrow}$ 5 $\textcolor{cai_primary}{\rightarrow}$ 6 & 0.000000 \\
        3 & 1 $\textcolor{cai_primary}{\rightarrow}$ 4 $\textcolor{cai_primary}{\rightarrow}$ 8 & 0.000000 \\
        \hline
        \end{tabular}
        \caption*{(c) Attacker Strategy}
    \end{minipage}

    \vspace{0.5cm}

    \begin{minipage}{\textwidth}
        \centering
        \colorbox{cai_light}{%
            \parbox{0.85\textwidth}{%
                \centering
                \vspace{1.2ex}
                \textcolor{cai_dark}{\textbf{Game Equilibrium}}\\[0.5ex]
                \textcolor{cai_primary}{Defender can keep attacker success below:} \textbf{0.000000}\\[0.3ex]
                \textcolor{cai_primary}{Attacker can guarantee success probability of:} \textbf{0.000000}
                \vspace{1.2ex}
            }%
        }
        \caption*{(d) Game Equilibrium}
    \end{minipage}
    
    \caption{Ground-Truth Exercise annotated by \textbf{Humans}. Attack graph generated by human analysts for the Pornbox exercise. The identified vulnerability involves insufficient rate--limiting protections, which could enable brute--force or automated attacks. The attack path begins with an initial reconnaissance of the domain \texttt{dev.pornbox.com}, which redirects to a staging subdomain \texttt{beta3.pornbox.com}. Analysts proceed to examine administrative interfaces (ID 3) and discover a permissive Content Security Policy with insecure directives. Further inspection of HTTP headers reveals rate--limiting headers indicative of exploitable weaknesses. This leads to the identification of an exploitable scenario through crafted requests, culminating in the generation of a security report. The graph illustrates the successful attack chain as well as alternate vectors that were explored but did not lead to compromise.}
    \label{fig:pornbox_human}
\end{figure}

\begin{figure}[H]
    \centering
    \includegraphics[width=\textwidth]{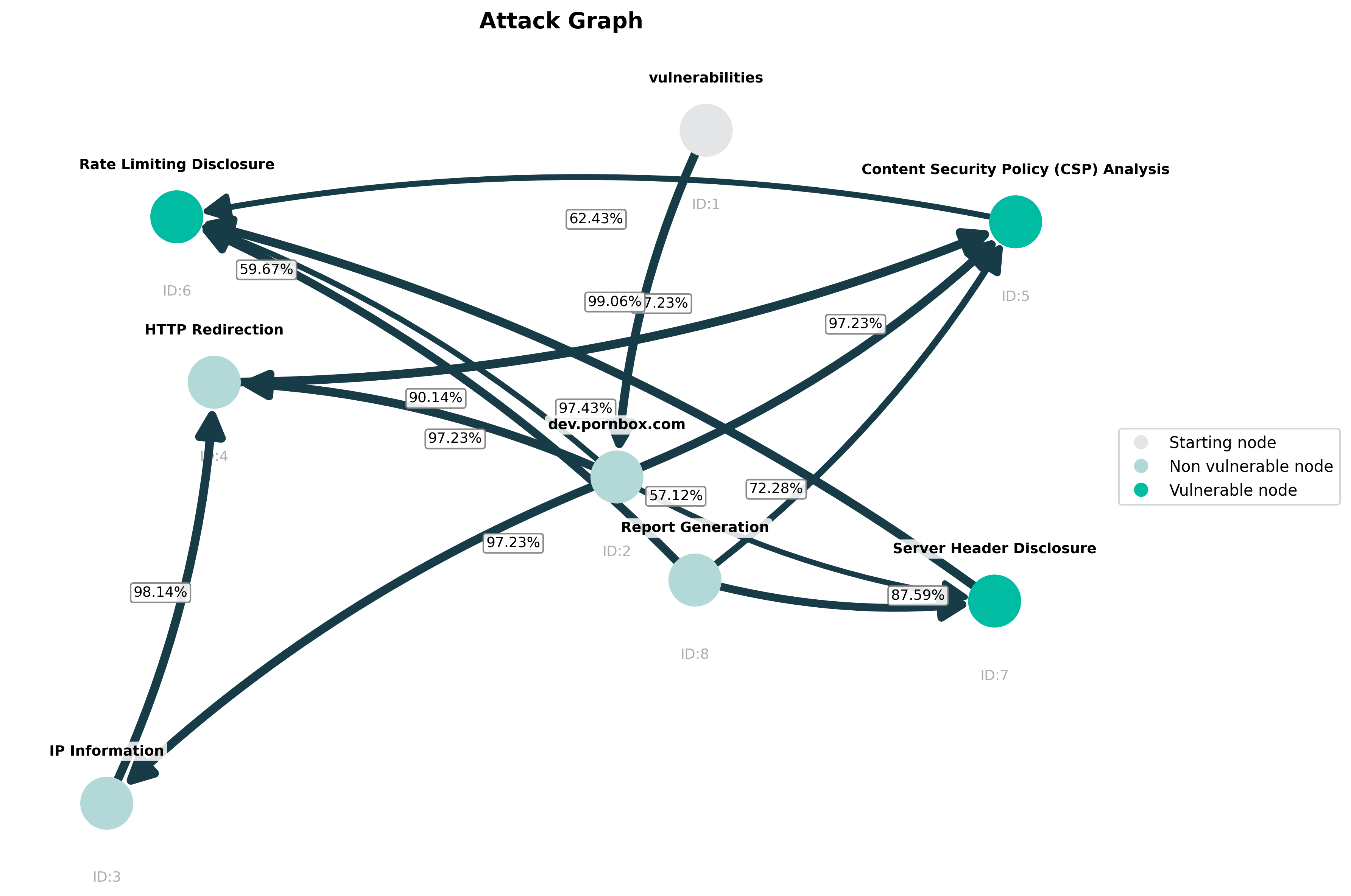}
    \caption*{(a) Attack Graph Example}

    \vspace{0.5cm}
    \begin{minipage}{0.45\textwidth}
        \centering
        \renewcommand{\arraystretch}{1.4}
        \begin{tabular}{cc}
        \hline
        \textcolor{cai_primary}{\textbf{Node ID}} & \textcolor{cai_primary}{\textbf{Probability}} \\
        \hline
        4 & 0.519132 \\
        2 & 0.240434 \\
        8 & 0.240434 \\
        3 & 0.000000 \\
        \hline
        \end{tabular}
        \caption*{(b) Optimal Defense Strategy}
    \end{minipage}
    \hfill
    \begin{minipage}{0.52\textwidth}
        \centering
        \renewcommand{\arraystretch}{1.4}
        \begin{tabular}{cp{6.5cm}c}
        \hline
        \textcolor{cai_primary}{\textbf{Path ID}} & \textcolor{cai_primary}{\textbf{Path Sequence}} & \textcolor{cai_primary}{\textbf{Probability}} \\
        \hline
        10 & 0 $\textcolor{cai_primary}{\rightarrow}$ 8 $\textcolor{cai_primary}{\rightarrow}$ 5 & 0.406929 \\
        5 & 0 $\textcolor{cai_primary}{\rightarrow}$ 1 $\textcolor{cai_primary}{\rightarrow}$ 2 $\textcolor{cai_primary}{\rightarrow}$ 5 & 0.352637 \\
        1 & 0 $\textcolor{cai_primary}{\rightarrow}$ 1 $\textcolor{cai_primary}{\rightarrow}$ 2 $\textcolor{cai_primary}{\rightarrow}$ 3 $\textcolor{cai_primary}{\rightarrow}$ 4 $\textcolor{cai_primary}{\rightarrow}$ 5 & 0.240434 \\
        2--14 & All other paths & 0.000000 \\
        \hline
        \end{tabular}
        \caption*{(c) Attacker Strategy}
    \end{minipage}

    \vspace{0.5cm}

    \begin{minipage}{\textwidth}
        \centering
        \colorbox{cai_light}{%
            \parbox{0.85\textwidth}{%
                \centering
                \vspace{1.2ex}
                \textcolor{cai_dark}{\textbf{Game Equilibrium}}\\[0.5ex]
                \textcolor{cai_primary}{Defender can keep attacker success below:} \textbf{0.027127}\\[0.3ex]
                \textcolor{cai_primary}{Attacker can guarantee success probability of:} \textbf{0.027127}
                \vspace{1.2ex}
            }%
        }
        \caption*{(d) Game Equilibrium}
    \end{minipage}
    \caption{Attack Graph and Equilibrium analysis for \texttt{gpt-4o}}
    \label{fig:pornbox_gpt}
\end{figure}

\begin{figure}[H]
    \centering
    \includegraphics[width=\textwidth]{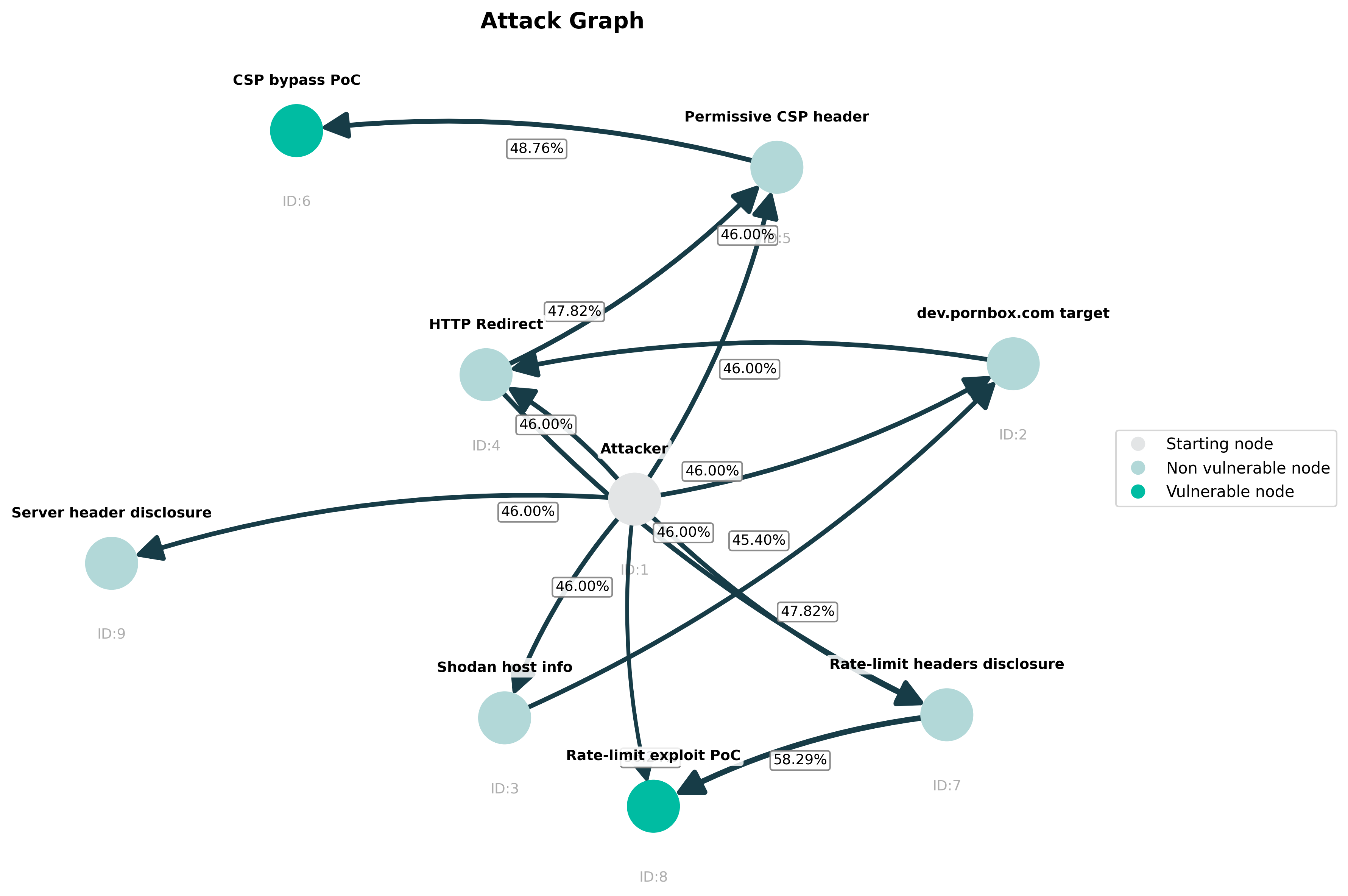}
    \caption*{(a) Attack Graph Example}

    \vspace{0.5cm}
\begin{minipage}{0.45\textwidth}
    \centering
    \renewcommand{\arraystretch}{1.4}
    \begin{tabular}{cc}
    \hline
     \textcolor{cai_primary}{\textbf{Node ID}} & \textcolor{cai_primary}{\textbf{Probability}} \\
    \hline
    7 & 0.500000 \\
    5 & 0.500000 \\
    2 & 0.000000 \\
    3 & 0.000000 \\
    4 & 0.000000 \\
    \hline
    \end{tabular}
    \caption*{(b) Optimal Defense Strategy}
\end{minipage}
\hfill
\begin{minipage}{0.50\textwidth}
    \centering
    \renewcommand{\arraystretch}{1.4}
    \begin{tabular}{ccc}
    \hline
\textcolor{cai_primary}{\textbf{Path ID}} & \textcolor{cai_primary}{\textbf{Path Sequence}} & \textcolor{cai_primary}{\textbf{Probability}} \\
    \hline
    3 & 1 $\textcolor{cai_primary}{\rightarrow}$ 3 $\textcolor{cai_primary}{\rightarrow}$ 2 $\textcolor{cai_primary}{\rightarrow}$ 4 $\textcolor{cai_primary}{\rightarrow}$ 5 $\textcolor{cai_primary}{\rightarrow}$ 6  & 0.500000 \\
    4 & 1 $\textcolor{cai_primary}{\rightarrow}$ 3 $\textcolor{cai_primary}{\rightarrow}$ 2 $\textcolor{cai_primary}{\rightarrow}$ 4 $\textcolor{cai_primary}{\rightarrow}$ 7 $\textcolor{cai_primary}{\rightarrow}$ 8  & 0.500000 \\
    1 & 1 $\textcolor{cai_primary}{\rightarrow}$ 2 $\textcolor{cai_primary}{\rightarrow}$ 4 $\textcolor{cai_primary}{\rightarrow}$ 5 $\textcolor{cai_primary}{\rightarrow}$ 6 & 0.000000 \\
    2 & 1 $\textcolor{cai_primary}{\rightarrow}$ 2 $\textcolor{cai_primary}{\rightarrow}$ 4 $\textcolor{cai_primary}{\rightarrow}$ 7 $\textcolor{cai_primary}{\rightarrow}$ 8  & 0.000000 \\
    5 & 1 $\textcolor{cai_primary}{\rightarrow}$ 4 $\textcolor{cai_primary}{\rightarrow}$ 5 $\textcolor{cai_primary}{\rightarrow}$ 6  & 0.000000 \\
    6 & 1 $\textcolor{cai_primary}{\rightarrow}$ 4 $\textcolor{cai_primary}{\rightarrow}$ 7 $\textcolor{cai_primary}{\rightarrow}$ 8  & 0.000000 \\
    7 & 1 $\textcolor{cai_primary}{\rightarrow}$ 5 $\textcolor{cai_primary}{\rightarrow}$ 6  & 0.000000 \\
    8 & 1 $\textcolor{cai_primary}{\rightarrow}$ 7 $\textcolor{cai_primary}{\rightarrow}$ 8 & 0.000000 \\
    9 & 1 $\textcolor{cai_primary}{\rightarrow}$ 8  & 0.000000 \\
    \hline
    \end{tabular}
    \caption*{(c) Attacker Strategy}
\end{minipage}

\vspace{0.5cm}

\begin{minipage}{\textwidth}
    \centering
    \colorbox{cai_light}{%
        \parbox{0.85\textwidth}{%
            \centering
            \vspace{1.2ex}
            \textcolor{cai_dark}{\textbf{Game Equilibrium}}\\[0.5ex]
            \textcolor{cai_primary}{Defender can keep attacker success below:} \textbf{0.025766}\\[0.3ex]
            \textcolor{cai_primary}{Attacker can guarantee success probability of:} \textbf{0.025766}
            \vspace{1.2ex}
        }%
    }
    \caption*{(d) Game Equilibrium}
\end{minipage}

    \caption{Attack Graph and Equilibrium analysis for \texttt{o3}}
    \label{fig:pornbox_o3}
\end{figure}

\begin{figure}[H]
    \centering
    \includegraphics[width=\textwidth]{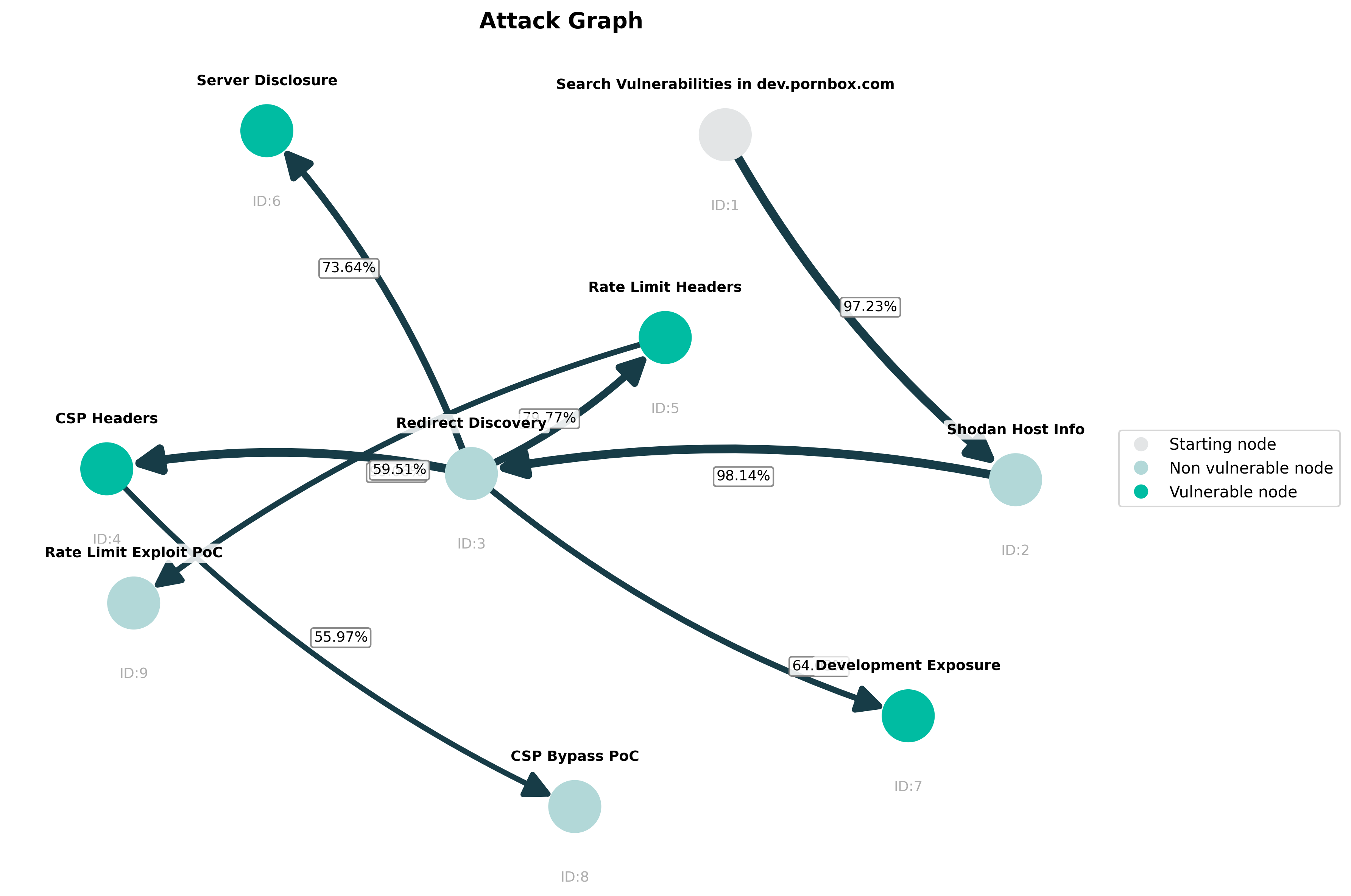}
    \caption*{(a) Attack Graph Example}

    \vspace{0.5cm}
    \begin{minipage}{0.45\textwidth}
    \centering
    \renewcommand{\arraystretch}{1.4}
    \begin{tabular}{cc}
    \hline
     \textcolor{cai_primary}{\textbf{Node ID}} & \textcolor{cai_primary}{\textbf{Probability}} \\
    \hline
    3 & 1.000000 \\
    2 & 0.000000 \\
    \hline
    \end{tabular}
    \caption*{(b) Optimal Defense Strategy}
\end{minipage}
\hfill
\begin{minipage}{0.52\textwidth}
    \centering
    \renewcommand{\arraystretch}{1.4}
    \begin{tabular}{ccc}
    \hline
\textcolor{cai_primary}{\textbf{Path ID}} & \textcolor{cai_primary}{\textbf{Path Sequence}} & \textcolor{cai_primary}{\textbf{Probability}} \\
    \hline
    1 & 1 $\textcolor{cai_primary}{\rightarrow}$ 2 $\textcolor{cai_primary}{\rightarrow}$ 3 $\textcolor{cai_primary}{\rightarrow}$ 4  & 1.000000 \\
    2 & 1 $\textcolor{cai_primary}{\rightarrow}$ 2 $\textcolor{cai_primary}{\rightarrow}$ 3 $\textcolor{cai_primary}{\rightarrow}$ 5   & 0.000000 \\
    3 & 1 $\textcolor{cai_primary}{\rightarrow}$ 2 $\textcolor{cai_primary}{\rightarrow}$ 3 $\textcolor{cai_primary}{\rightarrow}$ 6   & 0.000000 \\
    4 & 1 $\textcolor{cai_primary}{\rightarrow}$ 2 $\textcolor{cai_primary}{\rightarrow}$ 3 $\textcolor{cai_primary}{\rightarrow}$ 7   & 0.000000 \\
    \hline
    \end{tabular}
    \caption*{(c) Attacker Strategy}
\end{minipage}

\vspace{0.5cm}

\begin{minipage}{\textwidth}
    \centering
    \colorbox{cai_light}{%
        \parbox{0.85\textwidth}{%
            \centering
            \vspace{1.2ex}
            \textcolor{cai_dark}{\textbf{Game Equilibrium}}\\[0.5ex]
            \textcolor{cai_primary}{Defender can keep attacker success below:} \textbf{0.000000}\\[0.3ex]
            \textcolor{cai_primary}{Attacker can guarantee success probability of:} \textbf{0.000000}
            \vspace{1.2ex}
        }%
    }
    \caption*{(d) Game Equilibrium}
\end{minipage}
    \caption{Attack Graph and Equilibrium analysis for \texttt{grok--4}}
    \label{fig:pornbox_gork}
\end{figure}

\begin{figure}[H]
    \centering
    \includegraphics[width=\textwidth]{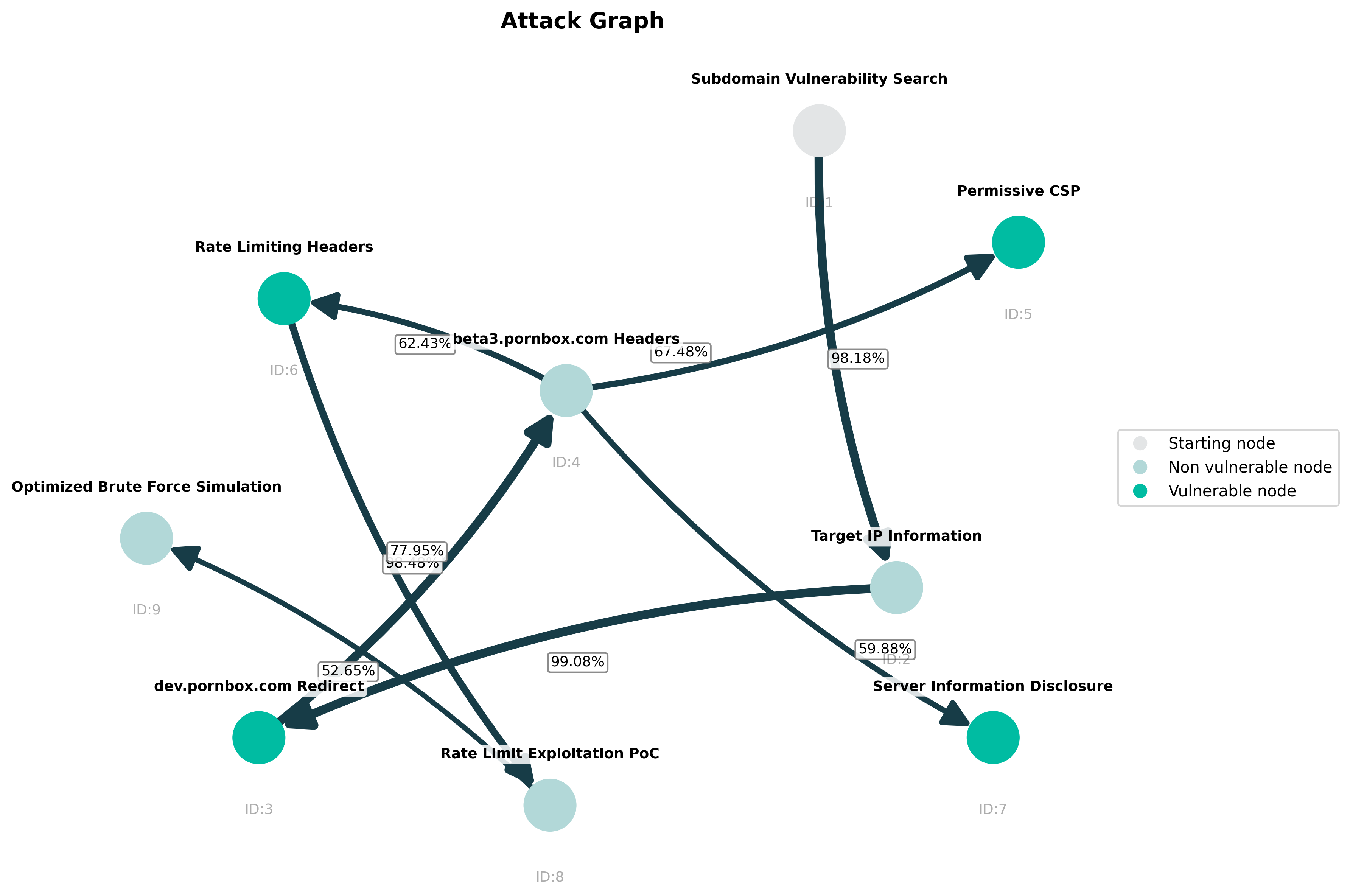}
    \caption*{(a) Attack Graph Example}

    \vspace{0.5cm}
    \begin{minipage}{0.45\textwidth}
    \centering
    \renewcommand{\arraystretch}{1.4}
    \begin{tabular}{cc}
    \hline
     \textcolor{cai_primary}{\textbf{Node ID}} & \textcolor{cai_primary}{\textbf{Probability}} \\
    \hline
    2 & 0.500000 \\
    4 & 0.500000 \\
    \hline
    \end{tabular}
    \caption*{(b) Optimal Defense Strategy}
\end{minipage}
\hfill
\begin{minipage}{0.52\textwidth}
    \centering
    \renewcommand{\arraystretch}{1.4}
    \begin{tabular}{ccc}
    \hline
\textcolor{cai_primary}{\textbf{Path ID}} & \textcolor{cai_primary}{\textbf{Path Sequence}} & \textcolor{cai_primary}{\textbf{Probability}} \\
    \hline
    2 & 1 $\textcolor{cai_primary}{\rightarrow}$ 2 $\textcolor{cai_primary}{\rightarrow}$ 3 $\textcolor{cai_primary}{\rightarrow}$ 4 $\textcolor{cai_primary}{\rightarrow}$ 5   & 0.500000 \\
    1 & 1 $\textcolor{cai_primary}{\rightarrow}$ 2 $\textcolor{cai_primary}{\rightarrow}$ 3   & 0.500000 \\
    3 & 1 $\textcolor{cai_primary}{\rightarrow}$ 2 $\textcolor{cai_primary}{\rightarrow}$ 3 $\textcolor{cai_primary}{\rightarrow}$ 4 $\textcolor{cai_primary}{\rightarrow}$ 6   & 0.000000 \\
    4 & 1 $\textcolor{cai_primary}{\rightarrow}$ 2 $\textcolor{cai_primary}{\rightarrow}$ 3 $\textcolor{cai_primary}{\rightarrow}$ 4 $\textcolor{cai_primary}{\rightarrow}$ 7   & 0.000000 \\
    \hline
    \end{tabular}
    \caption*{(c) Attacker Strategy}
\end{minipage}

\vspace{0.5cm}

\begin{minipage}{\textwidth}
    \centering
    \colorbox{cai_light}{%
        \parbox{0.85\textwidth}{%
            \centering
            \vspace{1.2ex}
            \textcolor{cai_dark}{\textbf{Game Equilibrium}}\\[0.5ex]
            \textcolor{cai_primary}{Defender can keep attacker success below:} \textbf{0.035714}\\[0.3ex]
            \textcolor{cai_primary}{Attacker can guarantee success probability of:} \textbf{0.035714}
            \vspace{1.2ex}
        }%
    }
    \caption*{(d) Game Equilibrium}
\end{minipage}
    \caption{Attack Graph and Equilibrium analysis for \texttt{claude--sonnet--4}}
    \label{fig:pornbox_claude}
\end{figure}

\begin{figure}[H]
    \centering
    \includegraphics[width=\textwidth]{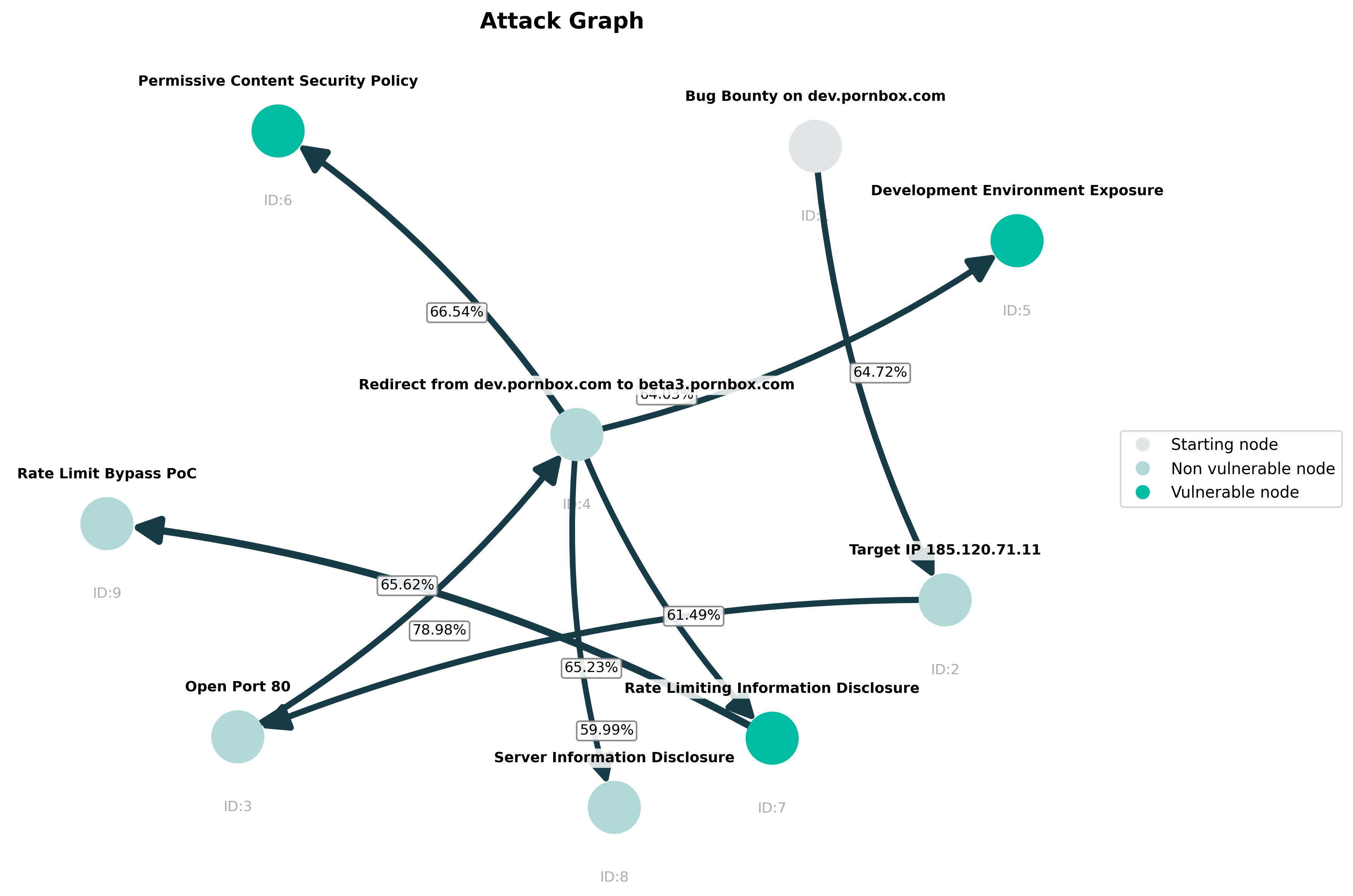}
    \caption*{(a) Attack Graph Example}

    \vspace{0.5cm}
    \begin{minipage}{0.45\textwidth}
        \centering
        \renewcommand{\arraystretch}{1.4}
        \begin{tabular}{cc}
        \hline
        \textcolor{cai_primary}{\textbf{Node ID}} & \textcolor{cai_primary}{\textbf{Probability}} \\
        \hline
        4 & 1.000000 \\
        2 & 0.000000 \\
        3 & 0.000000 \\
        \hline
        \end{tabular}
        \caption*{(b) Optimal Defense Strategy}
    \end{minipage}
    \hfill
    \begin{minipage}{0.52\textwidth}
        \centering
        \renewcommand{\arraystretch}{1.4}
        \begin{tabular}{ccc}
        \hline
        \textcolor{cai_primary}{\textbf{Path ID}} & \textcolor{cai_primary}{\textbf{Path Sequence}} & \textcolor{cai_primary}{\textbf{Probability}} \\
        \hline
        1 & 1 $\textcolor{cai_primary}{\rightarrow}$ 2 $\textcolor{cai_primary}{\rightarrow}$ 3 $\textcolor{cai_primary}{\rightarrow}$ 4 $\textcolor{cai_primary}{\rightarrow}$ 5 & 1.000000 \\
        2 & 1 $\textcolor{cai_primary}{\rightarrow}$ 2 $\textcolor{cai_primary}{\rightarrow}$ 3 $\textcolor{cai_primary}{\rightarrow}$ 4 $\textcolor{cai_primary}{\rightarrow}$ 6 & 0.000000 \\
        3 & 1 $\textcolor{cai_primary}{\rightarrow}$ 2 $\textcolor{cai_primary}{\rightarrow}$ 3 $\textcolor{cai_primary}{\rightarrow}$ 4 $\textcolor{cai_primary}{\rightarrow}$ 7 & 0.000000 \\
        \hline
        \end{tabular}
        \caption*{(c) Attacker Strategy}
    \end{minipage}

    \vspace{0.5cm}

    \begin{minipage}{\textwidth}
        \centering
        \colorbox{cai_light}{%
            \parbox{0.85\textwidth}{%
                \centering
                \vspace{1.2ex}
                \textcolor{cai_dark}{\textbf{Game Equilibrium}}\\[0.5ex]
                \textcolor{cai_primary}{Defender can keep attacker success below:} \textbf{0.000000}\\[0.3ex]
                \textcolor{cai_primary}{Attacker can guarantee success probability of:} \textbf{0.000000}
                \vspace{1.2ex}
            }%
        }
        \caption*{(d) Game Equilibrium}
    \end{minipage}
    \caption{Attack Graph and Equilibrium analysis for \texttt{alias0}}
    \label{fig:pornbox_alias0}
\end{figure}

\section{Appendix 4: hm.com}
\label{appendix_hm}

\begin{figure}[H]
    \centering
    \includegraphics[width=\textwidth]{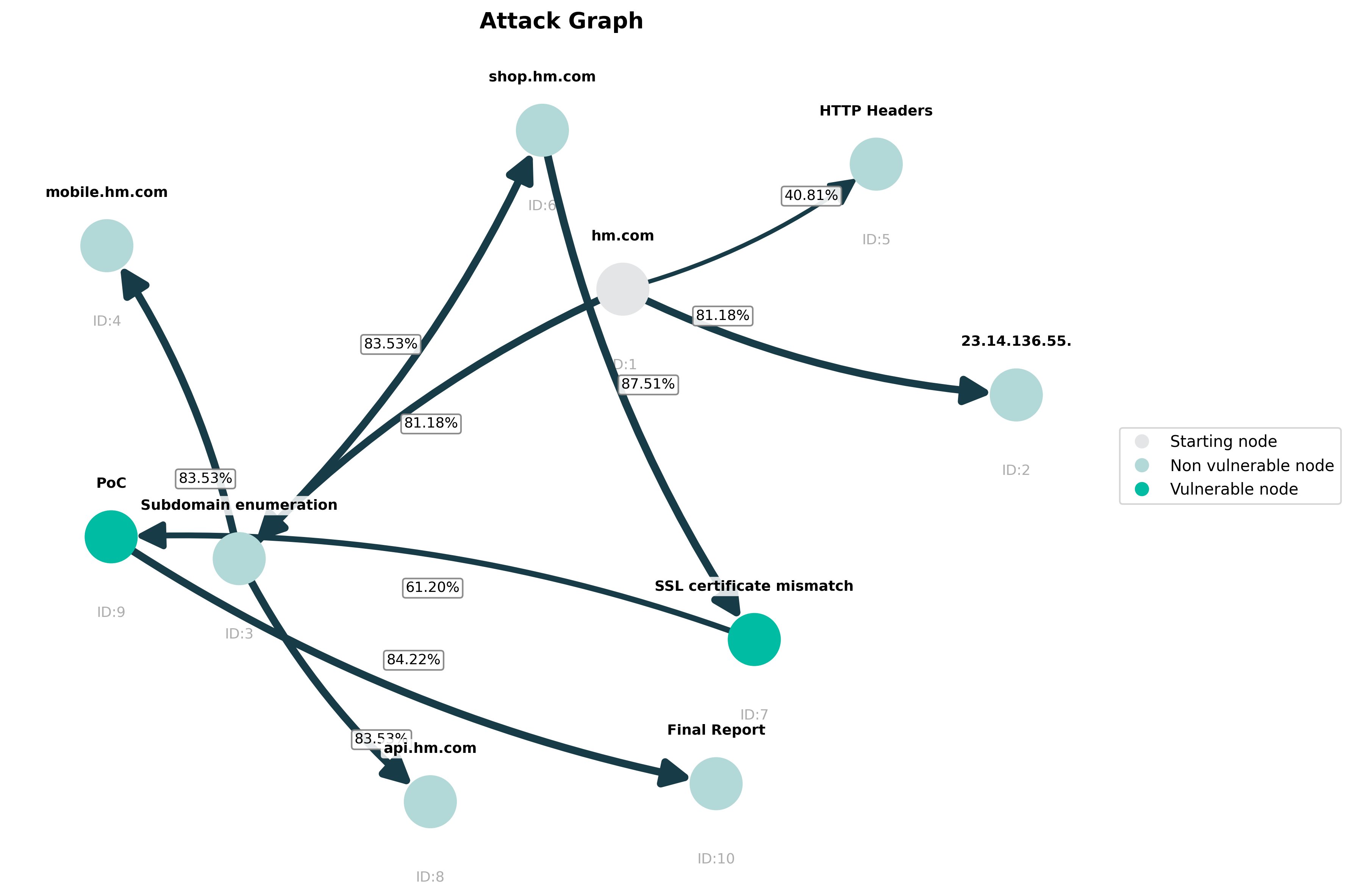}
    \caption*{(a) Human Attack Graph}

    \vspace{0.5cm}

    \begin{minipage}{0.45\textwidth}
        \centering
\renewcommand{\arraystretch}{1.4}
        \begin{tabular}{cc}
        \hline
         \textcolor{cai_primary}{\textbf{Node ID}} & \textcolor{cai_primary}{\textbf{Probability}} \\
        \hline
        6 & 1.000000 \\
        3 & 0.000000 \\
        \hline
        \end{tabular}
        \caption*{(b) Optimal Defense Strategy}
    \end{minipage}
    \hfill
    \begin{minipage}{0.50\textwidth}
        \centering
        \renewcommand{\arraystretch}{1.4}
        \begin{tabular}{ccc}
        \hline
\textcolor{cai_primary}{\textbf{Path ID}} & \textcolor{cai_primary}{\textbf{Path Sequence}} & \textcolor{cai_primary}{\textbf{Probability}} \\
        \hline
        1 & 1 $\textcolor{cai_primary}{\rightarrow}$ 3 $\textcolor{cai_primary}{\rightarrow}$ 6 $\textcolor{cai_primary}{\rightarrow}$ 7 & 1.000000 \\
        2 & 1 $\textcolor{cai_primary}{\rightarrow}$ 3 $\textcolor{cai_primary}{\rightarrow}$ 6 $\textcolor{cai_primary}{\rightarrow}$ 7 $\textcolor{cai_primary}{\rightarrow}$ 9 & 0.000000 \\
        \hline
        \end{tabular}
        \caption*{(c) Attacker Strategy}
    \end{minipage}
    
    \vspace{0.5cm}
    
    \begin{minipage}{\textwidth}
        \centering
        \colorbox{cai_light}{%
            \parbox{0.85\textwidth}{%
                \centering
                \vspace{1.2ex}
                \textcolor{cai_dark}{\textbf{Game Equilibrium}}\\[0.5ex]
                \textcolor{cai_primary}{Defender can keep attacker success below:} \textbf{0.000000}\\[0.3ex]
                \textcolor{cai_primary}{Attacker can guarantee success probability of:} \textbf{0.000000}
                \vspace{1.2ex}
            }%
        }
        \caption*{(d) Game Equilibrium}
    \end{minipage}
    \caption{\textbf{Ground-Truth attack graph annotated by Humans in the hm.com doamin}: This manually created graph outlines the step-by-step process followed by a human  during a real-world bug bounty assessment of the hm.com domain and its subdomains. The engagement began with subdomain enumeration and IP resolution, followed by inspection of HTTP headers and SSL configurations across endpoints such as mobile.hm.com, shop.hm.com, and api.hm.com. While the latter was protected by Akamai CDN, the researcher discovered an SSL certificate mismatch vulnerability on shop.hm.com, posing a potential risk to user trust and secure communication. This finding was validated and supported with a custom proof-of-concept (PoC) script. The flow concludes with a final report summarizing the issue and providing remediation guidance.}
    \label{fig:hm_human}
\end{figure}

\begin{figure}[H]
    \centering
    \includegraphics[width=\textwidth]{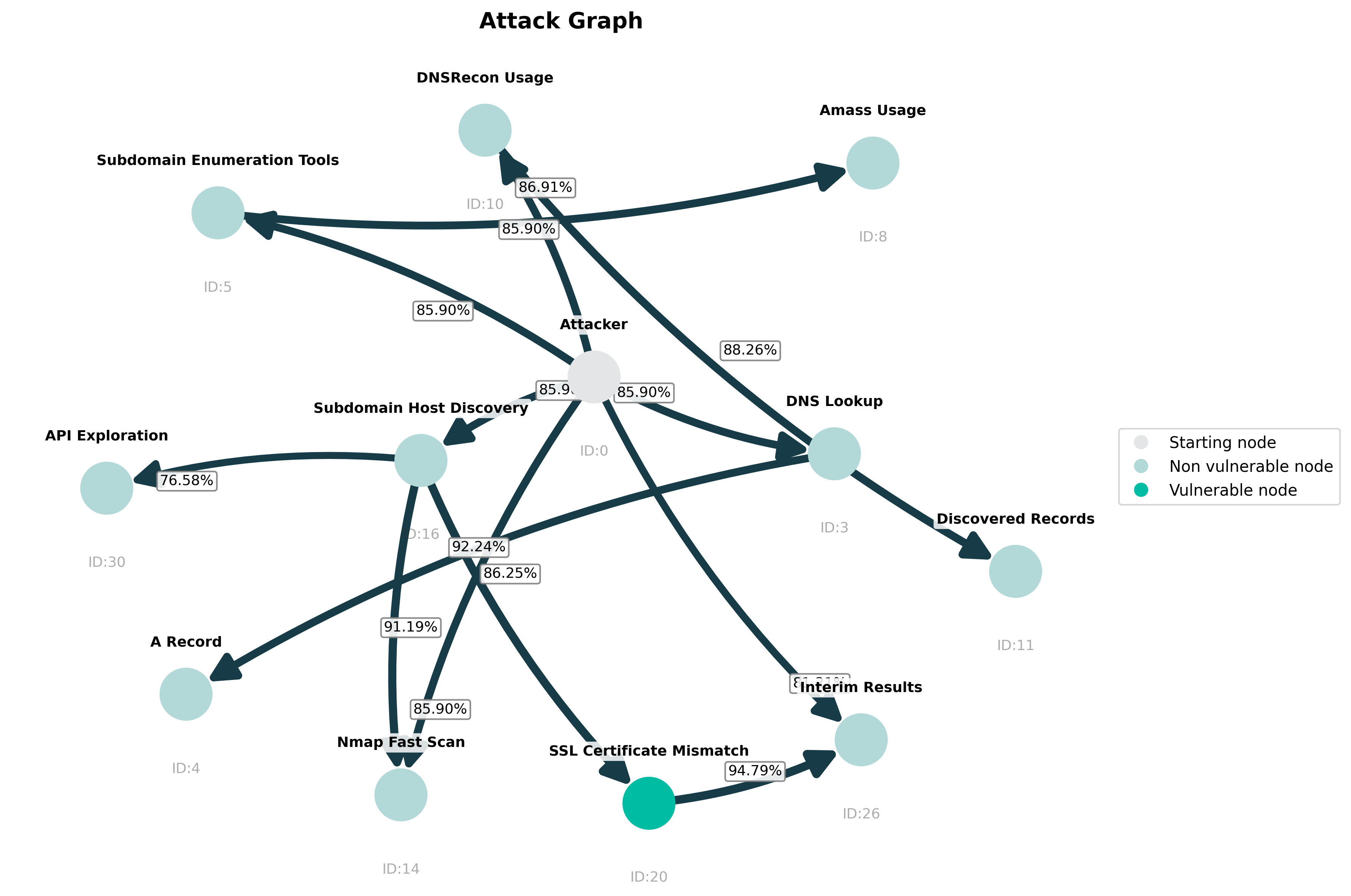}
    \caption*{(a) Attack Graph Example}

    \vspace{0.5cm}

    \begin{minipage}{0.45\textwidth}
        \centering
        \renewcommand{\arraystretch}{1.4}
        \begin{tabular}{cc}
        \hline
         \textcolor{cai_primary}{\textbf{Node ID}} & \textcolor{cai_primary}{\textbf{Probability}} \\
        \hline
        16 & 1.000000 \\
        \hline
        \end{tabular}
        \caption*{(b) Optimal Defense Strategy}
    \end{minipage}
    \hfill
    \begin{minipage}{0.50\textwidth}
        \centering
        \renewcommand{\arraystretch}{1.4}
        \begin{tabular}{ccc}
        \hline
\textcolor{cai_primary}{\textbf{Path ID}} & \textcolor{cai_primary}{\textbf{Path Sequence}} & \textcolor{cai_primary}{\textbf{Probability}} \\
        \hline
        1 & 0 $\textcolor{cai_primary}{\rightarrow}$ 16 $\textcolor{cai_primary}{\rightarrow}$ 20  & 1.000000 \\
        \hline
        \end{tabular}
        \caption*{(c) Attacker Strategy}
    \end{minipage}
    
    \vspace{0.5cm}
    
    \begin{minipage}{\textwidth}
        \centering
        \colorbox{cai_light}{%
            \parbox{0.85\textwidth}{%
                \centering
                \vspace{1.2ex}
                \textcolor{cai_dark}{\textbf{Game Equilibrium}}\\[0.5ex]
                \textcolor{cai_primary}{Defender can keep attacker success below:} \textbf{0.000000}\\[0.3ex]
                \textcolor{cai_primary}{Attacker can guarantee success probability of:} \textbf{0.000000}
                \vspace{1.2ex}
            }%
        }
        \caption*{(d) Game Equilibrium}
    \end{minipage}
    \caption{Attack Graph and Equilibrium analysis for \texttt{gpt-4o}.}
    \label{fig:hm_gpt4o}
\end{figure}

\begin{figure}[H]
    \centering
    \includegraphics[width=\textwidth]{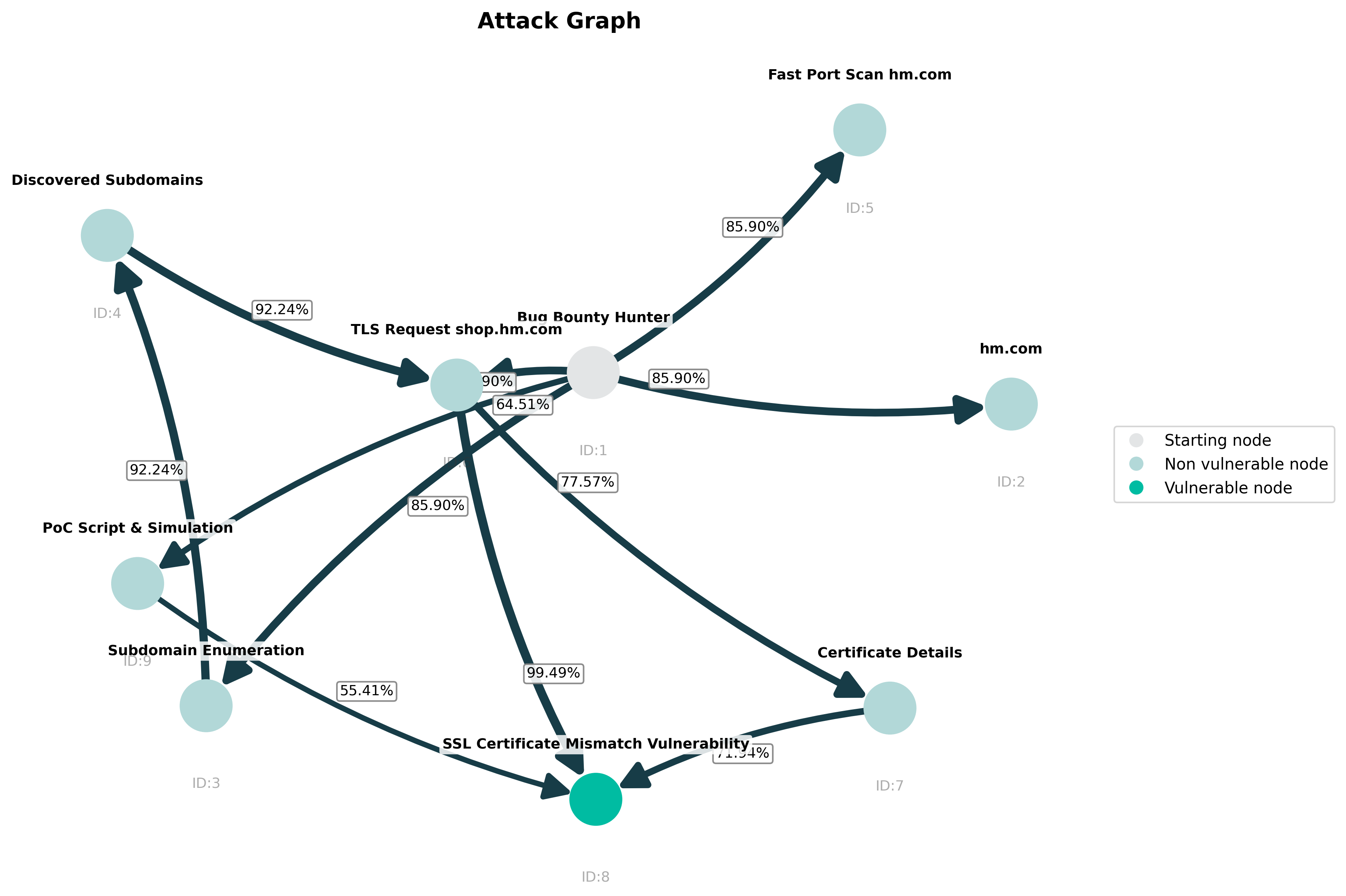}
    \caption*{(a) Attack Graph Example}

    \vspace{0.5cm}

    \begin{minipage}{0.45\textwidth}
        \centering
        \renewcommand{\arraystretch}{1.4}
        \begin{tabular}{cc}
        \hline
         \textcolor{cai_primary}{\textbf{Node ID}} & \textcolor{cai_primary}{\textbf{Probability}} \\
        \hline
        6 & 0.511996 \\
        7 & 0.313940 \\
        9 & 0.174064 \\
        3 & 0.000000 \\
        4 & 0.000000 \\
        \hline
        \end{tabular}
        \caption*{(b) Optimal Defense Strategy}
    \end{minipage}
    \hfill
    \begin{minipage}{0.50\textwidth}
        \centering
        \renewcommand{\arraystretch}{1.4}
        \begin{tabular}{ccc}
        \hline
\textcolor{cai_primary}{\textbf{Node ID}} & \textcolor{cai_primary}{\textbf{Probability}} \\
        \hline
        5 & 1 $\textcolor{cai_primary}{\rightarrow}$ 9 $\textcolor{cai_primary}{\rightarrow}$ 8 & 0.531336 \\
        1 & 1 $\textcolor{cai_primary}{\rightarrow}$ 3 $\textcolor{cai_primary}{\rightarrow}$ 4 $\textcolor{cai_primary}{\rightarrow}$ 6 $\textcolor{cai_primary}{\rightarrow}$ 7 $\textcolor{cai_primary}{\rightarrow}$ 8  & 0.294600 \\
        2 & 1 $\textcolor{cai_primary}{\rightarrow}$ 3 $\textcolor{cai_primary}{\rightarrow}$ 4 $\textcolor{cai_primary}{\rightarrow}$ 6 $\textcolor{cai_primary}{\rightarrow}$ 8  & 0.174064 \\
        3 & 1 $\textcolor{cai_primary}{\rightarrow}$ 6 $\textcolor{cai_primary}{\rightarrow}$ 7 $\textcolor{cai_primary}{\rightarrow}$ 8  & 0.000000 \\
        4 & 1 $\textcolor{cai_primary}{\rightarrow}$ 6 $\textcolor{cai_primary}{\rightarrow}$ 8  & 0.000000 \\
        \hline
        \end{tabular}
        \caption*{(c) Attacker Strategy}
    \end{minipage}
    
    \vspace{0.5cm}
    
    \begin{minipage}{\textwidth}
        \centering
        \colorbox{cai_light}{%
            \parbox{0.85\textwidth}{%
                \centering
                \vspace{1.2ex}
                \textcolor{cai_dark}{\textbf{Game Equilibrium}}\\[0.5ex]
                \textcolor{cai_primary}{Defender can keep attacker success below:} \textbf{0.023598}\\[0.3ex]
                \textcolor{cai_primary}{Attacker can guarantee success probability of:} \textbf{0.023598}
                \vspace{1.2ex}
            }%
        }
        \caption*{(d) Game Equilibrium}
    \end{minipage}
    \caption{Attack Graph and Equilibrium analysis for \texttt{o3}.}
    \label{fig:hm_o3}
\end{figure}

\begin{figure}[H]
    \centering
    \includegraphics[width=\textwidth]{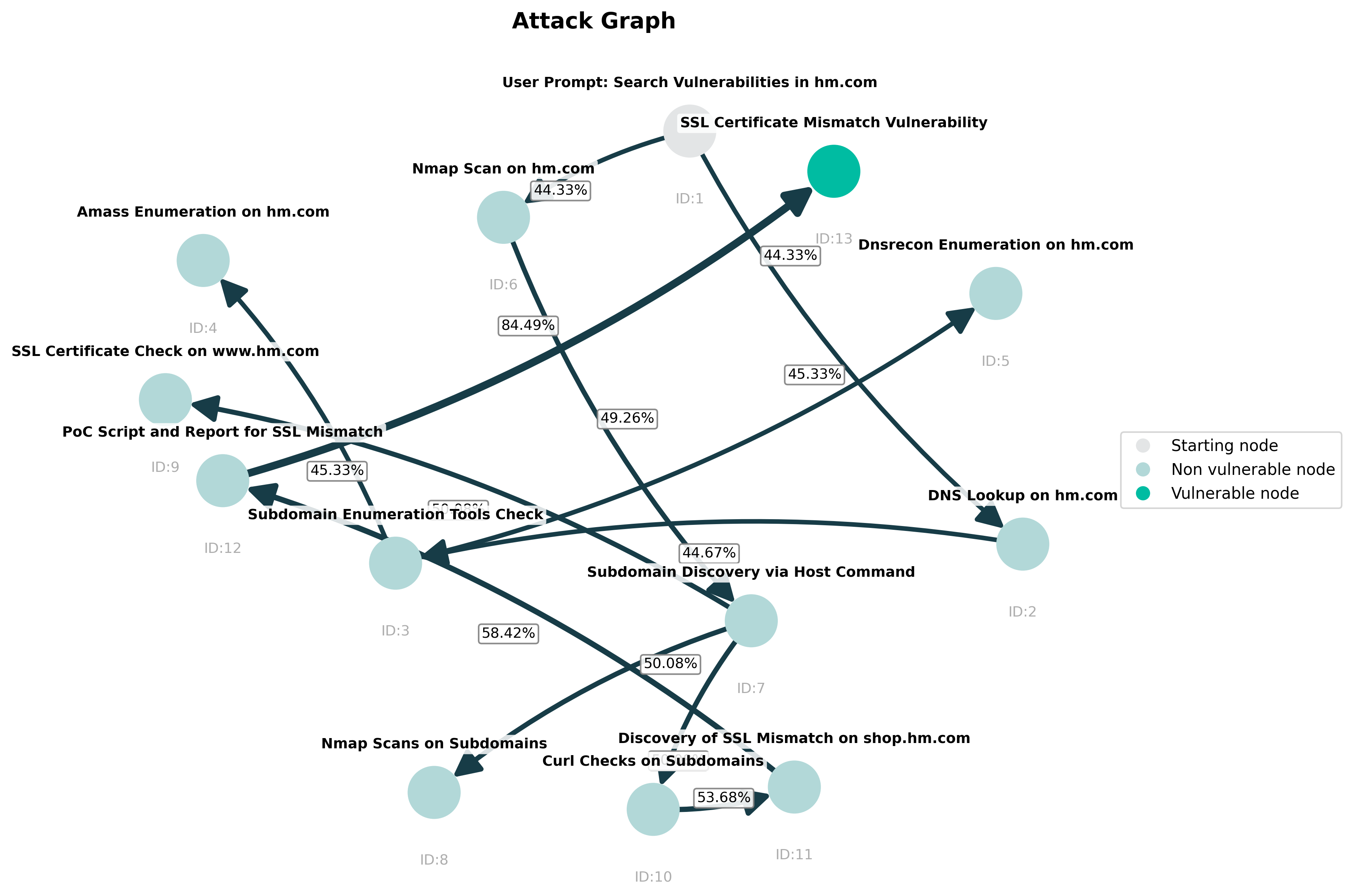}
    \caption*{(a) Attack Graph Example}

    \vspace{0.5cm}

    \begin{minipage}{\textwidth}
        \centering
        \renewcommand{\arraystretch}{1.4}
        \begin{tabular}{cc}
        \hline
         \textcolor{cai_primary}{\textbf{Node ID}} & \textcolor{cai_primary}{\textbf{Probability}} \\
        \hline
        12 & 1.000000 \\
        10 & 0.000000 \\
        11 & 0.000000 \\
        6 & 0.000000 \\
        7 & 0.000000 \\
        \hline
        \end{tabular}
        \caption*{(b) Optimal Defense Strategy}
    \end{minipage}
     
     \vspace{0.5cm}
     
    \begin{minipage}{\textwidth}
        \centering
        \renewcommand{\arraystretch}{1.4}
        \begin{tabular}{ccc}
        \hline
\textcolor{cai_primary}{\textbf{Path ID}} & \textcolor{cai_primary}{\textbf{Path Sequence}} & \textcolor{cai_primary}{\textbf{Probability}} \\
        \hline
        1 & 1 $\textcolor{cai_primary}{\rightarrow}$ 6 $\textcolor{cai_primary}{\rightarrow}$ 7 $\textcolor{cai_primary}{\rightarrow}$ 10 $\textcolor{cai_primary}{\rightarrow}$ 11 $\textcolor{cai_primary}{\rightarrow}$ 12 $\textcolor{cai_primary}{\rightarrow}$ 13  & 1.000000 \\
        \hline
        \end{tabular}
        \caption*{(c) Attacker Strategy}
    \end{minipage}
    
    \vspace{0.5cm}
    
    \begin{minipage}{\textwidth}
        \centering
        \colorbox{cai_light}{%
            \parbox{0.85\textwidth}{%
                \centering
                \vspace{1.2ex}
                \textcolor{cai_dark}{\textbf{Game Equilibrium}}\\[0.5ex]
                \textcolor{cai_primary}{Defender can keep attacker success below:} \textbf{0.000000}\\[0.3ex]
                \textcolor{cai_primary}{Attacker can guarantee success probability of:} \textbf{0.000000}
                \vspace{1.2ex}
            }%
        }
        \caption*{(d) Game Equilibrium}
    \end{minipage}
    \caption{Attack Graph and Equilibrium analysis for \texttt{grok--4}.}
    \label{fig:hm_grok}
\end{figure}

\begin{figure}[H]
    \centering
    \includegraphics[width=\textwidth]{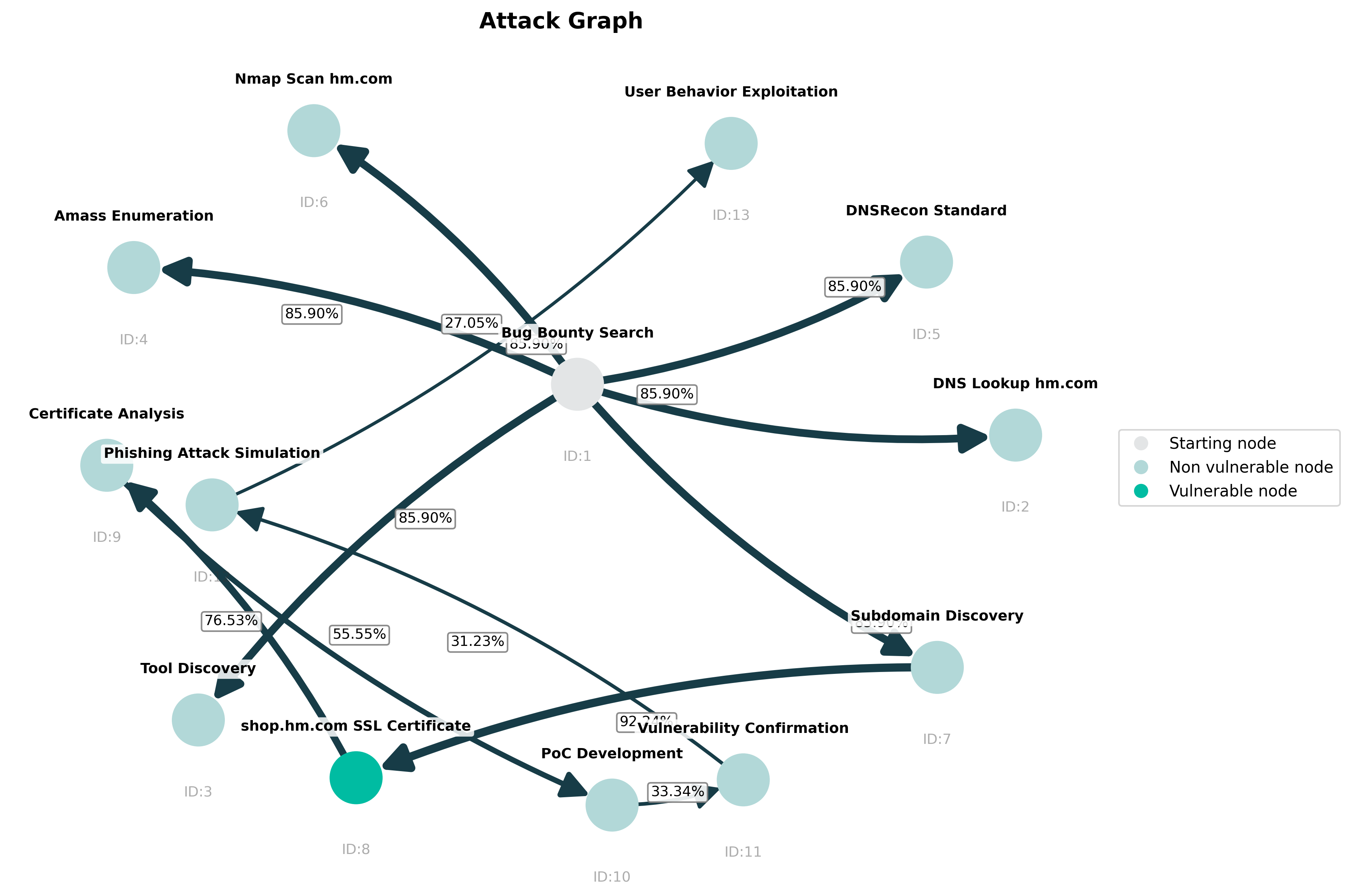}
    \caption*{(a) Attack Graph Example}

    \vspace{0.5cm}

    \begin{minipage}{0.45\textwidth}
        \centering
        \renewcommand{\arraystretch}{1.4}
        \begin{tabular}{cc}
        \hline
         \textcolor{cai_primary}{\textbf{Node ID}} & \textcolor{cai_primary}{\textbf{Probability}} \\
        \hline
        7 & 1.000000 \\
        \hline
        \end{tabular}
        \caption*{(b) Optimal Defense Strategy}
    \end{minipage}
    \hfill
    \begin{minipage}{0.50\textwidth}
        \centering
        \renewcommand{\arraystretch}{1.4}
        \begin{tabular}{ccc}
        \hline
\textcolor{cai_primary}{\textbf{Path ID}} & \textcolor{cai_primary}{\textbf{Path Sequence}} & \textcolor{cai_primary}{\textbf{Probability}} \\
        \hline
        1 & 1 $\textcolor{cai_primary}{\rightarrow}$ 7 $\textcolor{cai_primary}{\rightarrow}$ 8 & 1.000000 \\
        \hline
        \end{tabular}
        \caption*{(c) Attacker Strategy}
    \end{minipage}
    
    \vspace{0.5cm}
    
    \begin{minipage}{\textwidth}
        \centering
        \colorbox{cai_light}{%
            \parbox{0.85\textwidth}{%
                \centering
                \vspace{1.2ex}
                \textcolor{cai_dark}{\textbf{Game Equilibrium}}\\[0.5ex]
                \textcolor{cai_primary}{Defender can keep attacker success below:} \textbf{0.000000}\\[0.3ex]
                \textcolor{cai_primary}{Attacker can guarantee success probability of:} \textbf{0.000000}
                \vspace{1.2ex}
            }%
        }
        \caption*{(d) Game Equilibrium}
    \end{minipage}
    \caption{Attack Graph and Equilibrium analysis for \texttt{claude-sonnet-4}.}
    \label{fig:hm_claude}
\end{figure}

\begin{figure}[H]
    \centering
    \includegraphics[width=\textwidth]{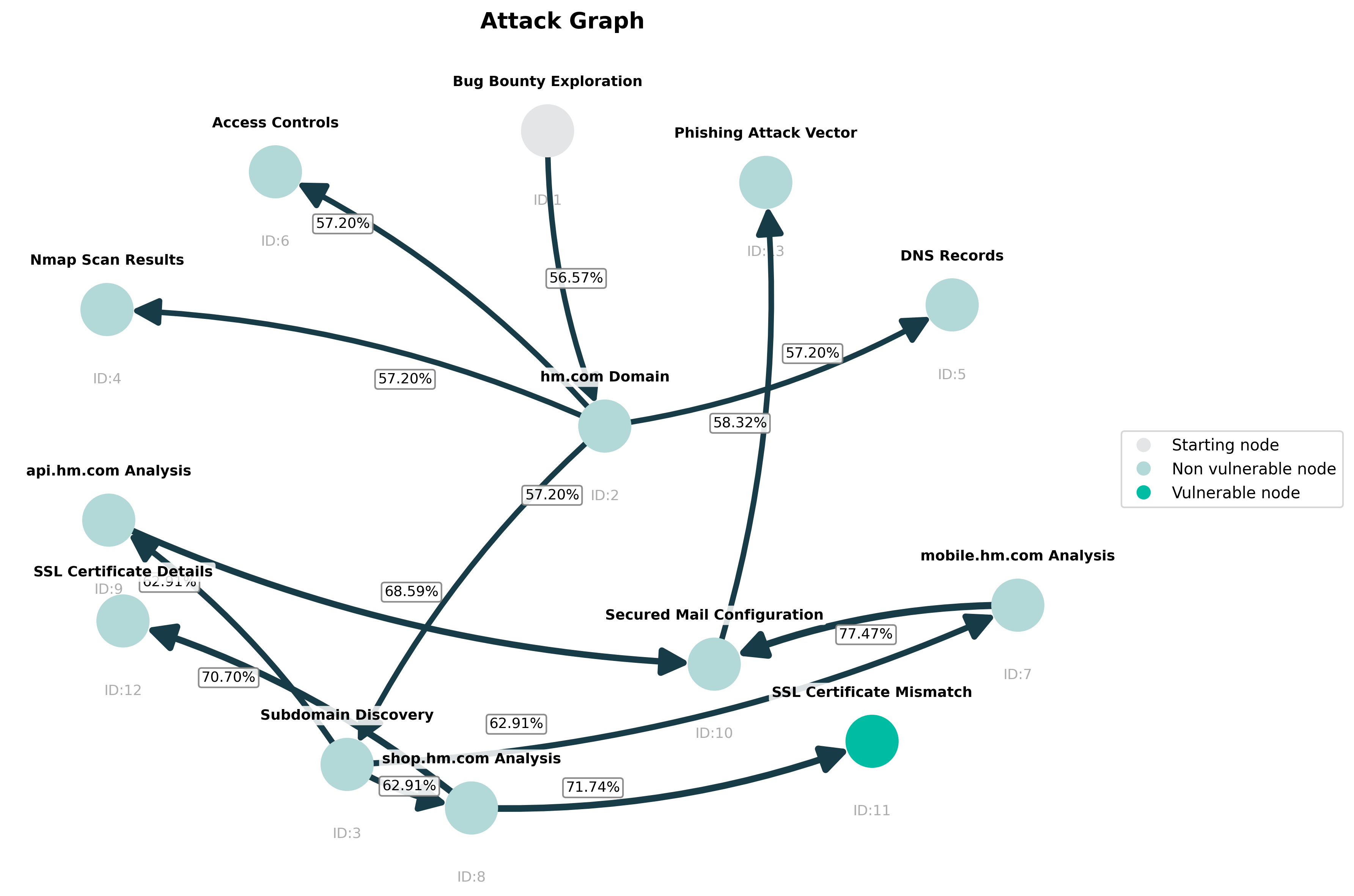}
    \caption*{(a) Attack Graph Example}

    \vspace{0.5cm}

    \begin{minipage}{0.45\textwidth}
        \centering
        \renewcommand{\arraystretch}{1.4}
        \begin{tabular}{cc}
        \hline
        \textcolor{cai_primary}{\textbf{Node ID}} & \textcolor{cai_primary}{\textbf{Probability}} \\
        \hline
        8 & 1.000000 \\
        2 & 0.000000 \\
        3 & 0.000000 \\
        \hline
        \end{tabular}
        \caption*{(b) Optimal Defense Strategy}
    \end{minipage}
    \hfill
    \begin{minipage}{0.50\textwidth}
        \centering
        \renewcommand{\arraystretch}{1.4}
        \begin{tabular}{ccc}
        \hline
        \textcolor{cai_primary}{\textbf{Path ID}} & \textcolor{cai_primary}{\textbf{Path Sequence}} & \textcolor{cai_primary}{\textbf{Probability}} \\
        \hline
        1 & 1 $\textcolor{cai_primary}{\rightarrow}$ 2 $\textcolor{cai_primary}{\rightarrow}$ 3 $\textcolor{cai_primary}{\rightarrow}$ 8 $\textcolor{cai_primary}{\rightarrow}$ 11 & 1.000000 \\
        \hline
        \end{tabular}
        \caption*{(c) Attacker Strategy}
    \end{minipage}
    
    \vspace{0.5cm}
    
    \begin{minipage}{\textwidth}
        \centering
        \colorbox{cai_light}{%
            \parbox{0.85\textwidth}{%
                \centering
                \vspace{1.2ex}
                \textcolor{cai_dark}{\textbf{Game Equilibrium}}\\[0.5ex]
                \textcolor{cai_primary}{Defender can keep attacker success below:} \textbf{0.000000}\\[0.3ex]
                \textcolor{cai_primary}{Attacker can guarantee success probability of:} \textbf{0.000000}
                \vspace{1.2ex}
            }%
        }
        \caption*{(d) Game Equilibrium}
    \end{minipage}
    \caption{Attack Graph and Equilibrium analysis for \texttt{alias0}.}
    \label{fig:hm_alias0}
\end{figure}

\section{Appendix 5: media.guilded.gg}
\label{appendix_mediaguildedgg}

\begin{figure}[H]
    \centering
    \includegraphics[width=\textwidth]{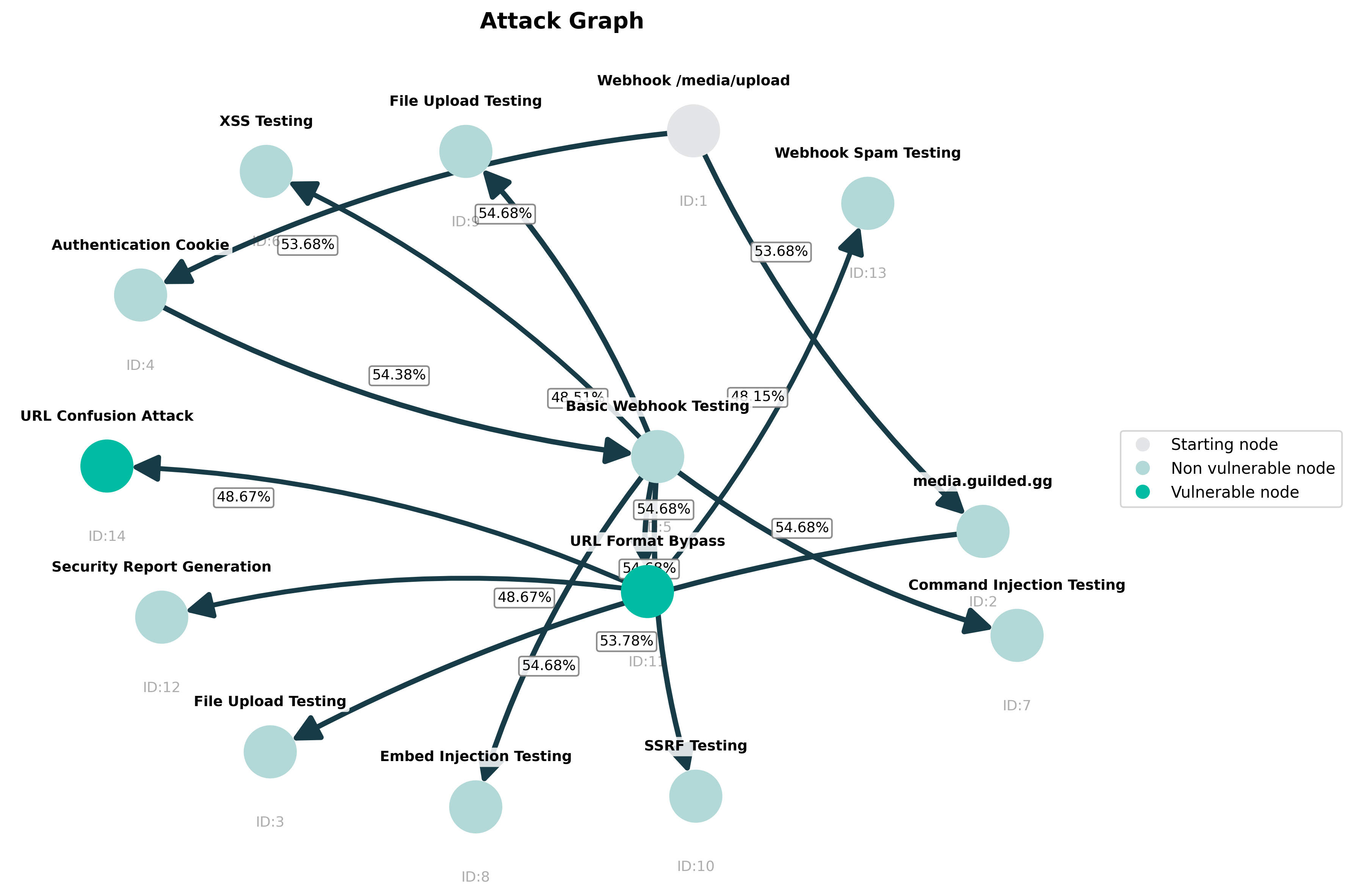}
    \caption*{(a) Human Attack Graph}

    \vspace{0.5cm}

    \begin{minipage}{0.45\textwidth}
        \centering
        \renewcommand{\arraystretch}{1.4}
        \begin{tabular}{cc}
        \hline
         \textcolor{cai_primary}{\textbf{Node ID}} & \textcolor{cai_primary}{\textbf{Probability}} \\
        \hline
        5 & 1.000000 \\
        4 & 0.000000 \\
        \hline
        \end{tabular}
        \caption*{(b) Optimal Defense Strategy}
    \end{minipage}
    \hfill
    \begin{minipage}{0.50\textwidth}
        \centering
        \renewcommand{\arraystretch}{1.4}
        \begin{tabular}{ccc}
        \hline
\textcolor{cai_primary}{\textbf{Path ID}} & \textcolor{cai_primary}{\textbf{Path Sequence}} & \textcolor{cai_primary}{\textbf{Probability}} \\
        \hline
        1 & 1 $\textcolor{cai_primary}{\rightarrow}$ 4 $\textcolor{cai_primary}{\rightarrow}$ 5 $\textcolor{cai_primary}{\rightarrow}$ 11 & 1.000000 \\
        2 & 1 $\textcolor{cai_primary}{\rightarrow}$ 4 $\textcolor{cai_primary}{\rightarrow}$ 5 $\textcolor{cai_primary}{\rightarrow}$ 11 $\textcolor{cai_primary}{\rightarrow}$ 14 & 0.000000 \\
        \hline
        \end{tabular}
        \caption*{(c) Attacker Strategy}
    \end{minipage}
    
    \vspace{0.5cm}
    
    \begin{minipage}{\textwidth}
        \centering
        \colorbox{cai_light}{%
            \parbox{0.85\textwidth}{%
                \centering
                \vspace{1.2ex}
                \textcolor{cai_dark}{\textbf{Game Equilibrium}}\\[0.5ex]
                \textcolor{cai_primary}{Defender can keep attacker success below:} \textbf{0.000000}\\[0.3ex]
                \textcolor{cai_primary}{Attacker can guarantee success probability of:} \textbf{0.000000}
                \vspace{1.2ex}
            }%
        }
        \caption*{(d) Game Equilibrium}
    \end{minipage}
    \caption{\textbf{Ground-Truth attack graph annotated by Humans in the media.guilded.gg domain}: This graph documents the real-world, manually performed security research process focused on the \texttt{/media/upload} webhook endpoint of \texttt{media.guilded.gg}. Starting with reconnaissance and authentication setup, the researcher tested a range of payloads, including malicious file uploads, XSS vectors, command and embed injections, and SSRF via the \texttt{avatar\_url} parameter. While many inputs were properly sanitized, the investigation uncovered a \textbf{URL format bypass} vulnerability where JavaScript-based URLs were allowed through encoding and Unicode tricks. This weakness enabled a URL confusion attack, crafting links like \texttt{https://facebook.com@github.com/...} that could mislead users and trigger unintended file downloads. The results led to the creation of a detailed security report and further experiments, including rate-limit testing via webhook spam.}
    \label{fig:media_human}
\end{figure}

\begin{figure}[H]
    \centering
    \includegraphics[width=\textwidth]{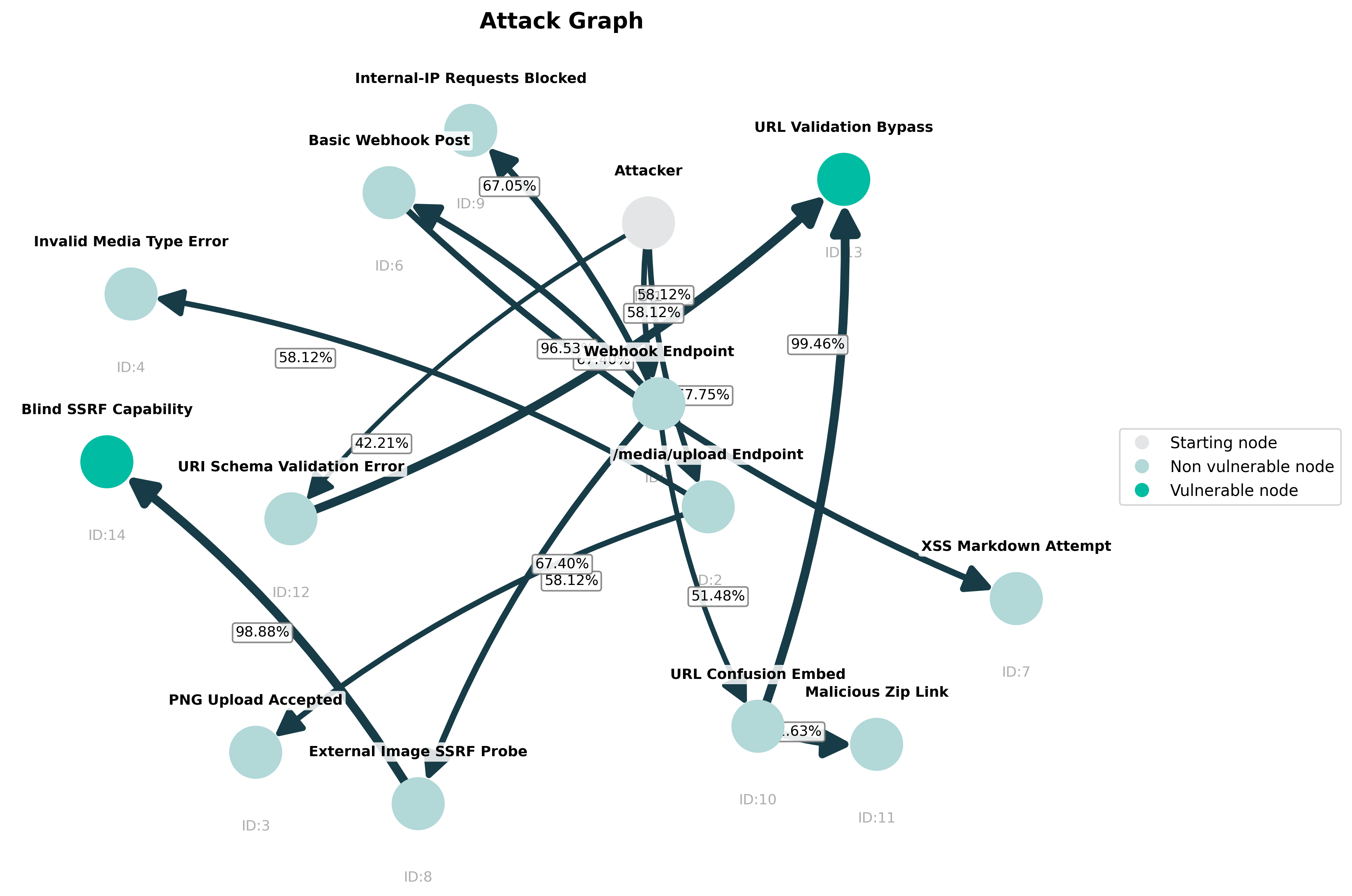}
    \caption*{(a) Attack Graph Example}

    \vspace{0.5cm}
\begin{minipage}{0.45\textwidth}
    \centering
    \renewcommand{\arraystretch}{1.4}
    \begin{tabular}{cc}
    \hline
     \textcolor{cai_primary}{\textbf{Node ID}} & \textcolor{cai_primary}{\textbf{Probability}} \\
    \hline
    8  & 0.423077 \\
    10 & 0.423077 \\
    12 & 0.153846 \\
    5  & 0.000000 \\
    \hline
    \end{tabular}
    \caption*{(b) Optimal Defense Strategy}
\end{minipage}
\hfill
\begin{minipage}{0.45\textwidth}
    \centering
    \renewcommand{\arraystretch}{1.4}
    \begin{tabular}{ccc}
    \hline
\textcolor{cai_primary}{\textbf{Path ID}} & \textcolor{cai_primary}{\textbf{Path Sequence}} & \textcolor{cai_primary}{\textbf{Probability}} \\
    \hline
    3 & 1 $\textcolor{cai_primary}{\rightarrow}$ 12 $\textcolor{cai_primary}{\rightarrow}$ 13  & 0.423077 \\
    1 & 1 $\textcolor{cai_primary}{\rightarrow}$ 5 $\textcolor{cai_primary}{\rightarrow}$ 8 $\textcolor{cai_primary}{\rightarrow}$ 14  & 0.288461 \\
    2 & 1 $\textcolor{cai_primary}{\rightarrow}$ 5 $\textcolor{cai_primary}{\rightarrow}$ 10 $\textcolor{cai_primary}{\rightarrow}$ 13  & 0.288461 \\
    \hline
    \end{tabular}
    \caption*{(c) Attacker Strategy}
\end{minipage}

\vspace{0.5cm}

\begin{minipage}{\textwidth}
    \centering
    \colorbox{cai_light}{%
        \parbox{0.85\textwidth}{%
            \centering
            \vspace{1.2ex}
            \textcolor{cai_dark}{\textbf{Game Equilibrium}}\\[0.5ex]
            \textcolor{cai_primary}{Defender can keep attacker success below:} \textbf{0.030220}\\[0.3ex]
            \textcolor{cai_primary}{Attacker can guarantee success probability of:} \textbf{0.030220}
            \vspace{1.2ex}
        }%
    }
    \caption*{(d) Game Equilibrium}
\end{minipage}

    \caption{Attack Graph and Equilibrium analysis for \texttt{gpt-4o}}
    \label{fig:media_gpt}
\end{figure}

\begin{figure}[H]
    \centering
    \includegraphics[width=\textwidth]{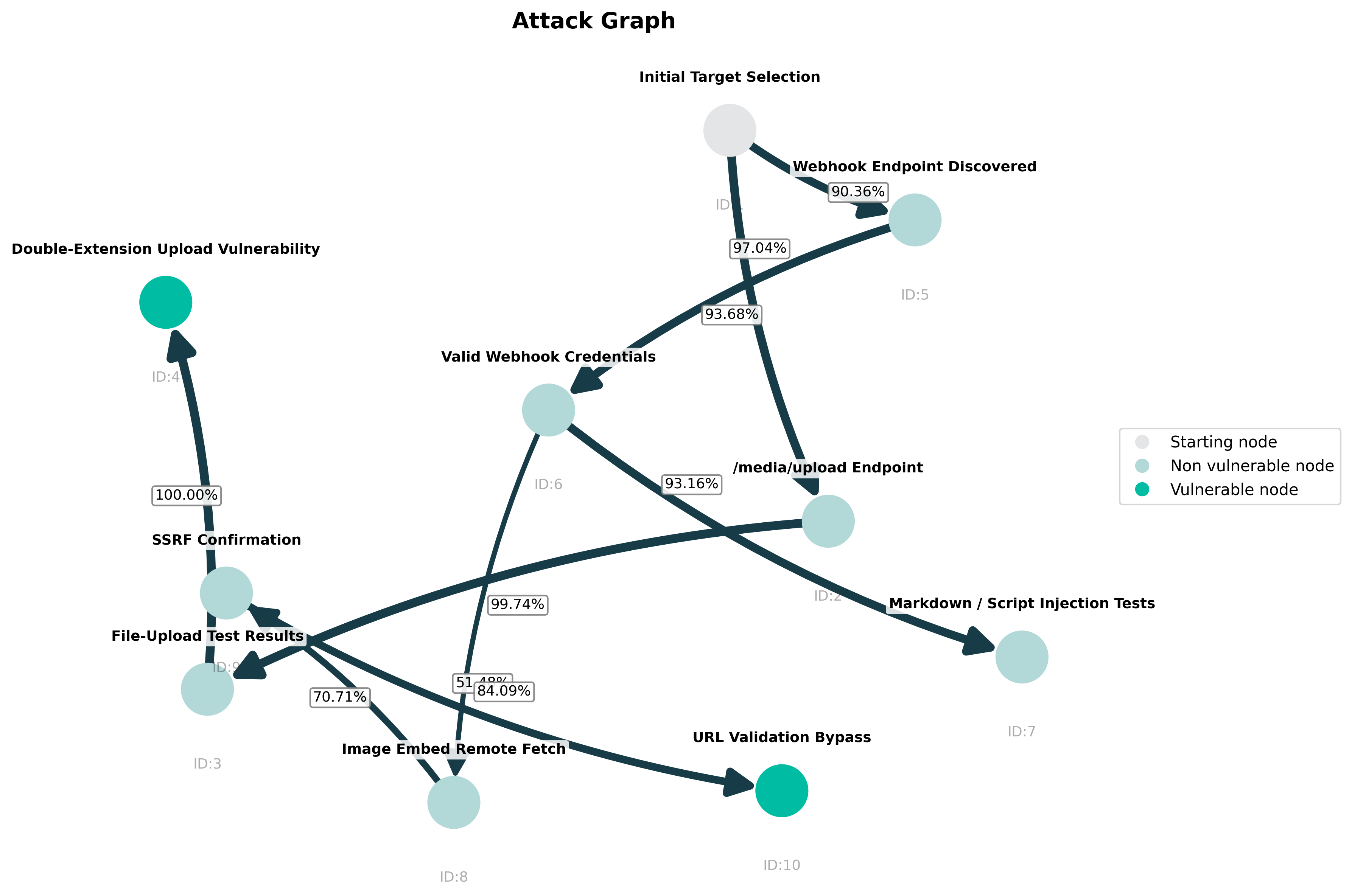}
    \caption*{(a) Attack Graph Example}

    \vspace{0.5cm}
\begin{minipage}{0.45\textwidth}
    \centering
    \renewcommand{\arraystretch}{1.4}
    \begin{tabular}{cc}
    \hline
     \textcolor{cai_primary}{\textbf{Node ID}} & \textcolor{cai_primary}{\textbf{Probability}} \\
    \hline
    9 & 0.551513 \\
    3 & 0.448487 \\
    2 & 0.000000 \\
    5 & 0.000000 \\
    6 & 0.000000 \\
    8 & 0.000000 \\
    \hline
    \end{tabular}
    \caption*{(b) Optimal Defense Strategy}
\end{minipage}
\begin{minipage}{0.45\textwidth}
    \centering
    \renewcommand{\arraystretch}{1.4}
    \begin{tabular}{ccc}
    \hline
\textcolor{cai_primary}{\textbf{Path ID}} & \textcolor{cai_primary}{\textbf{Path Sequence}} & \textcolor{cai_primary}{\textbf{Probability}} \\
    \hline
    1 & 1 $\textcolor{cai_primary}{\rightarrow}$ 2 $\textcolor{cai_primary}{\rightarrow}$ 3 $\textcolor{cai_primary}{\rightarrow}$ 4 & 0.551513 \\
    2 & 1 $\textcolor{cai_primary}{\rightarrow}$ 5 $\textcolor{cai_primary}{\rightarrow}$ 6 $\textcolor{cai_primary}{\rightarrow}$ 8 $\textcolor{cai_primary}{\rightarrow}$ 9 $\textcolor{cai_primary}{\rightarrow}$ 10 & 0.448487 \\
    \hline
    \end{tabular}
    \caption*{(c) Attacker Strategy}
\end{minipage}

\vspace{0.5cm}

\begin{minipage}{\textwidth}
    \centering
    \colorbox{cai_light}{%
        \parbox{0.85\textwidth}{%
            \centering
            \vspace{1.2ex}
            \textcolor{cai_dark}{\textbf{Game Equilibrium}}\\[0.5ex]
            \textcolor{cai_primary}{Defender can keep attacker success below:} \textbf{0.019259}\\[0.3ex]
            \textcolor{cai_primary}{Attacker can guarantee success probability of:} \textbf{0.019259}
            \vspace{1.2ex}
        }%
    }
    \caption*{(d) Game Equilibrium}
\end{minipage}

    \caption{Attack Graph and Equilibrium analysis for \texttt{o3}}
    \label{fig:media_o3}
\end{figure}

\begin{figure}[H]
    \centering
    \includegraphics[width=\textwidth]{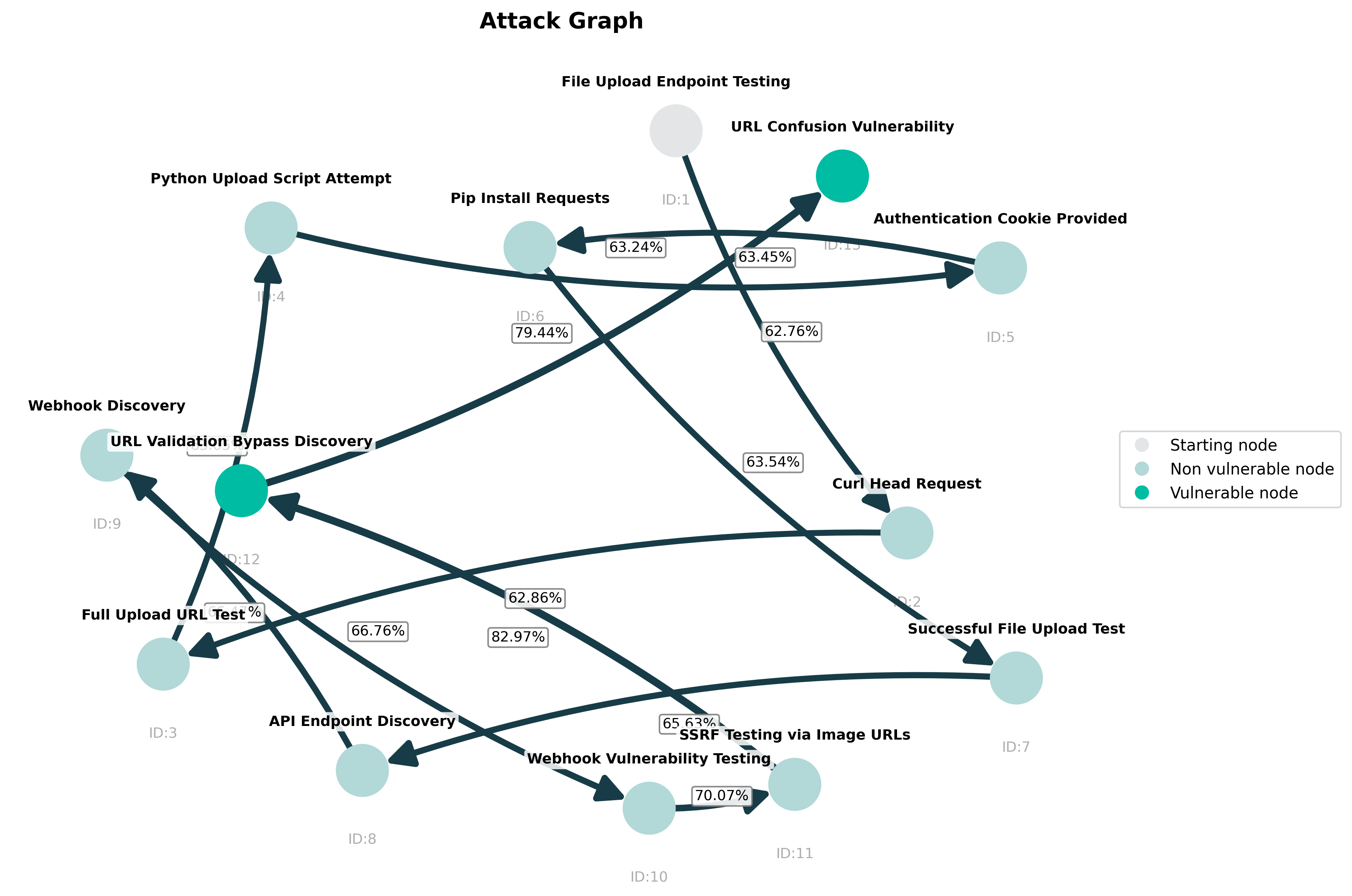}
    \caption*{(a) Attack Graph Example}

    \vspace{0.5cm}

    \begin{minipage}{\textwidth}
        \centering
        \renewcommand{\arraystretch}{1.4}
        \begin{tabular}{cc}
        \hline
         \textcolor{cai_primary}{\textbf{Node ID}} & \textcolor{cai_primary}{\textbf{Probability}} \\
        \hline
        11 & 1.000000 \\
        2-10 & 0.000000 \\
        \hline
        \end{tabular}
        \caption*{(b) Optimal Defense Strategy}
    \end{minipage}
    
    \vspace{0.5cm}
    
    \begin{minipage}{\textwidth}
        \centering
        \renewcommand{\arraystretch}{1.4}
        \begin{tabular}{cp{9cm}c}
        \hline
\textcolor{cai_primary}{\textbf{Path ID}} & \textcolor{cai_primary}{\textbf{Path Sequence}} & \textcolor{cai_primary}{\textbf{Probability}} \\
        \hline
        1 & 1 $\textcolor{cai_primary}{\rightarrow}$ 2 $\textcolor{cai_primary}{\rightarrow}$ 3 $\textcolor{cai_primary}{\rightarrow}$ 4 $\textcolor{cai_primary}{\rightarrow}$ 5 $\textcolor{cai_primary}{\rightarrow}$ 6 $\textcolor{cai_primary}{\rightarrow}$ 7 $\textcolor{cai_primary}{\rightarrow}$ 8 $\textcolor{cai_primary}{\rightarrow}$ 9 $\textcolor{cai_primary}{\rightarrow}$ 10 $\textcolor{cai_primary}{\rightarrow}$ 11 $\textcolor{cai_primary}{\rightarrow}$ 12  & 1.000000 \\
        2 & 1 $\textcolor{cai_primary}{\rightarrow}$ 2 $\textcolor{cai_primary}{\rightarrow}$ 3 $\textcolor{cai_primary}{\rightarrow}$ 4 $\textcolor{cai_primary}{\rightarrow}$ 5 $\textcolor{cai_primary}{\rightarrow}$ 6 $\textcolor{cai_primary}{\rightarrow}$ 7 $\textcolor{cai_primary}{\rightarrow}$ 8 $\textcolor{cai_primary}{\rightarrow}$ 9 $\textcolor{cai_primary}{\rightarrow}$ 10 $\textcolor{cai_primary}{\rightarrow}$ 11 $\textcolor{cai_primary}{\rightarrow}$ 12 $\textcolor{cai_primary}{\rightarrow}$ 13  & 0.000000 \\
        \hline
        \end{tabular}
        \caption*{(c) Attacker Strategy}
    \end{minipage}

    \vspace{0.5cm}
    
    \begin{minipage}{\textwidth}
        \centering
        \colorbox{cai_light}{%
            \parbox{0.85\textwidth}{%
                \centering
                \vspace{1.2ex}
                \textcolor{cai_dark}{\textbf{Game Equilibrium}}\\[0.5ex]
                \textcolor{cai_primary}{Defender can keep attacker success below:} \textbf{0.000000}\\[0.3ex]
                \textcolor{cai_primary}{Attacker can guarantee success probability of:} \textbf{0.000000}
                \vspace{1.2ex}
            }%
        }
        \caption*{(d) Game Equilibrium}
    \end{minipage}
    \caption{Attack Graph and Equilibrium analysis for \texttt{grok--4}}
    \label{fig:media_grok}
\end{figure}


\begin{figure}[H]
    \centering
    \includegraphics[width=\textwidth]{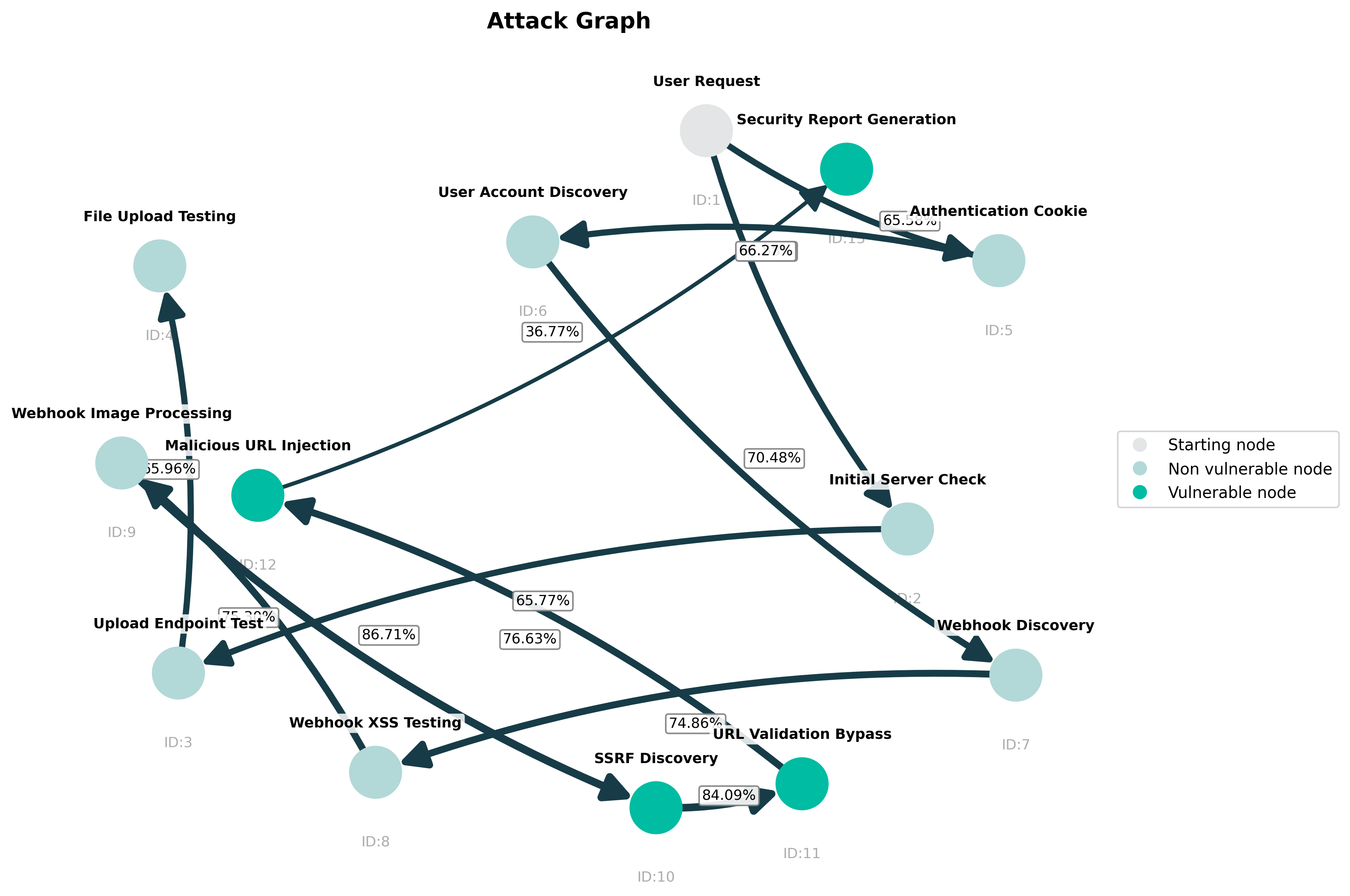}
    \caption*{(a) Attack Graph Example}

    \vspace{0.5cm}

    \begin{minipage}{\textwidth}
        \centering
        \renewcommand{\arraystretch}{1.4}
        \begin{tabular}{cc}
            \hline
             \textcolor{cai_primary}{\textbf{Node ID}} & \textcolor{cai_primary}{\textbf{Probability}} \\
            \hline
            9 & 1.000000 \\
            5-8 & 0.000000 \\
            \hline
        \end{tabular}
        \caption*{(b) Optimal Defense Strategy}
    \end{minipage}

    \vspace{0.5cm}

    \begin{minipage}{\textwidth}
        \centering
        \renewcommand{\arraystretch}{1.4}
        \begin{tabular}{ccc}
            \hline
    \textcolor{cai_primary}{\textbf{Node ID}} & \textcolor{cai_primary}{\textbf{Probability}} \\
            \hline
            1 & 1 $\textcolor{cai_primary}{\rightarrow}$ 5 $\textcolor{cai_primary}{\rightarrow}$ 6 $\textcolor{cai_primary}{\rightarrow}$ 7 $\textcolor{cai_primary}{\rightarrow}$ 8 $\textcolor{cai_primary}{\rightarrow}$ 9 $\textcolor{cai_primary}{\rightarrow}$ 10 $\textcolor{cai_primary}{\rightarrow}$ 11 $\textcolor{cai_primary}{\rightarrow}$ 12 $\textcolor{cai_primary}{\rightarrow}$ 13 & 1.000000 \\
            2 & 1 $\textcolor{cai_primary}{\rightarrow}$ 5 $\textcolor{cai_primary}{\rightarrow}$ 6 $\textcolor{cai_primary}{\rightarrow}$ 7 $\textcolor{cai_primary}{\rightarrow}$ 8 $\textcolor{cai_primary}{\rightarrow}$ 9 $\textcolor{cai_primary}{\rightarrow}$ 10 $\textcolor{cai_primary}{\rightarrow}$ 11 $\textcolor{cai_primary}{\rightarrow}$ 12 & 0.000000 \\
            3 & 1 $\textcolor{cai_primary}{\rightarrow}$ 5 $\textcolor{cai_primary}{\rightarrow}$ 6 $\textcolor{cai_primary}{\rightarrow}$ 7 $\textcolor{cai_primary}{\rightarrow}$ 8 $\textcolor{cai_primary}{\rightarrow}$ 9 $\textcolor{cai_primary}{\rightarrow}$ 10 $\textcolor{cai_primary}{\rightarrow}$ 11 & 0.000000 \\
            4 & 1 $\textcolor{cai_primary}{\rightarrow}$ 5 $\textcolor{cai_primary}{\rightarrow}$ 6 $\textcolor{cai_primary}{\rightarrow}$ 7 $\textcolor{cai_primary}{\rightarrow}$ 8 $\textcolor{cai_primary}{\rightarrow}$ 9 $\textcolor{cai_primary}{\rightarrow}$ 10 & 0.000000 \\
            \hline
        \end{tabular}

        \caption*{(c) Attacker Strategy}
    \end{minipage}

    \vspace{0.5cm}

    \begin{minipage}{\textwidth}
        \centering
        \colorbox{cai_light}{%
            \parbox{0.85\textwidth}{%
                \centering
                \vspace{1.2ex}
                \textcolor{cai_dark}{\textbf{Game Equilibrium}}\\[0.5ex]
                \textcolor{cai_primary}{Defender can keep attacker success below:} \textbf{0.000000}\\[0.3ex]
                \textcolor{cai_primary}{Attacker can guarantee success probability of:} \textbf{0.000000}
                \vspace{1.2ex}
            }%
        }
        \caption*{(d) Game Equilibrium}
    \end{minipage}

    \caption{Attack Graph and Equilibrium analysis for \texttt{claude--sonnet--4}}
    \label{fig:media_claude}
\end{figure}

\begin{figure}[H]
    \centering
    \includegraphics[width=\textwidth]{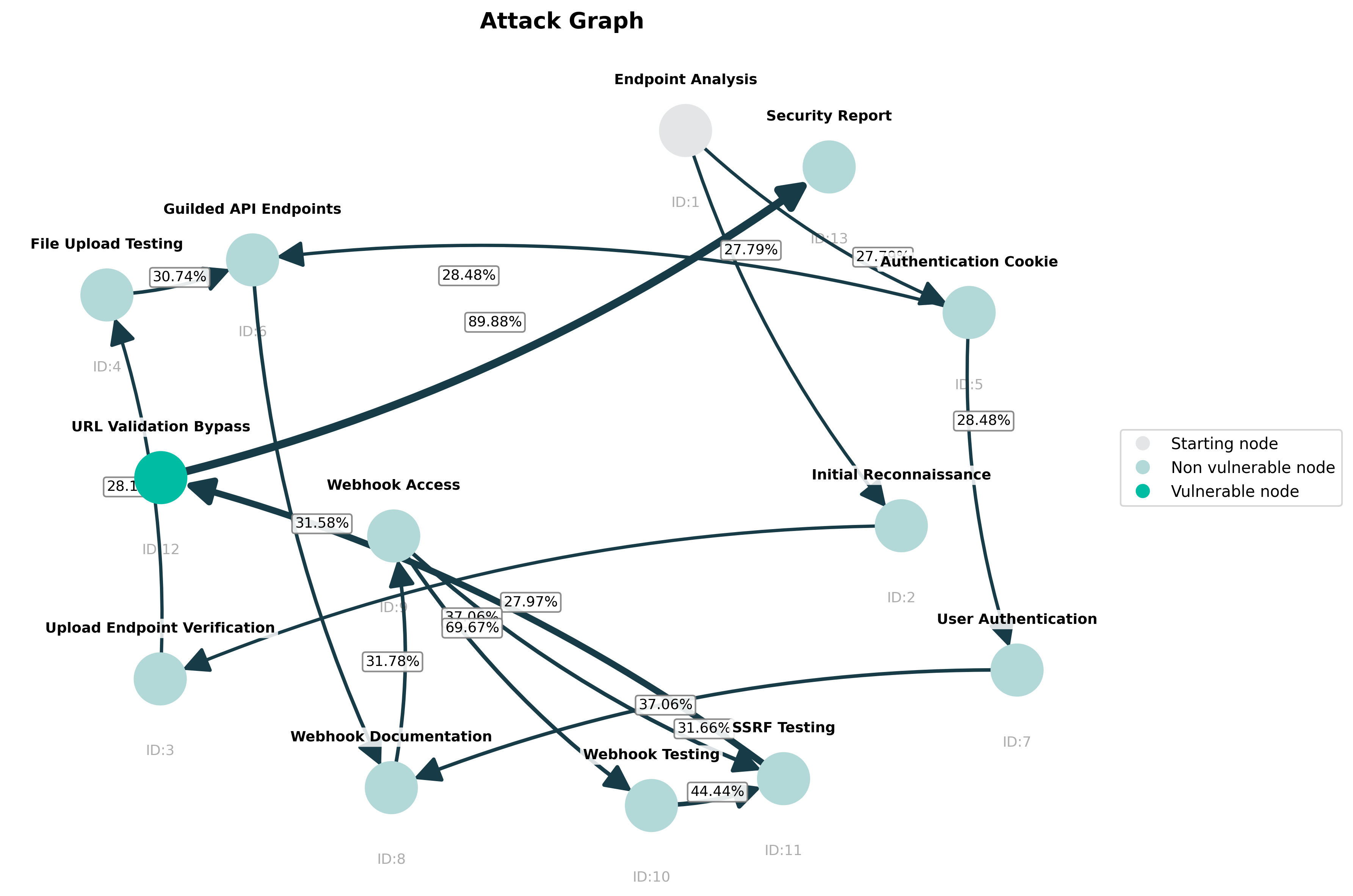}
    \caption*{(a) Attack Graph Example}

    \vspace{0.5cm}

    \begin{minipage}{0.45\textwidth}
        \centering
        \renewcommand{\arraystretch}{1.4}
        \begin{tabular}{cc}
        \hline
        \textcolor{cai_primary}{\textbf{Node ID}} & \textcolor{cai_primary}{\textbf{Probability}} \\
        \hline
        8 & 1.000000 \\
        2 & 0.000000 \\
        3 & 0.000000 \\
        \hline
        \end{tabular}
        \caption*{(b) Optimal Defense Strategy}
    \end{minipage}

    \hfill

    \begin{minipage}{0.52\textwidth}
        \centering
        \renewcommand{\arraystretch}{1.4}
        \begin{tabular}{ccc}
        \hline
        \textcolor{cai_primary}{\textbf{Path ID}} & \textcolor{cai_primary}{\textbf{Path Sequence}} & \textcolor{cai_primary}{\textbf{Probability}} \\
        \hline
        1 & 1 $\textcolor{cai_primary}{\rightarrow}$ 2 $\textcolor{cai_primary}{\rightarrow}$ 3 $\textcolor{cai_primary}{\rightarrow}$ 8 $\textcolor{cai_primary}{\rightarrow}$ 11 & 1.000000 \\
        \hline
        \end{tabular}
        \caption*{(c) Attacker Strategy}
    \end{minipage}

    \vspace{0.5cm}

    \begin{minipage}{\textwidth}
        \centering
        \colorbox{cai_light}{%
            \parbox{0.85\textwidth}{%
                \centering
                \vspace{1.2ex}
                \textcolor{cai_dark}{\textbf{Game Equilibrium}}\\[0.5ex]
                \textcolor{cai_primary}{Defender can keep attacker success below:} \textbf{0.000000}\\[0.3ex]
                \textcolor{cai_primary}{Attacker can guarantee success probability of:} \textbf{0.000000}
                \vspace{1.2ex}
            }%
        }
        \caption*{(d) Game Equilibrium}
    \end{minipage}

    \caption{Attack Graph and Equilibrium analysis for \texttt{alias0}}
    \label{fig:media_alias0}
\end{figure}

\end{document}